\documentclass[usenatbib, useAMS]{mn2e}
\input epsf

\newcommand{\lapprox}{\lower0.8ex\hbox{$\buildrel <\over\sim$}}
\newcommand{\gapprox}{\lower0.8ex\hbox{$\buildrel >\over\sim$}}
\usepackage{epsfig}
\usepackage{epsf}
\usepackage{times}
\input epsf
\def \ul #1:#2:{$^{+#1}_{-#2}$}

\title[3-D photoionization modelling of NGC~3918]{Three-dimensional photoionization modelling of the planetary nebula NGC~3918}
\author[Ercolano et al. ] {B. Ercolano$^1$, C. Morisset$^{2,3}$, M. J. Barlow$^1$, P. J. Storey$^1$, X.-W. Liu$^{1,4}$\\
$^1$University College London, Gower Street, London WC1E 6BT, UK\\
$^2$Laboratoire d'Astrophysique de Marseille, CNRS, BP 8, 13376 Marseille Cedex 12, France\\
$^3$Current address: Instituto de Astronomia, Universidad Nacional Aut\'onoma de México, M\'exico, D.F. 04510, M\'exico\\
$^4$Current address: Department of Astronomy, Peking University, Beijing 100871, P. R. China
}

\date{Received:}
\begin{document}
\maketitle
\begin{abstract}
\noindent
The three-dimensional Monte Carlo photoionization code Mocassin has been applied to construct a realistic model of the planetary nebula NGC~3918. Three different geometric models were tried, the first being the biconical density distribution already used by \citet{clegg87}. In this model the nebula is approximated by a biconical structure of enhanced density, embedded in a lower density spherical region. Spindle-like density distributions were used for the other two models (models~A and B). Model~A used a mass distribution slightly modified from one of Mellema's (1996) hydrodynamical models that had already been adopted by \citet{corradi99} for their observational analysis  of NGC~3918. Our spindle-like model~B instead used an analytical expression to describe the shape of the inner shell of this object as consisting of an ellipsoid embedded in a sphere. 

The effects of the interaction of the diffuse fields coming from two adjacent regions of different densities were investigated. These are found to be non-negligible, even for the relatively uncomplicated case of a biconical geometry. We found that the ionization structure of low ionization species near the boundaries is particularly affected. 

It is found that all three models provided acceptable matches to the integrated nebular optical and ultraviolet spectrum. Large discrepancies were found between all of the model predictions of infrared fine-structure line fluxes and {\it ISO~SWS} measurements. This was found to be largely due to an offset of $\approx$14~arcsec from the centre of the nebula that affected all of the {\it ISO} observations of NGC~3918. 

For each model, we also produced projected emission-line maps and position-velocity diagrams from synthetic long-slit spectra, which could be compared to recent {\it HST} images and ground-based long-slit echelle spectra. This comparison showed that spindle-like model~B provided the best match to the observations. Although the integrated emission line spectrum of NGC~3918 can be reproduced by all three of the three-dimensional models investigated in this work, the capability of creating projected emission-line maps and position-velocity diagrams from synthetic long-slit spectra was found to be crucial in allowing us to constrain the structure of this object. 

\end{abstract}


\section{Introduction}
\label{sec:ngc3918intro}

The southern planetary nebula NGC~3918 (PN G294.6+04.7) is a very well known and widely studied object. A detailed study, based on UV, optical and radio 
observations, was presented by \citet[][from now on C87]{clegg87}. The morphological and  kinematical information available at the time of their work, was, however,  
very limited, and, based on this, they constructed a photoionization model, using the Harrington code \citep[e.g.][]{harrington82}, assuming a biconical geometry for a nebula seen almost {\it pole-on}. A two-dimensional representation 
of their model is shown in Figure~10 of C87. They had first tried a spherical model, deriving the radial hydrogen density distribution from
the average radial intensity profile in the H$\beta$ map they used, which had been obtained at the Boyden Observatory, South Africa, in 1973 by Drs.~K.~Reay
and S.~P.~Worswick. The Zanstra temperature they derived for the central star was $117,000$\,K, corresponding to a luminosity of $4900\,L_{\odot}$, for an adopted distance of 1.5~kpc. This model did not succeed in reproducing the line strengths from some of the high ionization species observed, such as, for example,  Ne~{\sc v}, O~{\sc v} 
and O~{\sc iv}. They interpreted this as an indication that the nebula could be optically thin in some directions as seen from the star; 
this led to the formulation of a biconical model. The presence of an optically thin phase required an upward correction to the original
Zanstra temperature and, in the final model, they adopted the ionizing spectrum described by a non-LTE model atmosphere for a central star having an effective temperature of 140,000~K, a surface gravity of ${\rm log}~g~=~6.5$ and and a photospheric ${\rm He}/{\rm H}~=~0.10$ by number. This model atmosphere was calculated by C87 using the program of \citet{mihalas75}. The resulting nebular model seemed to 
reproduce the main spectroscopic features observed. 

\begin{figure}
\begin{center}
\psfig{file=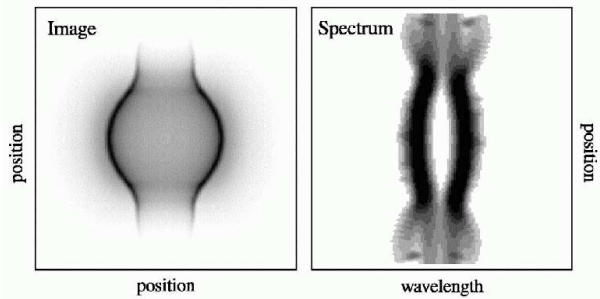, height=40mm, width=80mm}
\caption[Hydrodynamic model by \citet{mellema96} used for NGC~3918]
{Model image (left) and long axis synthetic slit spectrum (right) from the hydrodynamical models of \citet{mellema96}. Figure adapted by C99 from Mellema's original thesis work.}
\label{fig:mellemadist}
\end{center}
\end{figure}

Present observational data, such as for example images of NGC~3918 taken by the {\it HST} (see Figure~\ref{fig:hstimage}) and the echellograms obtained in several emission lines by \citet[][from now on C99]{corradi99} (Figure~\ref{fig:longslit}), show, however, that a biconical model is inconsistent with the spatio-kinematical structure of this planetary nebula. C99 presented an analysis of optical images and high resolution long slit spectra of three planetary nebulae, including 
NGC~3918. They concluded that the large scale structure of this object consists of a bright inner shell of roughly elliptical shape, from which two fainter 
protrusions extend in the polar directions, giving what was described in their paper as an overall spindle-like appearance. This inner shell, which has a size of 
$12'' \times\,20''$, measured along its major and minor axes, is surrounded by an elliptical shell with a diameter of $16''$. From the images and from the long 
slit spectra they obtained the basic kinematical, geometrical and orientation parameters of the inner shell. They adopted a hydrodynamical model  by 
G.~Mellema (1996) to reproduce, at least qualitatively, the observations. The model which gave the best fit was one, from Mellema's set, which posits an
initial mass distribution strongly enhanced in the equatorial region, with a density contrast between the equatorial and the polar regions as large as 10 (see Figure~\ref{fig:mellemadist}). The effects of an expanding shock driven by a strong wind would give a spindle-like structure similar to the one observed in the inner shell of 
NGC~3918. C99 derived the inclination and the kinematical parameters of the inner shell by using the spatio-kinematical model of \citet{solf85} (which was also used to 
obtain $H\alpha$ position-velocity diagrams, as well as the shape of the inner shell) to match the observational data. Their final model still showed some deviation from the observations, particularly in the long-axis position-velocity plot, which they attributed to simplified assumptions in the spatio-kinematical model. 
Photoionization calculations, however, were not carried out in their work, and therefore no comparison with the observed spectrum was available to them. 

Given the large amount of observational data available for this object and the existence of the two different models described above, NGC~3918 seemed  an excellent candidate for a detailed three-dimensional photoionization study using the Mocassin code described by \citet{ercolano02} and \citet{ercolanoI}. Three photoionization models were constructed, using different 
density distribution descriptions, in order to try to reproduce the main spectroscopic features, as well as the projected maps published by C99. The models used a 23$\times$23$\times$23 Cartesian grid, with the ionizing source being placed in a corner in order to utilize the symmetry of the geometries used. The complete model nebulae were, therefore, contained in 45$\times$45$\times$45 cubic grids. All the grid cells have the same size. Velocity fields were then applied to the final converged grids in order to produce position-velocity diagrams to compare with the observations. In Section~2 the biconical density distribution model of C87 is described and the results obtained for it with Mocassin are presented and compared with those of C87, who used the one-dimensional Harrington code. A consistency test for the diffuse radiation field is also carried out in this section, in order to investigate the effects of discontinuities in the diffuse field transport in one-dimensional codes. The results obtained from the spindle-like models of NGC~3918 are presented in Section~3. Discrepancies were found between the predictions of all the models and the {\it ISO~SWS} measurements of the infrared fine-structure lines; the possible reasons of these discrepancies are discussed in Section~4.

\section{The Biconical Density Distribution Model}

\begin{table}
\begin{center}
\caption{Input parameters for the NGC~3918 biconical model. }
\begin{tabular}{lc|lc}
\multicolumn{4}{c}{} \\
\hline
$L_*$	& 6900 $L_{\odot}$ &He/H	& 0.107	\\
$R_{in}$ 	& 0.0 pc	& C/H	& 8.0$\times$10$^{-4}$  \\
$R_{out}$& 0.106 pc 	& N/H	& 1.5$\times$10$^{-4}$  \\
$T_{eff}$ & 140,000 K	& O/H	& 5.0$\times$10$^{-4}$  \\
$R_1$	& 0.0333 pc	& Ne/H	& 1.2$\times$10$^{-4}$  \\
$n$	& 2.3		& Mg/H	& 1.4$\times$10$^{-5}$	\\
$f$	& 3		& Si/H	& 1.0$\times$10$^{-5}$\\
$\Phi$	& 45$^\circ$	& S/H	& 1.6$\times$10$^{-5}$	\\
$N_0$	& 2381 cm$^{-3}$& Ar/H	& 2.0$\times$10$^{-6}$	\\
$a/b$	&  1		& Fe/H	& 3.7$\times$10$^{-7}$	\\
\hline
\label{tab:cleggpar}
\end{tabular}
\\
\small{The abundances are given by number, relative to H}
\end{center}
\end{table}

The first nebular model to be run was the pole-on biconical distribution described by C87. This consisted of an optically thick biconical structure embedded into an optically thin sphere. The bicones are aligned along the polar axis of the nebula, which describes an angle of 15$^{\circ}$ with the line of sight. Exactly the same parameters were used for the central star and, 
apart from helium, argon and iron, 
the same nebular abundances  as in their model; these are summarised in Table~\ref{tab:cleggpar}. The main reason for using exactly the same parameters
is to be able to distinguish between the causes of any discrepancies between our model results and the original C87 model results. 
However, in the case of helium the empirical value derived by C87 was used instead of their model value, since it was commented in their paper that this is more reliable. In the case of argon and iron their empirical abundances also had to be used, since these elements were not included in their code (that of J.P. Harrington) and, therefore, photoionization model results for them did not exist. 
The biconical radial density law that was adopted by C87, and which was mapped onto the three-dimensional Cartesian grid used in the current model, was constructed  by them by deconvolving the observed nebular surface brightness, using the formalism of \citet{harrington83}, and has the form
\begin{equation}
N_H(R)=N_0\,(R/R_1)^n\,{\rm exp}[-0.5(R/R_1)^2]
\end{equation}
where $N_H(R)$ is the hydrogen number density at distance $R$ from the centre of the nebula, $N_0$ is a normalization constant and $R_1$ and $n$ are 
parameters which may be varied. The values used for these parameters are also given in Table~\ref{tab:cleggpar}, together with $\Phi$, the half angle of the density enhanced cones, the density enhancement factor between the two regions, $f$, and the axial ratio, $a/b$. The optical depths at 1. Ryd, calculated at the outer edge of the nebula in the polar (optically thick) and equatorial (optically thin) directions are 29.4 and 0.94, respectively. 

\subsection{The Biconical Density Distribution Model: Results}
\label{sub:biconicalresults}

\renewcommand{\baselinestretch}{0.9}
\begin{table*}
\begin{center}
\caption{Observed and predicted emission lines fluxes for NGC~3918 and predictions for the C87 biconical models. }
\begin{tabular}{lcccccccc}
\multicolumn{9}{c}{}\\ 
\hline
Line			& \multicolumn{3}{c}{This Work}  & \multicolumn{3}{c}{Clegg et al.} & Observed & Ref	\\
\cline{2-7}
				& Thin	& Thick	& Total	& Thin	& Thick	& Total	& 	&	\\
\hline
H$\beta$/10$^{-10}$$\frac{{\rm erg}}{{\rm cm}^{2}{\rm s}}$ 
				& 0.45	& 1.88	& 2.33	& 0.45	& 1.86	& 2.32	& 2.29 	& b	\\
H$\beta$ 4861			& 100	& 100	& 100	& 100 	& 100 	& 100	& 100 	& -	\\
He~{\sc i} 5876     		& 0.08	& 8.5 	& 8.5	& 0.06	& 7.7	& 7.8	& 10.81 & a	\\
He~{\sc ii} 4686		& 24	& 30	& 54	& 22	& 26	& 48	& 41.65	& a	\\
C~{\sc ii}$]$ 2326		& 0.21	& 57 	& 57	& 0.72	& 81	& 82	& 31.24	& a 	\\
C~{\sc ii} 4267			& 0.004	& 0.32	& 0.32	& 0.006	& 0.32	& 0.32	& 0.50 	& a	\\
$[$C~{\sc ii}$]$ 157.74$\mu$m	& 0.0	& 0.35	& 0.35	& -	& -	& -	& 0.21	& c	\\
C~{\sc iii}$]$ 1908 		& 33	& 577	& 610	& 58.7	& 577	& 637	& 492.5	& a	\\
C~{\sc iii} 4648 		& 0.10	& 0.50	& 0.61 	& 0.10	& 0.42	& 0.52	& 0.42	& a 	\\
C~{\sc iii}$]$ 2297		& 1.11	& 5.9	& 7.0	& 1.0	& 4.9	& 5.9	& 5:	& e	\\		
C~{\sc iv} 1550$^a$		& 596	& 964	& 1560	& 659	& 779	& 1439	& 457.6	& a	\\
C~{\sc i
v} 4658			& 0.25	& 0.16	& 0.40 	& 0.24	& 0.16	& 0.40	& 0.33	& a	\\ 		
$[$N~{\sc i}$]$ 5199		& 0.0	& 0.21	& 0.21	& -	& -	& -	& 0.63	& a	\\ 
$[$N~{\sc ii}$]$ 6584		& 0.06	& 73	& 73	& 0.2	& 73	& 73	& 84	& a	\\	
$[$N~{\sc iii}$]$ 1751		& 2.0	& 22	& 24	& 4.6	& 25	& 29	& 26.7	& a	\\ 
$[$N~{\sc iii}$]$ 57.3 $\mu$m 	& 0.64	& 12	&  13	& 1.0	& 12	& 13	& 13.5	& c	\\
N~{\sc iv}$]$ 1486		& 19	& 26	& 45	& 31	& 26	& 57	& 46.4	& a	\\
 N~{\sc v} 1240$^a$		& 42	& 22	& 64	& 47	& 19	& 66	& 41.4	& a	\\
$[$O~{\sc i}$]$ 6300		& 0.0	& 2.7	& 2.7	& 0.0	& 0.56	& 0.56	& 5.24	& a	\\
$[$O~{\sc i}$]$ 63.12 $\mu$m	& 0.0	& 0.23	& 0.23	& -	& -	& -	& 6.72	& c	\\
$[$O~{\sc i}$]$ 145.5 $\mu$m 	& 0.0	& 0.018	& 0.018	& -	& -	& -	& 0.23	& c	\\		
$[$O~{\sc ii}$]$ 3727		& 0.08	& 104	& 104	& 0.2	& 89	& 89	& 94.5	& a	\\
$[$O~{\sc iii}$]$ 51.8 $\mu$m 	& 2.8	& 90	& 93	& 4.0	& 94	& 98	& 88.6	& c	\\
$[$O~{\sc iii}$]$ 88.4 $\mu$m 	& 0.91	& 16	& 17	& 1.3	& 17	& 19	& 27.51	& c	\\ 
$[$O~{\sc iii}$]$ 5007		& 40	& 1543	& 1583	& 63	& 1539	& 1602	& 1657	& a	\\
$[$O~{\sc iii}$]$ 4363		& 0.85	& 19	& 20	& 1.4	& 18.9	& 20	& 21.6	& a	\\
$[$O~{\sc iv}$]$ 25.9 $\mu$m	& 190	& 161	& 352	& 274	& 177	& 451	& 89	& d	\\
O~{\sc iv}$]$ 1402		& 16	& 14	& 30	& 28	& 16	& 43	& 48.9	& a	\\
O~{\sc iv} 1342			& 1.4	& 0.83	& 2.2	& 1.0	& 0.6	& 1.6	& 2:	& e 	\\
$[$Ne~{\sc ii}$]$ 12.8 $\mu$m 	& 0.001	& 0.94	& 0.94	&  0.0	& 0.77	& 0.77	& 12:	& f	\\
$[$Ne~{\sc iii}$]$ 15.6 $\mu$m 	& 1.8	& 130	& 132	& 1.3	& 167	& 168	& 46.3	& d	\\
$[$Ne~{\sc iii}$]$ 3869		& 4.1	& 161	& 165	& 3	& 172	& 175	& 135.5	& a	\\
$[$Ne~{\sc iv}$]$ 2423		& 35	& 42	& 77	& 50	& 52	& 102	& 132	& e	\\
$[$Ne~{\sc v}$]$ 3426		& 75	& 45	& 119	& 70	& 40	& 110	& 80	& e	\\
$[$Ne~{\sc v}$]$ 14.3 $\mu$m 	& 179	& 107	& 286	& 18	& 13	& 31	& 19.6	& d 	\\
$[$Ne~{\sc v}$]$ 24.3 $\mu$m 	& 156	& 64	& 220	& 30	& 19	& 49	& 23.1	& d	\\
$[$Ne~{\sc vi}$]$ 7.6 $\mu$m 	& 49	& 22	& 70	& 11	& 4.9	& 15.8	& -	& -	\\
Mg~{\sc i}$]$ 4565		& 0.00	& 0.73	& 0.73	& 0.00	& 0.28	& 0.28	& 0.13	& a	\\
Mg~{\sc ii} 2800$^a$		& 0.26 	& 78	& 78	& 0.01	& 43	& 43	&$\leq$1& e	\\
$[$Mg~{\sc iv}$]$ 4.49 $\mu$m 	& 0.42	& 3.1	& 3.5	& 0.96	& 4.4	& 5.4	& -	& -	\\
$[$Mg~{\sc v}$]$ 2783	 	& 2.2	& 2.4	& 4.6	& 3.2	& 2.6	& 5.8	& 5.6	& e	\\
$[$Mg~{\sc v}$]$ 5.6 $\mu$m 	& 4.5	& 6.3	& 10.9	& 4.9	& 5.4	& 10.3	& -	& -	\\
Si~{\sc iii}$]$ 1887		& 0.97	& 15	& 16	& 1.2	& 12.5	& 13.7	& 8	& e	\\
Si~{\sc iv} 1399$^a$		& 2.7	& 7.9	& 11	& 2.8	& 5.4	& 8.2	& 9	& e	\\
$[$S~{\sc ii}$]$ 4070		& 0.0	& 2.1	& 2.1	& 0.0	& 2.8	& 2.8	& 2.693	& a	\\
$[$S~{\sc iii}$]$ 18.7 $\mu$m	& 0.03	& 38	& 38	& 0.31	& 38	& 38	& 8.7	& d	\\
$[$S~{\sc iii}$]$ 33.6 $\mu$m	& 0.02	& 16	& 16	& 0.2	& 15.6	& 16	& - 	& -	\\
$[$S~{\sc iii}$]$ 6312		& 0.006	& 3.7	& 3.7	& 0.005	& 3.4	& 3.4	& 2.27	& a	\\
$[$S~{\sc iv}$]$ 10.5$\mu$m	& 7.23	& 207	& 214	& 24	& 210	& 234	& 35.0	& d	\\
\hline
\label{tab:clegglines}
\end{tabular}
\end{center}
\small{$^a$ Attenuated by dust absorption.\\           
           $^b$~references: a:~\citet{tsamis02}; b:~\citet{cahn92}; c:~\citet{liu01}; d:~\citet{bower01}; e:~C87; f:~\citet{pottasch86}.}
\end{table*}
\renewcommand{\baselinestretch}{1.5}

\renewcommand{\baselinestretch}{1.2}
\begin{table}
\begin{center}
\caption{Diagnostic ratios for the electron density, $N_e$ and electron temperature, $T_e$ (C87 biconical model).}
\begin{tabular}{lccccc}
\multicolumn{6}{c}{} \\
\hline
Ion$^a$		& Lines 	& Ratio	 	& Ratio 	& Ratio		& obs. 	\\
		& ({\AA})	& (Moc.)	& (C87)		& (obs.) 	& ref.$^d$\\	 
\hline
		&		& $N_e$		&		&		&	\\
\hline
Mg{\sc i}	& 4562/4571	& 0.54 		& 0.42:		& 0.28 	 	& 1	\\
S~{\sc ii}	& 6716/6731	& 0.72		& 0.70		& 0.57 		& 1	\\
O~{\sc ii}	& 3726/3729	& 1.63		& 1.61		& 1.98 		& 2	\\
O~{\sc ii}	& 7325/3727$^b$	& 0.071		& -		& 0.078 	& 1	\\
C~{\sc iii}	& 1906/1909	& 1.34		& 1.36		& 1.31 		& 2	\\
Si~{\sc iii}	& 1883/1892	& 1.28		& 1.30		& 1.37 		& 2	\\
Ne~{\sc iv}	& 2421/2424	& 0.94		& 0.93		& 1.00 		& 2	\\
N~{\sc iv}	& 1483/1486	& 1.50		& 1.87		& 1.85 		& 2	\\			
O~{\sc iv}	& 1401/1404	& 1.21		& 1.14		& 1.50 		& 2	\\
\hline
		&		& $T_e$		&		&		&	\\
\hline
N~{\sc ii}	& 5755/6584	& 0.022		& 0.025		& 0.021 	& 1	\\
S~{\sc ii}	& 4073/6724$^c$	& 0.17		&-		& 0.34	 	& 1 	\\
C~$^{2+}$	& 4267/1908	& 0.0005	& 0.0005	& 0.0008 	& 2	\\
S~{\sc iii}	& 6312/9532	& 0.032		& 0.029		& 0.032 	& 2	\\
O~{\sc iii}	& 4363/5007	& 0.013		& 0.013		& 0.013 	& 1	\\
Ne~{\sc iv}	& 1602/2423	& 0.049		& 0.049		& 0.047 	& 2	\\
Ne~{\sc v}	& 1575/3426	& 0.0104	&0.012		& 0.007: 	& 2	\\
\hline
\label{tab:diagnosticsbiconical}
\end{tabular}
\end{center}
\small{	Moc. = Mocassin results \\
	   $^a$ In order of increasing ionization potential.\\
           $^b$ (7321+7332)/(3726+3729)\\
           $^c$ (4068+4076)/(6717+6731)\\
           $^d$ references: 1: \citet{tsamis02}; 2: \citet{clegg87}}
\end{table}
\renewcommand{\baselinestretch}{1.5}

\renewcommand{\baselinestretch}{1.2}
\begin{table*}
\begin{center}
\caption{Nebular averaged fractional ionic abundances for NGC~3918 (Mocassin: C87 biconical model) }
\begin{tabular}{lccccccc}
\multicolumn{8}{c}{} \\
\hline
  	&	&	&	& Ion	 &	&	&	\\
\cline{2-8}
Element	& {\sc i}&{\sc ii}&{\sc iii}&{\sc iv}&{\sc v}&{\sc vi}&{\sc vii}\\
\hline
H	& 0.14(-1)& 0.986& 	& 	& 	& 	& 	\\
	& 0.51(-3)& 0.999& 	&  	& 	& 	& 	\\
	&	&	&	&	&	&	&	\\
He	& 0.44(-2)& 0.671& 0.325&  	& 	& 	& 	\\
	& 0.95(-5)& 0.028& 0.972&  	& 	& 	& 	\\
	&	&	&	&	&	&	&	\\
C	& 0.12(-3)& 0.053& 0.487& 0.333	& 0.128	&	&	\\
	& 0.46(-7)& 0.21(-3)& 0.028& 0.260& 0.711&	&	\\
	&	&	&	&	&	&	&	\\
N	& 0.15(-2)& 0.065& 0.464& 0.365	& 0.071	& 0.033	& 	\\
	& 0.12(-7)& 0.12(-3)&0.039&0.342& 0.365 & 0.254	& 	\\
	&	&	&	&	&	&	&	\\
O	& 0.80(-2)& 0.067& 0.674& 0.156	& 0.070	& 0.021	& 0.43(-2)\\
	& 0.21(-7)& 0.83(-4)& 0.039& 0.368 & 0.395&0.164& 0.034	\\
	&	&	&	&	&	&	&	\\
Ne	& 0.54(-4)& 0.11(-1)& 0.733& 0.133 & 0.102 & 0.021 & 0.65(-3)\\
	& 0.32(-8)& 0.55(-4)& 0.038 &0.258 & 0.536 & 0.162 & 0.48(-2)\\
	&	&	&	&	&	&	&	\\
S	& 0.17(-4)& 0.027 &0.294& 0.362	& 0.214	& 0.081	& 0.022	\\
	& 0.30(-9)& 0.27(-5)& 0.84(-3)& 0.039	& 0.348	& 0.438	& 0.173	\\
	&	&	&	&	&	&	&	\\
Si	&0.21(-3)& 0.325 & 0.195 & 0.256 & 0.127 & 0.078& 0.019\\	
	&0.14(-6)&0.62(-3)& 0.015& 0.092& 0.322	& 0.425	& 0.144\\
	&	&	&	&	&	&	&	\\
Mg	& 0.32(-2)& 0.112& 0.532& 0.119	& 0.118	& 0.098	& 0.018	\\
	& 0.61(-6)& 0.59(-4)&0.34(-2)& 0.056& 0.298& 0.507 & 0.136\\
\hline
\label{tab:mocassinionratio}
\end{tabular}
\end{center}
\small{For each element the upper row is for the optically thick phase and the lower row is for the optically thin phase.}
\end{table*}
\renewcommand{\baselinestretch}{1.5}

\renewcommand{\baselinestretch}{1.2}
\begin{table*}
\begin{center}
\caption{Mean temperatures (K) weighted by ionic species for NGC~3918 (Mocassin: C87 biconical model). }
\begin{tabular}{lccccccc}
\multicolumn{8}{c}{} \\
\hline
	&	&	&	& Ion	 &	&	&	\\
\cline{2-8}
Element	& {\sc i}	&{\sc ii}	&{\sc iii}&{\sc iv}&{\sc v}&{\sc vi}&{\sc vii}\\
\hline
H	& 11,386	& 12,925	& 	&	&	&	&	\\
	& 15,881 	& 16,326 	& 	&  	& 	& 	& 	\\
	&		&		&	&	&	&	&	\\
He	& 11,247	& 12,044	& 14,703	&  	& 	& 	& 	\\
	& 15,495	& 15,722	& 16,344	&  	& 	& 	& 	\\
	&		&		&	&	&	&	&	\\
C	& 11,582	& 11,585	& 12,116	& 13,133	& 15,858	& 	&	\\
	& 15,300	& 15,459	& 15,603	& 15,861	& 16,526	&	&	\\
	&	&	&	&	&	&	&	\\
N	& 10,339	& 11,522	& 12,182	& 13,186	& 15,435	& 17,312	& 	\\
	& 15,348	& 15,506	& 15,631	& 15,842	& 16,234	& 17,217	& 	\\
	&	&	&	&	&	&	&	\\
O	& 10,834	& 11,618	& 12,221	& 14,507	& 15,680	& 17,081	& 20,296	\\
	& 14,874	& 15,298	& 15,509	& 15,887	& 16,260	& 17,020	& 19,477	\\
	&	&	&	&	&	&	&	\\
Ne	& 10,136	& 11,636	& 12,156	& 14,340	& 15,561	& 17,360	& 22,911	\\
	& 14,990	& 15,243	& 15,467	& 15,859	& 16,269	& 17,301	& 21,769	\\
	&	&	&	&	&	&	&	\\
S	& 11,549	& 11,412	& 11,845	& 12,409	& 13,887	& 15,5584	& 17,637	\\
	& 15,016	& 15,115	& 15,353	& 15,603	& 15,910	& 16,285	& 17,437	\\
	&	&	&	&	&	&	&	\\
Si	& 11,817	& 11,794	& 12,349	& 12,875	& 14,333	& 15,543	& 17,684	\\
	& 15,362	& 15,398	& 15,561	& 15,677	& 15,949	& 16,364	& 17,563	\\
	&	&	&	&	&	&	&	\\
Mg	& 11,747	& 11,690	& 12,013	& 13,491	& 14,605	& 15,547	& 17,563	\\
	& 14,517	& 15,530	& 14,963	& 15,632	& 15,919	& 16,329	& 17,526	\\
\hline
\label{tab:mocassintemperatures}
\end{tabular}
\end{center}
\small{For each element the upper row is for the optically thick phase and the lower row is for the optically thin phase.}
\end{table*}
\renewcommand{\baselinestretch}{1.5}

\begin{figure*}
\begin{center}
\begin{minipage}[t]{6.cm} 
\psfig{file=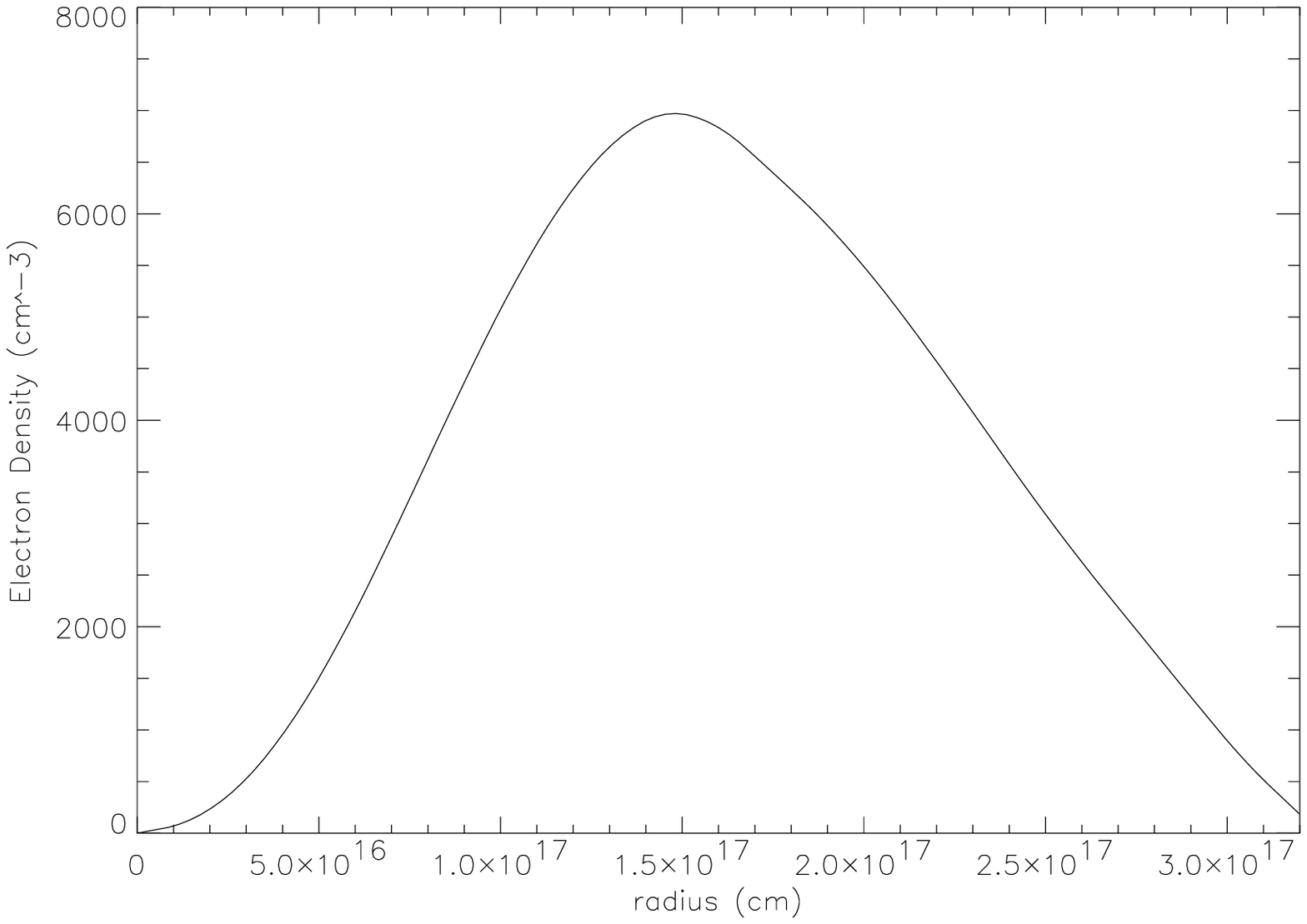, height=60mm, width=60mm}
\end{minipage}
\begin{minipage}[t]{6.cm}
\psfig{file=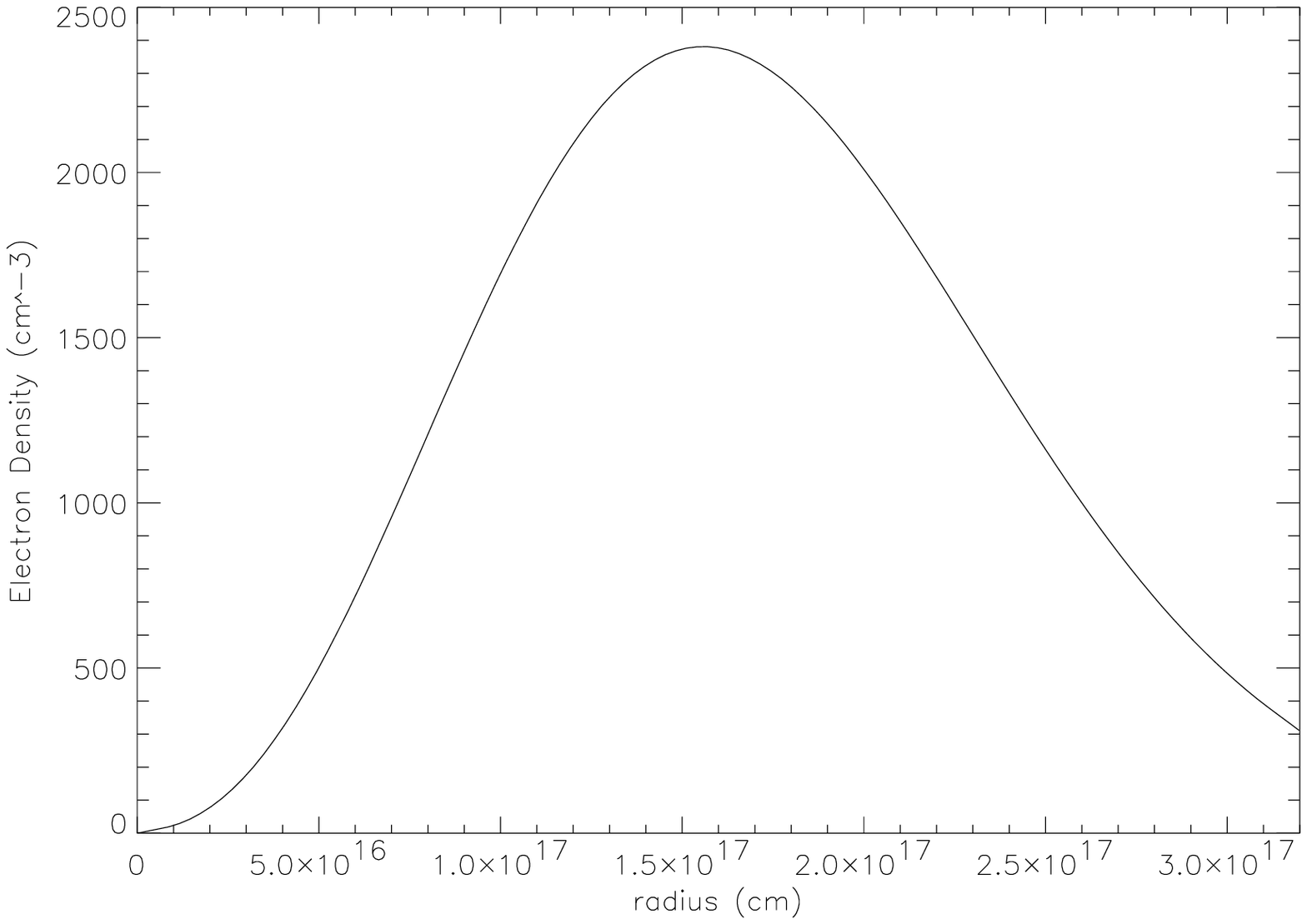, height=60mm, width=60mm}
\end{minipage}
\begin{minipage}[t]{6.cm} 
\psfig{file=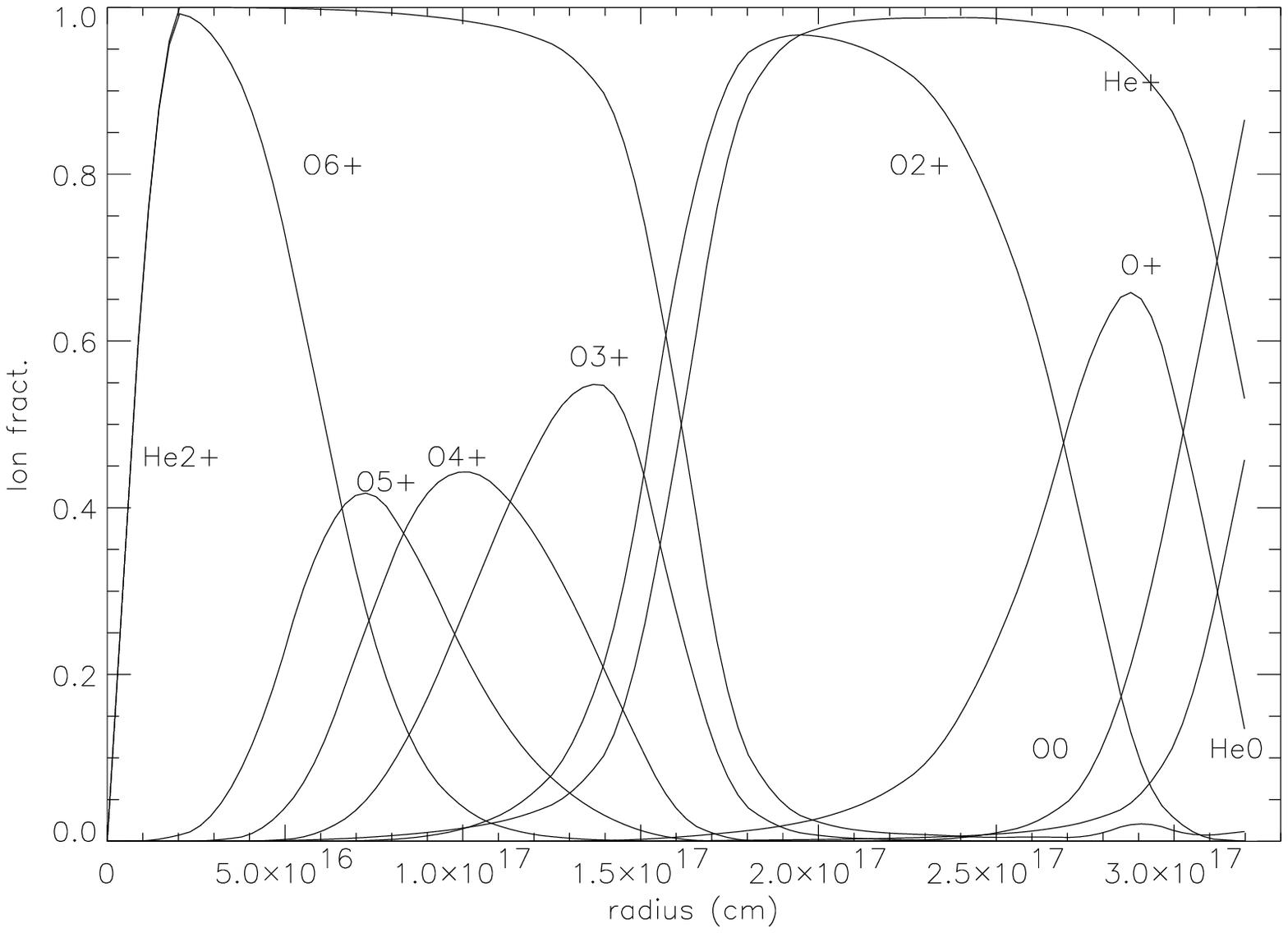, height=60mm, width=60mm}
\end{minipage}
\begin{minipage}[t]{6.cm}
\psfig{file=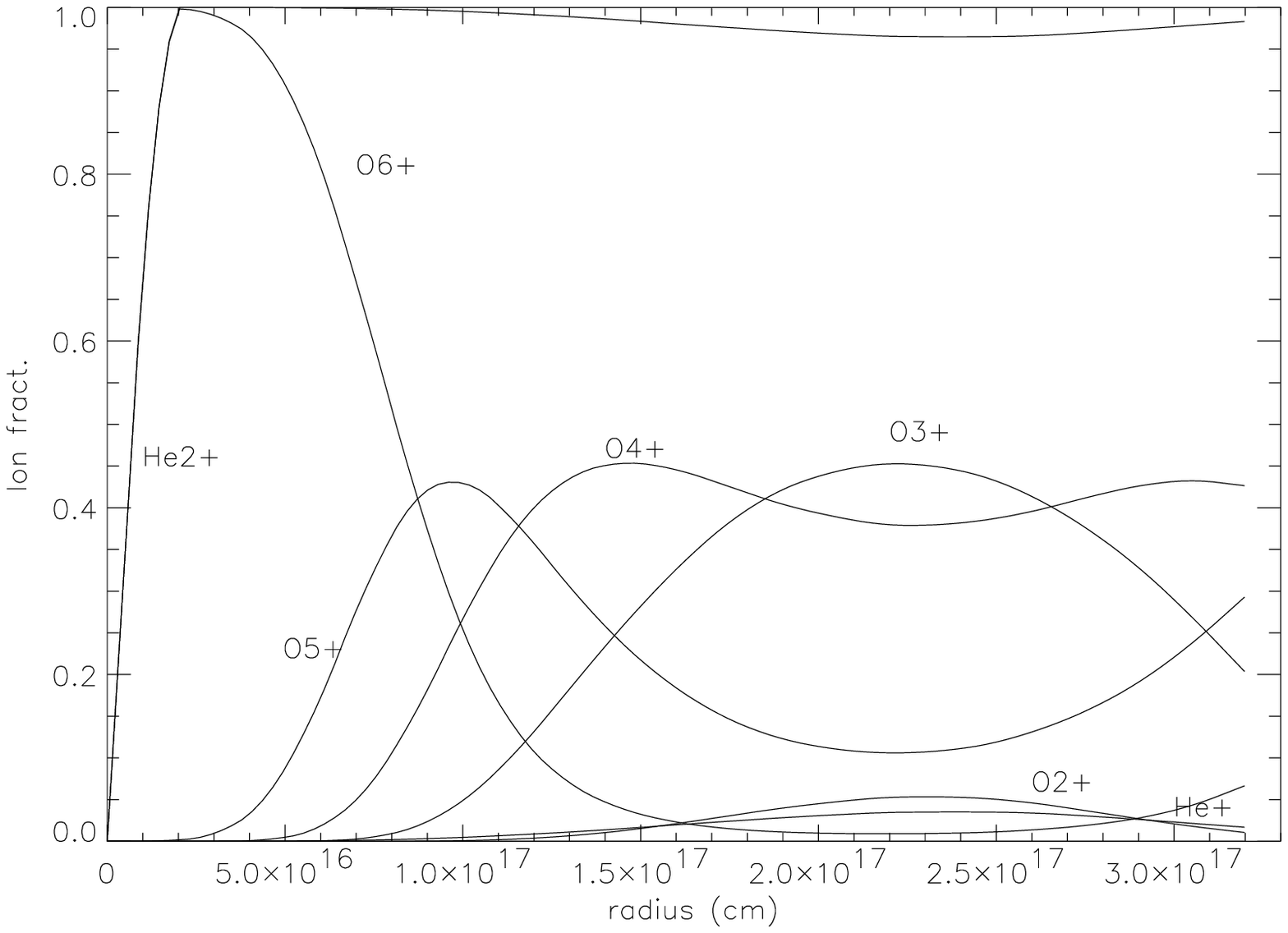, height=60mm, width=60mm}
\end{minipage}
\begin{minipage}[t]{6.cm} 
\psfig{file=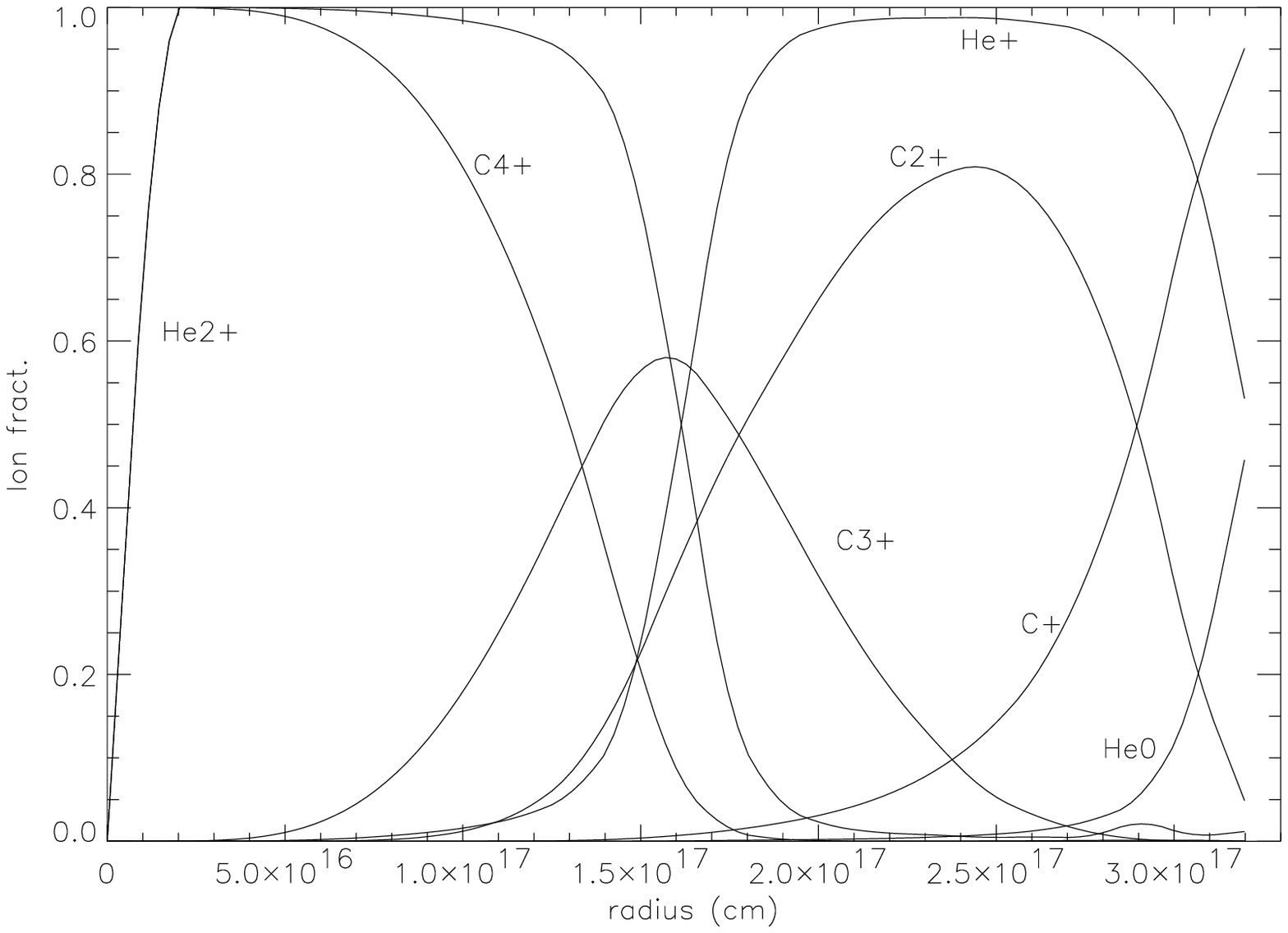, height=60mm, width=60mm}
\end{minipage}
\begin{minipage}[t]{6.cm}
\psfig{file=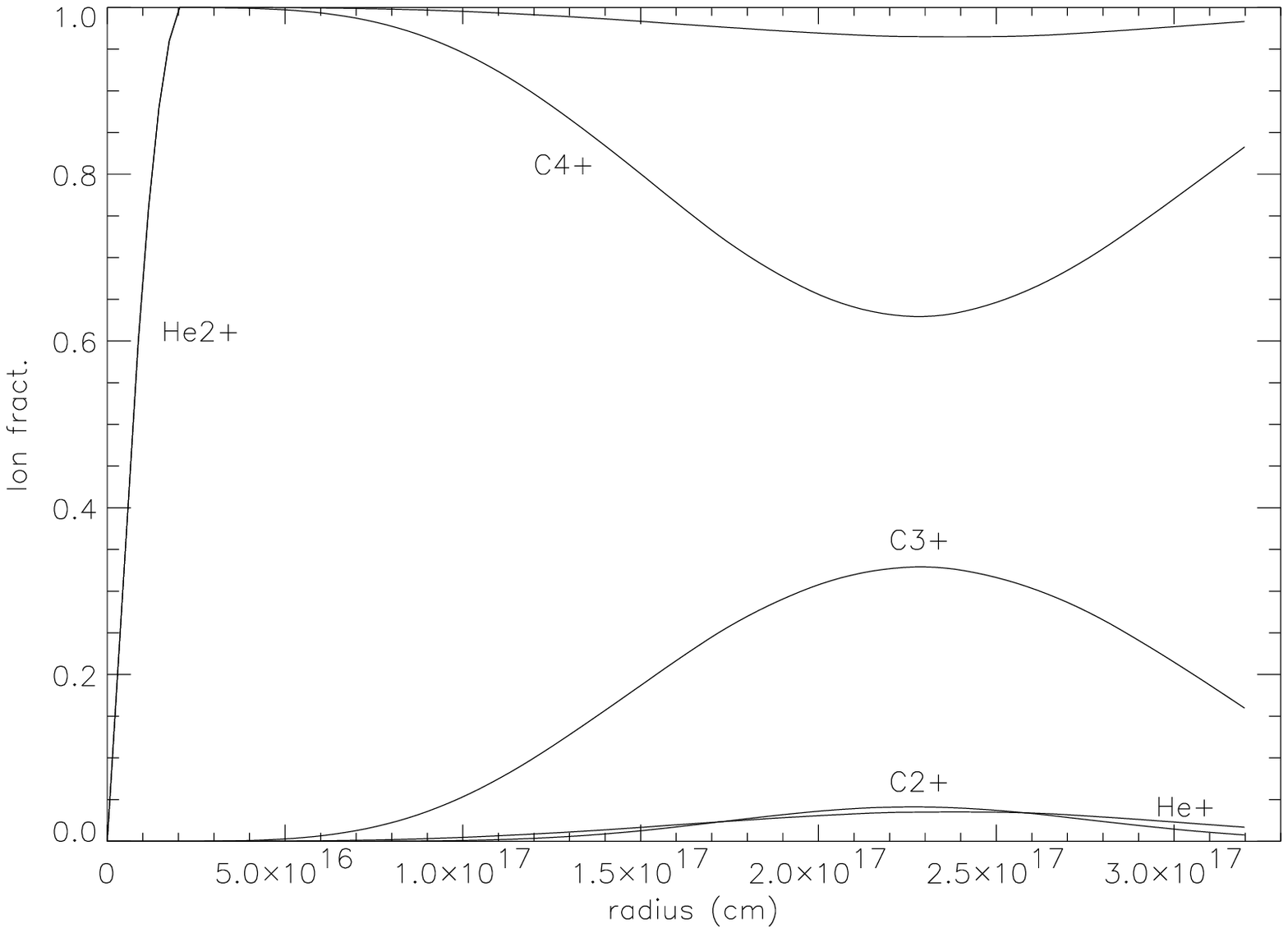, height=60mm, width=60mm}
\end{minipage}
\caption[Ionization fractions for oxygen and carbon along the polar and equatorial directions in NGC3918. Mocassin biconical model.]{Mocassin ionic fractions for oxygen (middle panels) and carbon (bottom panels) along the polar (left panels) and equatorial (right panels) directions for the C87 biconical model for NGC~3918. The ionization structure of helium is also plotted in each panel. The top panels show the electron density plotted as a function of radius along the polar (top left panel) and equatorial (top right panel) directions.}
\label{fig:iondistribution}
\end{center}
\end{figure*}

Following the C87 analysis, emission line fluxes are given relative to the total dereddened $H{\beta}$ flux integrated over the entire nebula, on a scale 
where $H{\beta}$ is equal to 100. The contributions from each sector and the total relative fluxes are listed in Table~\ref{tab:clegglines}, together with C87's results for both sectors, and the total observed values. The last column of the table contains the references for the observational data used. The fractional ionic abundances were also calculated for the thin and the thick regions and are given in Table~\ref{tab:mocassinionratio}. The mean temperatures weighted by the ionic abundances are calculated and given in Table~\ref{tab:mocassintemperatures}. The definitions for both the fractional ionic abundances, $f(N_{\rm ij})$, and for the mean temperatures, $T(N_{\rm ij}$, are given by \citet{harrington82} and are repeated here for convenience:
\begin{equation}
f(N_{ij}) = \frac{<N_{\rm ij}>}{\sum_{\rm k}N_{\rm ik}}
\label{eq:fnij}
\end{equation}

{\noindent} and 
\begin{equation}
T(N_{\rm ij}) = \frac{\int{N_{\rm e}N_{\rm ij}T_{\rm e}dV}}{\int{N_{\rm e}N_{\rm ij}dV}}
\label{eq:tij}
\end{equation}
{\noindent} where the average density of an ion over the nebular volume is defined as $<N_{\rm ij}>~=~\frac{\int{N_{\rm e}N_{\rm ij}dV}}{\int{N_{\rm e}dV}}$; $N_{\rm e}$ and $N_{\rm ij}$ are the electron density and the the density of the ion $j$ of element $i$, respectively, and the summation is over all $k$ ionization stages; $T_{\rm e}$ is the electron temperature and $dV$ is the volume element.

 The fractional ionic abundances and the mean temperatures weighted by ionic species were also presented by C87 (Tables~18 and~19 in their paper), these are of great use in this work as they provide important parameters for the comparison of the Mocassin model to that of C87.
The observed value of the integrated H$\beta$ flux was obtained from \citet{cahn92} and it was dereddened by adopting an interstellar reddening constant, $c({\rm H}\beta)$, equal to 0.40. This value, which is within the error bars of the value estimated by C87,  $c(H\beta) = 0.43 \pm 0.05$,  was derived by \citet{tsamis02} and has been adopted in this work because it was derived from observations of the optical Balmer emission lines, also published by \citet{tsamis02} and listed in our Table~\ref{tab:clegglines} (as reference $a$). It was not necessary to adjust the infrared line fluxes as these are not affected significantly by interstellar reddening. The optical observations presented by \citet{tsamis02} were obtained by uniformly scanning a long-slit across the nebula, which method, when combined with the total H$\beta$ flux measured using a large entrance aperture \citep[e.g.][]{cahn92}, yields the total fluxes from the whole nebula for all emission lines detected in the optical spectrum. These fluxes can thus be directly compared to those measured with spaceborne facilities using large apertures, such as the {\it IUE} in the ultraviolet and {\it ISO} in the infrared. 

Most of the results shown in Table~\ref{tab:clegglines} are in reasonable agreement with the results published by C87,  also reported in Table~\ref{tab:clegglines}. For both models, however,  the doubly ionized helium recombination line at 4686\,{\AA} is too strong, while the singly ionized helium triplet line at 5876\,{\AA} is too weak, when compared with the observations. This result is consistent with the ionic fractions listed in Table~18 of C87 and in Table~\ref{tab:mocassinionratio} of this work for these ions. It is worth noting, however, that the values of the relative fluxes that C87 observed for these lines are 45.0 for He~{\sc ii}\,4686\,{\AA} and 9.5  for He~{\sc i}\,~5876{\AA}, on a scale where H$\beta~=~100$. These values are closer to their model values of 48.0 and 7.8 than are the more recent observations of 41.7 and 10.8 by \citet{tsamis02} quoted in Table~\ref{tab:clegglines}, the former being the values they had tried to reproduce in their model. A better agreement with the more recent observations can be obtained by using a different model atmosphere for the ionizing source. As has already been mentioned, a non-LTE model atmosphere for a central star with an effective temperature of 140,000~K, ${\rm log}\,g~=~6.5$ and a photospheric ${\rm He}/{\rm H}~=~0.10$ by number was used by C87 and here; however by increasing log~$g$, the depth of the doubly ionized helium jump is increased. This will result in fewer ${\rm He}^+$ ionizing photons and, therefore, lower He$^{2+}$ and higher He$^+$ nebular ionic fractions. The values obtained by using a non-LTE model for a central star of the same luminosity,  effective temperature and He/H ratio, but with log~$g=7.5$, are 9.2 and 48.0 for the He~{\sc i}~5876{\AA} and He~{\sc ii}~4686{\AA} line fluxes (relative to H$\beta$, on a scale where H$\beta$ = 100), respectively. These values are closer to the observations than those obtained with log\,$g=6.5$ and reported in Table~\ref{tab:clegglines}. One final point to be made about the nebular helium results is that, as was mentioned at the start of this section, the nebular helium abundance used in this work is the empirical value calculated by C87, He/H=0.107 by number, which is slightly higher than the value they used in their model, He/H=0.10; this also contributes to the difference between the results we obtained for the He~{\sc ii}~4686\,{\AA} and He~{\sc i}~5876\,{\AA} line fluxes and those presented by C87.

Large discrepancies also exist between the two models' predictions and the {\it ISO~SWS} fluxes measured by Bower (2001~-~reference $d$ in Table~~\ref{tab:clegglines}). This discrepancy is mainly due to a pointing error which affected all {\it ISO} observations of NGC~3918; this is discussed in more detail in Section~4 and Appendix~C. 

With regard to the other emission line fluxes listed in Table~\ref{tab:clegglines}, some discrepancies between the two models do exist, but they are, in the majority of cases within 30\% and the overall ionization structure of the nebula coincides. In particular, large discrepancies were found between the current work and that of C87 in the predictions of the [Ne~{\sc v}] fine-structure line fluxes, which will be discussed in Appendix~B. Considering the significant progress in atomic physics made in the past fifteen years, on the one hand, and the completely different treatment of the radiative transfer adopted by the two codes, on the other, that there are some differences in the results obtained is hardly surprising. Neither the Opacity Project \citep{verner96} nor the Iron Project \citep{hummer93} existed in 1987 and the rates for other important processes have changed since then. 

Mocassin's predictions for  electron density and temperature diagnostic line ratios can be compared to those obtained by C87, and to the observed ratios, in Table~\ref{tab:diagnosticsbiconical}. There is very good agreement between the two models' predictions of line ratios, which is to be expected, since the density distributions used were the same. The small differences observed are reflected in the predicted emission line fluxes, listed in Table~\ref{tab:clegglines}, and, as has already been discussed, can probably be mostly assigned to differences in the atomic data sets. The diffuse radiation field, however, might also have some  effect on the ionization structure, especially at the edges between the two sectors and in the outer regions of the nebula.

\subsubsection{Resonance lines}
\label{subsub:resonance}

There are some large discrepancies between the predictions of both the Mocassin and C87 models and the observations for the fluxes of the  C~{\sc iv}~1550\,{\AA} and Mg~{\sc ii}~2800\,{\AA} resonance lines. The reason for this is that these lines are attenuated by nebular dust absorption, which was not accounted for in this work, or by C87. Moreover, Mg~{\sc ii}~2800\,{\AA} also suffers from absorption by interstellar Mg$^+$ and this is probably the main reason for the almost total absence of this line in the observed spectrum. Resonance lines must travel a much longer effective distance to escape from a nebula than do forbidden lines or the continuum radiation, because of the high optical depth in them; this translates into an enhanced probability of absorption by dust. The N~{\sc v}~1240\,{\AA} resonance doublet is also affected by this process, although in a minor way. The Si~{\sc iv} resonance line doublet at 1393~{\AA} and 1401~{\AA} can also be subject to attenuation by dust absorption, however this line appears to be only slightly overestimated by the Mocassin model and is in agreement with the observations in C87.  Harrington, Monk \& Clegg published a dust model for NGC~3918 which confirms that  the intensities of the resonance lines are indeed reduced. In their dust model \citet{harrington88} followed the same method as in the C87 photoionization model for this object; treating the thin and the thick sectors as two separate, spherically symmetric objects and then combining the results, weighted by the respective volume of each sector, i.e. 29\% for the thick sector and 71\% for the thin sector. In their model they found that the total flux of the C~{\sc iv}~1550~{\AA} doublet, relative to H$\beta$ = 100, was reduced from 1435 to 950, with 45\% of photons destroyed in the optically thick sector and 21\% in the optically thin sector. Their attenuated flux was still 85\% too strong compared to the observed value that they used (F(1550)\,=\,512) and 107\% too strong compared to the more recently re-calibrated and extracted {\it IUE} archive spectra presented by \citet{tsamis02} (F(1550)\,=\,458). \citet{harrington88} did not give a final answer as to whether the cause of this discrepancy lies in the photoionization model producing too much C~{\sc iv} line radiation or in the dust model being too ineffective in destroying it, however they speculate that if the grain opacity near 1550~{\AA} was greater in their dust model, more line photons would be destroyed hence improving the fit. The flux of N~{\sc v}~1240~{\AA},  relative to H$\beta$~=~100, is reduced from 66 to 52, with 25\% of photons being destroyed in the thick sector and 20\% in the thin sector. The value measured by \citet{tsamis02} for this line is 41.4. In the case of Si~{\sc iv}~1400~{\AA} dust actually makes the fit worse, by attenuating the flux, relative to H$\beta$~=~100, from 8.2 to 5.7, whereas the observed value is 9.0 (C87). This is, however, a weak line and the observed flux might not be accurate. Finally, \citet{harrington88} predicted that the flux of Mg~{\sc ii}~2800{\AA} is also reduced by dust attenuation, from 43 to 29. However, as mentioned above, this line shows evidence of interstellar absorption, so it cannot be measured properly. Aside from this, the dusty photoionization model of \citet{harrington88} is nearly identical to the one presented by C87.

\subsubsection{The ionization structure}

The general ionization structures of the C87 biconical nebula calculated by the two codes, and shown in Table~18 of C87 and in Table~\ref{tab:mocassinionratio} of this work, are in satisfactory agreement, with small differences which, again,  can mainly be attributed to improvements in the atomic data and the different treatments of the radiative transfer. In fact, although both codes treat the diffuse field exactly, the C87 model could not allow for transfer of diffuse radiation between the thick and the thin sectors. As shown by the scatter of values obtained for the benchmark problems \citep{pequignot01,ercolano02,ercolanoI}, even amongst one-dimensional classical codes it is usual for different codes to return slightly different predictions for the same problem, even in the extremely simplified case of homogeneous, spherically symmetric nebulae. Figure~\ref{fig:iondistribution} shows plots of the ionization structure of oxygen (middle panels) and carbon (bottom panels) as a function of radius along the optically thin polar direction (left panels) and along the optically thick equatorial direction (right panels). The ionization structure of helium is also plotted in each panel. The electron density distribution is also plotted as a function of radius for the optically thin polar direction (top left panel) and for the optically thick equatorial direction (top right panels). The effects of the charge exchange reaction O$^{3+}$+H$^0$$\rightarrow$O$^{2+}$+H$^+$ are evident, particularly in the optically thick region, where {\it f}(He$^{2+}$) is about 10\% larger than the sum of {\it f}(O$^{3+}$)+{\it f}(O$^{4+}$)+{\it f}(O$^{5+}$)+...

\subsubsection{The temperature structure}

The mean temperatures weighted by ionic species calculated by Mocassin for the C87 biconical model are listed in Table~\ref{tab:mocassintemperatures}; these can be compared to those calculated by C87, as listed in Table~19 of their paper. The most obvious difference between the two sets of results is that Mocassin's mean temperatures are systematically lower than Clegg et al.'s. This effect is even more pronounced for the optically thin region. First of all, the higher collision strengths for the [Ne~{\sc v}] fine structure lines, as calculated by \citet{aggarwal83} and by \citet{lennon91}, and used in this work (see Appendix~B), will make [Ne~{\sc v}] an important coolant. Furthermore, the fact that in the optically thin region the fractional ionic abundance of Ne$^{4+}$ is higher could also explain why  this region seems to be more affected by the cooling.
It is also worth noticing that the mean ionic temperatures predicted by C87 show very little variation for the neutral species in each sector. These temperatures are all concentrated around $12,065\,\pm\,35\,$K for the neutral species in the optically thick sector, and around $16,120\,\pm\,210\,$K for the optically thin sector (if $T_e(H^0)\,=\,16,330\,$K is excluded then the rest of the neutral ion temperatures in the optically thin sector are grouped around the value 16,005$\pm$95\,K). The scatter gets larger for the higher ionization species. Mocassin's results, on the other hand, show a larger scatter, even for the mean temperatures of the neutral species, with their temperatures grouped around $10960\,\pm\,896\,$K for the optically thick region and $15105\,\pm\,682$~K for the optically thin region. One probable explanation for this effect is  the fact that, in C87's work, the biconical geometry is reproduced by combining two separate spherically symmetric models, hence taking no account of the interaction between the diffuse fields from the two phases, which could lead to fluctuations in local temperatures, particularly at the boundaries between the two regions. This is discussed in more detail in the next subsection.

\subsection{Diffuse Radiation Field Consistency Test}
\label{sub:test}
\renewcommand{\baselinestretch}{0.9}
\begin{table}
\begin{center}
\caption{Predicted emission lines fluxes for NGC~3918 (composite versus self-consistent model.) }
\begin{tabular}{lcccccc}
\multicolumn{7}{c}{} \\
\hline
Line			& \multicolumn{3}{c}{Composite}& \multicolumn{3}{c}{Self-Consistent}\\ 
\cline{2-7}
			& Thin	& Thick	& Tot.	& Thin	& Thick	& Tot.	\\
\hline
H$\beta$/10$^{-10}\frac{{\rm erg}}{{\rm cm}^{2}{\rm s}}$
				& 0.45	& 1.90	& 2.35	& 0.45	& 1.88	& 2.33	\\
H$\beta$ 4861			& 100	& 100	& 100	& 100	& 100	& 100	\\
He~{\sc i} 5876     		&  0.07	& 8.7	& 8.8	& 0.08	& 8.5	& 8.5	\\
He~{\sc ii} 4686		& 23	& 28	& 51	& 24	& 30	& 54	\\
C~{\sc ii}$]$ 2326		& 0.22	& 34	& 34	& 0.21	& 57	& 57	\\
C~{\sc ii} 4267			& 0.004	& 0.31	& 0.31	& 0.004	& 0.32	& 0.32	\\
$[$C~{\sc ii}$]$ 157.74$\mu$m 	& .0005	& .12	& .12	& 0.0	& 0.35	& 0.35	\\
C~{\sc iii}$]$ 1908 		& 33	& 551	& 584	& 33	& 577	& 610	\\
C~{\sc iii} 4648 		& 0.10	& 0.56	& 0.66	& 0.10 	& 0.50	& 0.61	\\
C~{\sc iii}$]$ 2297		& 1.05	& 6.6	& 7.7	& 1.11	& 5.9	& 7.0	\\
C~{\sc iv} 1550			& 607	& 968	& 1575	& 596	& 964	& 1560	\\
C~{\sc iv} 4658			& 0.24	& 0.16	& 0.40	& 0.25	& 0.16	& 0.40	\\
$[$N~{\sc i}$]$ 5199		& 0.0	& 0.006	& 0.006	& 0.0	& 0.21	& 0.21	\\
$[$N~{\sc ii}$]$ 6584		& 0.06	& 27	& 28	& 0.06	& 73	& 73	\\
N~{\sc iii}$]$ 1751		& 2.0	& 22	& 22	& 2.0	& 22	& 24	\\
$[$N~{\sc iii}$]$ 57.3 $\mu$m 	& 0.60	& 14	& 15	& 0.64	&  12	& 13	\\
N~{\sc iv}$]$ 1486		& 19	& 26	& 45	& 19	& 26	& 45	\\
 N~{\sc v} 1240			& 44	& 21	& 65	& 42	& 22	& 64	\\
$[$O~{\sc i}$]$ 6300		& 0.0	& 0.18	& 0.18	& 0.0	& 2.7	& 2.7	\\
$[$O~{\sc i}$]$ 63.12 $\mu$m 	& 0.0	& 0.011	& 0.012	& 0.0	& 0.23	& 0.23	\\
$[$O~{\sc i}$]$ 145.5 $\mu$m 	& 0.0	& .0008 & .0008 & 0.0	& 0.018	& 0.018	\\
$[$O~{\sc ii}$]$ 3727		& 0.08	& 47.5	& 48	& 0.08	& 104	& 104	\\
$[$O~{\sc iii}$]$ 51.8 $\mu$m 	& 2.4	& 101	& 103	& 2.8	& 90	& 93	\\
$[$O~{\sc iii}$]$ 88.4 $\mu$m 	& 0.75	& 19.	& 20	& 0.91	& 16	& 17	\\
$[$O~{\sc iii}$]$ 5007		& 35	& 1624	& 1659	& 40	& 1543	& 1583	\\
$[$O~{\sc iii}$]$ 4363		& 0.78	& 20	& 21	& 0.85	& 20	& 20	\\
$[$O~{\sc iv}$]$ 25.9 $\mu$m 	& 186	& 157	& 143	& 190	& 161	& 352	\\
O~{\sc iv}$]$ 1402		& 16	& 14	& 29	& 16	& 14	& 30	\\ 
O~{\sc iv} 1342			& 1.3	& 0.84	& 2.1	& 1.4	& 0.83	& 2.2	\\
$[$Ne~{\sc ii}$]$ 12.8 $\mu$m 	& 0.001	& 0.62	& 0.64	& 0.001	& 0.94	& 0.94	\\
$[$Ne~{\sc iii}$]$ 15.6 $\mu$m 	& 1.6	& 130	& 133	& 1.8	& 130	& 132	\\ 
$[$Ne~{\sc iii}$]$ 3869		& 3.7	& 161	& 165	& 4.1	& 161	& 165	\\
$[$Ne~{\sc iv}$]$ 2423		& 36	& 41	& 77	& 35	& 42	& 77	\\
$[$Ne~{\sc v}$]$ 3426		& 76	& 44	& 120	& 75	& 45	& 119	\\
$[$Ne~{\sc v}$]$ 14.3 $\mu$m 	& 176	& 110	& 286	& 179	& 107	& 286	\\
$[$Ne~{\sc v}$]$ 24.3 $\mu$m 	& 154	& 65	& 219	& 156	& 64	& 220	\\
$[$Ne~{\sc vi}$]$ 7.6 $\mu$m 	& 49	& 21	& 70	& 49	& 22	& 70	\\
Mg~{\sc i}$]$ 4565		& 0.0	&0.34	& 0.38	& 0.0	& 0.73	& 0.73	\\
Mg~{\sc ii} 2800		& 0.20	& 57	& 57	& 0.26	& 78	& 78	\\
$[$Mg~{\sc iv}$]$ 4.49 $\mu$m 	& 0.39	& 3.1	& 3.5	& 0.42	& 3.1	& 3.5	\\
$[$Mg~{\sc v}$]$ 2783	 	& 2.2	& 2.4	& 4.6	& 2.2	& 2.4	& 4.6	\\
$[$Mg~{\sc v}$]$ 5.6 $\mu$m 	& 4.4	& 6.6	& 11	& 4.5	& 6.3	& 11	\\
Si~{\sc iii}$]$ 1887		& 1.1	& 14	& 15	& 0.97	& 15	& 16	\\
Si~{\sc iv} 1399		& 2.9	& 7.6	& 11	& 2.7	& 7.9	& 11	\\
$[$S~{\sc ii}$]$ 4070		& 0.	& 1.3	& 1.3	& 0.0	& 2.1	& 2.1	\\
$[$S~{\sc iii}$]$ 18.7 $\mu$m	& 0.03	& 32	& 32	& 0.03	& 38	& 38	\\
$[$S~{\sc iii}$]$ 33.6 $\mu$m	& 0.020	& 14	& 14	& 0.02	& 16	& 16	\\
$[$S~{\sc iii}$]$ 6312		& 0.006 & 3.2	& 3.2	& 0.006	& 3.7	& 3.7	\\
$[$S~{\sc iv}$]$ 10.5$\mu$m 	& 6.8 & 223	& 235	& 7.23	& 207	& 214	\\
\hline
\label{tab:testlines}
\end{tabular}
\end{center}
\end{table}
\renewcommand{\baselinestretch}{1.5}

\renewcommand{\baselinestretch}{1.2}
\begin{table*}
\begin{center}
\caption{Fractional ionic abundances for NGC~3918 (Mocassin: composite biconical model) }
\begin{tabular}{lccccccc}
\multicolumn{8}{c}{}\\
\hline
	&	&	&	& Ion	 &	&	&	\\
\cline{2-8}
Element	& {\sc i}	&{\sc ii}	&{\sc iii}&{\sc iv}&{\sc v}	&{\sc vi}&{\sc vii}\\
\hline
H	& 0.61(-2)& 0.994	& 	& 	& 	& 	& 	\\
	& 0.54(-3)& 0.999	& 	&  	& 	& 	& 	\\
	&	&	&	&	&	&	&	\\
He	& 0.19(-2)& 0.697	& 0.301	&  	& 	& 	& 	\\
	& 0.88(-5)& 0.025	& 0.975	&  	& 	& 	& 	\\
	&	&	&	&	&	&	&	\\
C	& 0.61(-4)& 0.029	& 0.481	& 0.366	& 0.124	&	&	\\
	& 0.56(-7)& 0.20(-3)& 0.027	& 0.254	& 0.718	&	&	\\
	&	&	&	&	&	&	&	\\
N	& 0.97(-4)& 0.025	& 0.476	& 0.399	& 0.068	& 0.032	& 	\\
	& 0.26(-7)& 0.12(-3)& 0.037	& 0.334	& 0.369	& 0.259	& 	\\
	&	&	&	&	&	&	&	\\
O	& 0.49(-3)& 0.031 	& 0.730	& 0.149	& 0.066	& 0.020	& 0.42(-2)\\
	& 0.16(-7)& 0.72(-4)	& 0.034	& 0.360	& 0.403	& 0.167	& 0.035	\\
	&	&	&	&	&	&	&	\\
Ne	& 0.38(-5)& 0.76(-2)	& 0.747	& 0.126	& 0.098	& 0.020	& 0.61(-3)\\
	& 0.24(-8)&0.45(-4)	& 0.033	& 0.257	& 0.539	& 0.165	& 0.49(-2)\\
	&	&	&	&	&	&	&	\\
S	& 0.32(-5)& 0.036 	& 0.368 & 0.387 & 0.127 & 0.061 & 0.019	\\
	& 0.16(-7)& 0.14(-3) 	& 0.015 & 0.126 & 0.275 & 0.413 & 0.170 \\
	&	&	&	&	&	&	&	\\
Si	& 0.13(-3)& 0.268 	& 0.207 & 0.282 & 0.143 & 0.080 & 0.019	\\
	& 0.19(-6)& 0.66(-3)& 0.016	& 0.094	& 0.319	& 0.425	& 0.144 \\
	&	&	&	&	&	&	&	\\
Mg	& 0.17(-2)& 0.082 	& 0.638 & 0.133 & 0.066 & 0.065 & 0.014	\\
	& 0.40(-5)& 0.34(-3) 	&0.024 & 0.162 & 0.248 & 0.441 & 0.123 \\
\hline
\label{tab:testionratio}
\end{tabular}
\end{center}
\small{For each element the upper row is for the optically thick phase and the lower row is for the optically thin phase.}
\end{table*}
\renewcommand{\baselinestretch}{1.5}

\renewcommand{\baselinestretch}{1.2}
\begin{table*}
\begin{center}
\caption{Mean temperatures (K) weighted by ionic species for NGC~3918 (Mocassin, composite biconical  model). }
\begin{tabular}{lccccccc}
\multicolumn{8}{c}{} \\
\hline
	&	&	&	& Ion	 &	&	&	\\
\cline{2-8}
Element	& {\sc i}	&{\sc ii}	&{\sc iii}&{\sc iv}&{\sc v}	&{\sc vi}&{\sc vii}\\
\hline
H	& 12,060	& 12,880	& 	&	&	&	&	\\
	& 16,061 	& 16,486 	& 	&  	& 	& 	& 	\\
	&	&	&	&	&	&	&	\\
He	& 11,965	& 12,052	& 14,770	&  	& 	& 	& 	\\
	& 15,844	& 16,001	& 16,498	&  	& 	& 	& 	\\
	&	&	&	&	&	&	&	\\
C	& 11,964	& 11,985	& 12,108	& 12,931	& 15,661	& 	&	\\
	& 15,702	& 15,757	& 15,842	& 16,039	& 16,667	&	&	\\
	&	&	&	&	&	&	&	\\
N	& 12,022	& 12,002	& 12,147	& 12,992	& 15,279	& 17,065	& 	\\
	& 15,704	& 15,761	& 15,842	& 16,006	& 16,369	& 17,360	& 	\\
	&	&	&	&	&	&	&	\\
O	& 12,032	& 11,989	& 12,155	& 14,481	& 15,507	& 16,862	& 20,001	\\
	& 15,678	& 15,749	& 15,838	& 16,036	& 16,388	& 17,153	& 19,615	\\
	&	&	&	&	&	&	&	\\
Ne	& 11,958	& 11,990	& 12,146	& 14,301	& 15,394	& 17,122	& 22,502	\\
	& 15,686	& 15,736	& 15,817	& 16,010	& 16,406	& 17,449	& 21,943	\\
	&	&	&	&	&	&	&	\\
S	& 11,986	& 11,982	& 12,946	& 12,233	& 13,638	& 15,402	& 17,367	\\
	& 15,675	& 15,689	& 15,742	& 15,850	& 16,075	& 16,419	& 17,573	\\
	&	&	&	&	&	&	&	\\
Si	& 11,919	& 11,947	& 12,169	& 12,624	& 14,066	& 15,293	& 17,397	\\
	& 15,674	& 15,704	& 15,773	& 15,858	& 16,100	& 16,518	& 17,744	\\
	&	&	&	&	&	&	&	\\
Mg	& 11,919	& 11,929	& 11,957	& 13,408	& 14,490	& 15,372	& 17,322	\\
	& 15,650	& 15,671	& 15,717	& 15,841	& 16,066	& 16,473	& 17,680	\\
\hline
\label{tab:testtemperatures}
\end{tabular}
\end{center}
\small{For each element the upper row is for the optically thick phase and the lower row is for the optically thin phase.}
\end{table*}
\renewcommand{\baselinestretch}{1.5}

C87 combined two appropriately weighted, spherically symmetric models to reproduce the biconical geometry.  A consequence of this approach was that the diffuse radiation field was not treated self-consistently, especially near the boundaries of the two zones. The effects of the diffuse radiation field are expected to  be small, but they could account for some of the differences between Moccassin's results and those obtained with the spherically symmetric Harrington code.  In order to investigate the magnitude of these effects, the biconical distribution was modelled again using Moccassin, but this time each sector was treated as if it were spherically symmetric.  The results from each sector were then weighted by the volume of the same sector and added together to yield the total flux relative to the total H$\beta$ flux from the whole nebula. From now on this model will be referred to as the {\it composite model}. 

Table~\ref{tab:testlines} lists the fluxes of the emission lines relative to H$\beta$, on a scale where H$\beta$ is equal to 100, for each sector.  For ease of comparison, the results from the previous, self-consistent model are also given in the same table.  Tables~\ref{tab:testionratio} and~\ref{tab:testtemperatures} list the fractional ionic abundances and the weighted mean electron temperatures for individual ionic species for the composite model. It becomes apparent from inspection of Table~\ref{tab:testlines} that all of the collisional lines coming from lower ionization species are much weaker in the optically thick region of the composite model than in the same region of the self-consistent model. This is consistent with the fractional ionic abundances of low ionization species being lower in the optically thick region of the composite model (Table~\ref{tab:testtemperatures}), hence yielding lower line fluxes. The helium recombination line intensities calculated for the optically thick region of the composite model also indicate a lower abundance of He$^0$ and He$^{2+}$, but a higher He$^+$ abundance, which is confirmed by the results in Table~\ref{tab:testionratio}. The optically thin region results do not seem to be affected as much and they are almost identical for the composite and for the self-consistent model. There are two main causes for the discrepancies observed between the results for the optically thick regions of the composite and self-consistent models. First of all, the diffuse radiation field coming from the optically thin region of the self-consistent model is weaker than the diffuse field coming from the same region in the optically thick spherically symmetric nebula of the composite-model, since the  diffuse field is approximately proportional to the square of the density and the densities in the two regions of the spherically symmetric optically thick composite model are, obviously, the same and are higher than the density in the optically thin region of the self-consistent model. This means that more diffuse photons are available to ionize the low ionization potential species in the composite model than there are in the self-consistent one. Secondly, the shape of the diffuse continuum at the edges between the two regions is different in the two models. In the self-consistent model the diffuse photons coming from the optically thin region will have higher frequencies, on average, than those coming from the same region in the optically thick spherically symmetric nebula of the composite model; this is due to the fact that in the optically thin medium the gas is in a higher ionization state. The photoionization cross-sections of the lower ionization species in the optically thick region of the self-consistent model will be, in general, quite small at the higher frequencies typical of  the diffuse photons coming from the optically thin region and so these species will be less efficiently ionized by them. In the composite model, however, since the density distribution is spherically symmetric, the ionization structure is also the same in all radial directions; this means that the diffuse photons coming from adjacent regions will have frequencies close to the ionization thresholds of the lower ionization species which will, therefore, be ionized to a greater degree than in the self-consistent model. On the other hand, the optically thin region of the self-consistent model will not be affected by the lower energy diffuse photons coming from the optically thick region since the gas in the optically thin region is already in a higher ionization stage. The combination of the two effects discussed above explains why the abundances of lower ionization species are lower in the optically thick region of the composite model than in the same region of the self-consistent one.

Another striking feature which emerges from running the composite spherically symmetric model is in the prediction of mean temperatures weighted by ionic species shown in Table~\ref{tab:testtemperatures}. As mentioned in the previous subsection, the mean ionic temperatures predicted by C87 show very little variation amongst the neutral species in each sector, whereas the scatter of the results from Mocassin's self-consistent model is much larger. It was anticipated that a possible explanation for this effect could be the fact that in Clegg et al.'s work the diffuse radiation field transfer was not self-consistent, particularly near the boundaries between the two density regions and that the interaction between the two phases' diffuse radiation fields could lead to the fluctuations in the kinetic temperatures obtained in Mocassin's self-consistent model. The results obtained from Mocassin's composite model indeed confirm this hypothesis -- within each sector the fluctuations in the mean ionic temperatures weighted by ionic abundances for neutral species disappear and their distribution resembles that obtained by C87 although the Mocassin temperatures are still slightly lower, particularly in the thin sector.

In conclusion, it is clear that the effects of the diffuse radiation field near the boundaries between the two phases are not negligible, even for the simple biconical distribution used in this case. These effects are particularly important for the abundances of the lower ionization species. It is not clear, however, why the ionization structure calculated by C87 (their Table~18), who used a composite model, bears more resemblance to Mocassin's self-consistent model results (see Table~\ref{tab:mocassinionratio}), than to Mocassin's composite model results (see Table~\ref{tab:testionratio}). Unfortunately, due to the completely different approach to the radiative transfer used by the two codes and the fact that fifteen years have passed since C87's work, over which period the atomic database has evolved considerably, any further comparison between the two codes becomes difficult.

\section{A Spindle-like Density Distribution Model for NGC~3918}
\label{sec:spindle}

\begin{figure}
\begin{center}
\psfig{file=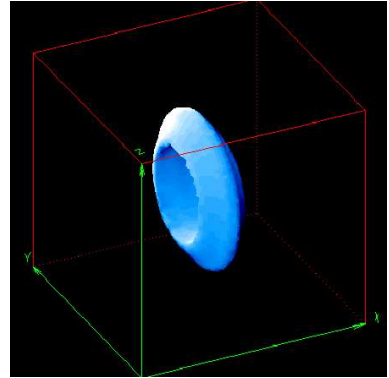, height=50mm, width=50mm}
\caption[3D iso-density plot of the spindle-like density distribution for model~A]
{Three-dimensional iso-density plot of the dense torus component of spindle-like model~A.}
\label{fig:garreltdensity}
\end{center}
\end{figure}

\begin{figure}
\begin{center}
\psfig{file=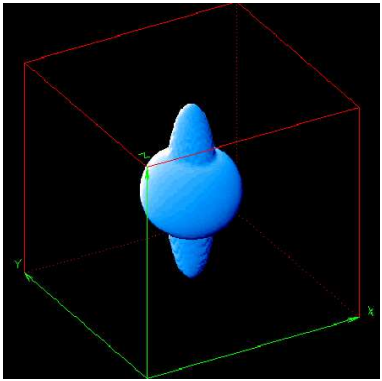, height=50mm, width=50mm}
\caption[3D iso-density plot of the spindle-like density distribution for model~B]
{Three-dimensional iso-density plot of the spindle-like density distribution for model~B.The density cut was chosen in order to show the shape of the inner shell of the nebula.}
\label{fig:chrisdensity}
\end{center}
\end{figure}

\begin{table}
\begin{center}
\caption{Input shape and thickness parameters for the analytical fit to the density distribution of spindle-like model~B (see Appendix~A). }
\begin{tabular}{lc|lc}
\multicolumn{4}{c}{}     \\
\hline
$n_1$ [cm$^{-3}$]         & 18000 & $\Delta\,\theta$ [$\deg]$ & 10  \\
$\theta_1$ [$\deg$]       & 25    & $\Delta\,r_{eq}^{in}$     & 0.06 \\
$r_{eq}$                  & 0.3   & $\Delta\,r_{eq}^{out}$    & 0.2 \\
$r_{po}^*$                & 0.2   & $\Delta\,r_{po}^{in}$     & 0.0032 \\
e                         & 0.25  & $\Delta\,r_{po}^{out}$    & 0.5 \\
\hline
\label{tab:spindleparametersan}
\end{tabular}
\\
\small{$^*$ normalized to the outer radius of the nebula}
\end{center}
\end{table}

\begin{table}
\begin{center}
\caption{Input parameters for the spindle-like models of NGC~3918. }
\begin{tabular}{lc|lc}
\multicolumn{4}{c}{}     \\
\hline
$L_*$(A)&6900$L_{\odot}$ 	 & N/H	& 1.5$\times$10$^{-4}$   \\
$L_*$(B)&5780$L_{\odot}$ 	 & O/H	& 5.0$\times$10$^{-4}$    \\ 
$R_{in}$& 0.0 pc             	 & Ne/H	& 1.2$\times$10$^{-4}$ 	 \\
$R_{out}$& 0.106 pc      	 & Mg/H	& 1.4$\times$10$^{-4}$   \\
$T_{eff}$& 140,000 K     	 & Si/H	& 1.0$\times$10$^{-4}$ 	 \\ 
log\,g  & 7.5            	 & S/H	& 1.6$\times$10$^{-4}$	 \\
He/H	& 0.107	         	 & Ar/H	& 2.0$\times$10$^{-4}$ 	 \\
C/H	& 8.0$\times$10$^{-4}$   & Fe/H	& 3.7$\times$10$^{-4}$	 \\    
\hline
\label{tab:garreltpar}
\end{tabular}
\\
\small{The abundances are given by number, relative to H}
\end{center}
\end{table}

\renewcommand{\baselinestretch}{0.9}
\begin{table}
\begin{center}
\caption{Observed and predicted emission lines fluxes for NGC~3918 (spindle-like models). }
\begin{tabular}{lcccc}
\multicolumn{5}{c}{}\\
\hline
			&  \multicolumn{2}{c}{Predicted}  &  &  \\
Line			& A 	& B	& Observed & ref$^b$ \\
\hline
H$_{\beta}$/10$^{-10}$ $\frac{erg}{cm^2s}$
			& 2.19	& 2.58	& 2.34 	& b	\\
H$_{\beta}$ 4861	& 100	& 100 	& 100	& - 	\\
He~{\sc i} 5876     	& 8.15	& 7.56	& 10.81	& a 	\\
He~{\sc ii} 4686	& 57.2	& 53.6	& 41.65	& a	\\
C~{\sc ii}$]$ 2326	& 34.6	& 27.8	& 31.24	& a 	\\
C~{\sc ii} 4267		& 0.23	& 0.21	& 0.50 	& a	\\
$[$C~{\sc ii}$]$ 157.74$\mu$m
			& 0.22	& 0.10	& 0.21	& c	\\
C~{\sc iii}$]$ 1908 	& 543	& 480	& 492.5	& a	\\
C~{\sc iii} 4648 	& 0.89	& 0.98	& 0.42	& a 	\\
C~{\sc iii}$]$ 2297	& 10.2	& 11.3	& 5:	& e	\\		
C~{\sc iv} 1550$^a$	& 2267	& 2220	& 457.6	& a	\\
C~{\sc iv} 4658	& 0.37	& 0.37	& 0.35	& a	\\ 		
$[$N~{\sc i}$]$ 5199	& 1.03	& 0.46	& 0.63	& a	\\ 
$[$N~{\sc ii}$]$ 6584	& 41.8	& 33.9	& 84	& a	\\	
N~{\sc iii}$]$ 1751	& 22.9	& 20.2	& 26.7	& a	\\ 
$[$N~{\sc iii}$]$ 57.3 $\mu$m 
			& 7.41	& 9.16	& 13.5	& c	\\
N~{\sc iv}$]$ 1486	& 61.4	& 59.3	& 46.4	& a	\\
 N~{\sc v} 1240$^a$	& 45.6	& 38.7	& 41.4	& a	\\
$[$O~{\sc i}$]$ 6300	& 3.95	& 2.57	& 5.24	& a	\\
$[$O~{\sc i}$]$ 63.12 $\mu$m	
                        & 0.40	& 0.26	& 6.72	& c	\\
$[$O~{\sc i}$]$ 145.5 $\mu$m
                  	& .029	& .014	& 0.23	& c	\\
$[$O~{\sc ii}$]$ 3727	& 42.7	& 37.9	& 94.5	& a	\\
$[$O~{\sc iii}$]$ 51.8 $\mu$m 
			& 79.4	& 112	& 88.6	& c	\\
$[$O~{\sc iii}$]$ 88.4 $\mu$m 
			& 14.2	& 22.2	& 27.5	& c	\\ 
$[$O~{\sc iii}$]$ 5007	& 1760	& 1789	& 1657	& a	\\
$[$O~{\sc iii}$]$ 4363	& 25.0	& 24.8	& 21.6	& a	\\
$[$O~{\sc iv}$]$ 25.9 $\mu$m	
			& 352	& 365	& 89	& d	\\
O~{\sc iv}$]$ 1402	& 34.9	& 31.8	& 48.9	& a	\\
O~{\sc iv} 1342		& 1.87	& 1.72	& 2:	& e 	\\
$[$Ne~{\sc ii}$]$ 12.8 $\mu$m 
			& 1.61	& 0.75	& 12:	& f	\\
$[$Ne~{\sc iii}$]$ 15.6 $\mu$m 
			& 123	& 127	& 46.3	& d	\\
$[$Ne~{\sc iii}$]$ 3869	& 174	& 173	& 135.5	& a	\\
$[$Ne~{\sc iv}$]$ 2423	& 89.2	& 87.5	& 132	& e	\\
$[$Ne~{\sc v}$]$ 3426	& 108	& 99.1	& 80	& e	\\
$[$Ne~{\sc v}$]$ 14.3 $\mu$m 
			& 267	& 256	& 19.6	& d 	\\
$[$Ne~{\sc v}$]$ 24.3 $\mu$m 
			& 170	& 146	& 23.1	& d	\\
$[$Ne~{\sc vi}$]$ 7.6 $\mu$m 
			& 62.0 	& 54.4	& -	& -	\\
Mg~{\sc i}$]$ 4565	& 0.32	& 0.33	&  0.13	& a	\\
Mg~{\sc ii} 2800$^a$	& 47.2	& 36.7	& $\leq$1& e	\\
$[$Mg~{\sc iv}$]$ 4.49 $\mu$m 
			& 4.25	& 5.10	& -	& -	\\
$[$Mg~{\sc v}$]$ 2783	& 4.84	& 5.02	& 5.6	& e	\\
$[$Mg~{\sc v}$]$ 5.6 $\mu$m 
			& 12.5	& 14.1	& -	& -	\\
Si~{\sc iii}$]$ 1887	& 13.9	& 11.2	& 8	& e	\\
Si~{\sc iv} 1399$^a$	& 11.5	& 10.2	& 9	& e	\\
$[$S~{\sc ii}$]$ 4070	& 1.90	& 1.42	& 2.69	& a	\\
$[$S~{\sc iii}$]$ 18.7 $\mu$m	
			& 16.6	& 14.3	& 8.7	& d	\\
$[$S~{\sc iii}$]$ 33.6 $\mu$m	
			& 5.21	& 4.68	& - 	& -	\\
$[$S~{\sc iii}$]$ 6312	& 1.98	& 1.55	& 2.27	& a	\\
$[$S~{\sc iv}$]$ 10.5$\mu$m	
			& 198	& 213	& 35.0	& d	\\
\hline
\label{tab:garreltlines}
\end{tabular}
\end{center}
\small{$^a$ Attenuated by dust absorption.\\           
	$^b$~references: a:~\citet{tsamis02}; b:~\citet{cahn92}; c:~\citet{liu01}; d:~\citet{bower01}; e:~C87; f:~\citet{pottasch86}.}
\end{table}
\renewcommand{\baselinestretch}{1.5}

\renewcommand{\baselinestretch}{1.2}
\begin{table}
\begin{center}
\caption{Diagnostic ratios for the electron density, $N_e$, and electron temperature, $T_e$ (NGC~3918 spindle-like models).}
\begin{tabular}{lcccccc}
\multicolumn{7}{c}{} \\
\hline
Ion$^a$		& Lines 		& Ratio	 	& Ratio		& Ratio 	& Ratio		& obs. \\
		& ({\AA})		& Moc. A	& Moc. B	& C87		& obs. 	& ref.$^d$	\\	 
\hline
		&			& $N_e$		&		&		&		&	\\
\hline
Mg~{\sc i}	& 4562/4571		& 0.38		& 0.40		& 0.42:		& 0.28 	 	& 1	\\
S~{\sc ii}	& 6716/6731		& 0.65		& 0.60		& 0.70		& 0.57 		& 1	\\
O~{\sc ii}	& 3726/3729		& 1.96		& 2.02		& 1.61		& 1.98 		& 2	\\
O~{\sc ii}	& 7325/3727$^b$		& 0.11		& 0.11		& -		& 0.0783 	& 1	\\
C~{\sc iii}	& 1906/1909		& 1.25		& 1.34		& 1.36		& 1.31 		& 2	\\
Si~{\sc iii}	& 1883/1892		& 1.15		& 1.31		& 1.30		& 1.37 		& 2	\\
Ne~{\sc iv}	& 2421/2424		& 1.09		& 1.06		& 0.93		& 1.00 		& 2	\\
N~{\sc iv}	& 1483/1486		& 1.45		& 1.47		& 1.87		& 1.85 		& 2	\\
O~{\sc iv}	& 1401/1404		& 1.32		& 1.36		& 1.14		& 1.50 		& 2	\\
\hline
		&			& $T_e$		&		&		&	\\
\hline
N~{\sc ii}	& 5755/6584		& 0.021		& 0.023		& 0.025		& 0.021 	& 1	\\
S~{\sc ii}	& 4073/6724$^c$		& 0.200		& 0.245 	&-		& 0.339 	& 1 	\\
C$^{2+}$	& 4267/1908		& 0.0004	& 0.0004	& 0.0005	& 0.0008 	& 2	\\
S~{\sc iii}	& 6312/9532		& 0.032		& 0.032		& 0.029		& 0.032 	& 2	\\
O~{\sc iii}	& 4363/5007		& 0.014		& 0.014		& 0.013		& 0.013 	& 1	\\
Ne~{\sc iv}	& 1602/2423		& 0.053		& 0.050		& 0.049		& 0.047 	& 2	\\
Ne~{\sc v}	& 1575/3426		& 0.009		& 0.009		& 0.012		& 0.007: 	& 2	\\
\hline
\label{tab:diagnosticsgarrelt}
\end{tabular}
\end{center}
\small{$^a$ In order of increasing ionization potential.\\
           $^b$ (7321+7332)/(3726+3729)\\
           $^c$ (4068+4076)/(6717+6731)\\
           $^d$ references: 1: \citet{tsamis02}; 2: C87}
\end{table}
\renewcommand{\baselinestretch}{1.5}

\renewcommand{\baselinestretch}{1.2}
\begin{table*}
\begin{center}
\caption{Fractional ionic abundances for NGC~3918 (Spindle-like models of NGC~3918) }
\begin{tabular}{lccccccc}
\multicolumn{8}{c}{} \\
\hline
 	&	&	&	& Ion	 &	&	&	\\
\cline{2-8}
Element	& {\sc i}&{\sc ii}&{\sc iii}&{\sc iv}&{\sc v}&{\sc vi}&{\sc vii}\\
\hline
H	& 0.14(-1)& 0.986	& 	& 	& 	& 	& 	\\
	& 0.85(-2)& 0.991	& 	&  	& 	& 	& 	\\
	&	  &		&	&	&	&	&	\\
He	& 0.64(-2)& 0.509	& 0.485	&  	& 	& 	& 	\\
	& 0.32(-2)& 0.513	& 0.483	&  	& 	& 	& 	\\
	&	&		&	&	&	&	&	\\
C	& 0.73(-4)& 0.33(-1)	& 0.284	& 0.462	& 0.221	&	&	\\
	& 0.96(-4)& 0.23(-1)	& 0.258	& 0.509	& 0.210	&	&	\\
	&	&	&	&	&	&	&	\\
N	& 0.53(-2)& 0.32(-1)	& 0.279	& 0.501	& 0.129	& 0.053	& 	\\
	& 0.23(-2)& 0.25(-1)	& 0.256	& 0.544	& 0.125	& 0.047	& 	\\
	&	&	&	&	&	&	&	\\
O	& 0.13(-1)& 0.28(-1)	& 0.544	& 0.250	& 0.120	& 0.038	& 0.62(-2)\\
	& 0.71(-2)& 0.24(-1)	& 0.562	& 0.257	& 0.112	& 0.033	& 0.52(-2)\\
	&	&	&	&	&	&	&	\\
Ne	& 0.68(-3)& 0.14(-1)	& 0.553	& 0.195	& 0.192	& 0.045	& 0.99(-3)\\
	& 0.12(-3)& 0.66(-2)	& 0.560	& 0.207	& 0.186	& 0.039	& 0.92(-3)\\
	&	&	&	&	&	&	&	\\
S	& 0.11(-4)& 0.21(-1)	& 0.116	& 0.295	& 0.386	& 0.149	& 0.032	\\
	& 0.15(-4)& 0.13(-1)	& 0.094 & 0.276	& 0.445	& 0.144	& 0.028	\\
	&	&	&	&	&	&	&	\\
Si	& 0.81(-4)& 0.133	& 0.128	& 0.247	& 0.211	& 0.211	& 0.070	\\
	& 0.84(-4)& 0.101	& 0.110	& 0.243	& 0.249	& 0.230	& 0.068 \\
	&	  &		&	&	&	&	&	\\
Mg	& 0.15(-2)& 0.59(-1)	& 0.351	& 0.123	& 0.179	& 0.225	& 0.060	\\
	& 0.15(-2)& 0.44(-1)	& 0.313 & 0.149 & 0.203	& 0.233	& 0.056 \\
\hline
\label{tab:spindleionratio}
\end{tabular}
\end{center}
\small{For each element the upper row is for model~A and the lower row is for model~B }
\end{table*}
\renewcommand{\baselinestretch}{1.5}

\renewcommand{\baselinestretch}{1.2}
\begin{table*}
\begin{center}
\caption{Mean temperatures (K) weighted by ionic species for NGC~3918 (Spindle-like models of NGC~3918) }
\begin{tabular}{lccccccc}
\multicolumn{8}{c}{} \\
\hline
  	&		&		&		& Ion	 	&		&		&	\\
\cline{2-8}
Element	& {\sc i}	&{\sc ii}	&{\sc iii}	&{\sc iv}	&{\sc v}	&{\sc vi}	&{\sc vii}\\
\hline
H	& 10,142	& 13,739	& 		&		&		&		&	\\
	& 10,763 	& 13,593 	& 		&  		& 		& 		& 	\\
	&		&		&		&		&		&		&	\\
He	&  9,795	& 12,430	& 15,061	&  		& 		& 		& 	\\
	& 10,519	& 12,457	& 14,770	&  		& 		& 		& 	\\
	&		&		&		&		&		&		&	\\
C	& 10,683	& 10,715	& 12,521	& 13,733	& 15,539	& 		&	\\
	& 10,945	& 11,193	& 12,484	& 13,518	& 15,278	&		&	\\
	&		&		&		&		&		&		&	\\
N	&  9,254	& 10,853	& 12,653	& 13,805	& 15,381	& 16,088	& 	\\
	&  9,805	& 11,239	& 12,606	& 13,594	& 15,085	& 15,924	& 	\\
	&		&		&		&		&		&		&	\\
O	&  9,606	& 11,256	& 12,729	& 14,939	& 15,524	& 16,020	& 17,278 \\
	& 10,188	& 11,518	& 12,689	& 14,656	& 15,288	& 15,792	& 17,839 \\
	&	&	&	&	&	&	&	\\
Ne	&  9,093	& 10,202	& 12,603	& 14,771	& 15,422	& 16,004	& 17,881 \\
	&  9,663	& 11,186	& 12,574	& 14,464	& 15,149	& 15,785	& 19,909 \\
	&	&	&	&	&	&	&	\\
S	& 10,487	& 10,207	& 11,871	& 12,801	& 14,221	& 15,434	& 16,218 \\
	& 10,872	& 10,796	& 11,840	& 12,712	& 13,875	& 15,133	& 16,117 \\
	&	&	&	&	&	&	&	\\
Si	& 11,636	& 11,557	& 12,579	& 13,117	& 14,308	& 15,081	& 15,736 \\
	& 11,628	& 11,652	& 12,443	& 12,968	& 13,935	& 14,655	& 15,381 \\
	&	&	&	&	&	&	&	\\
Mg	& 11,403	& 11,189	& 12,197	& 13,730	& 14,759	& 15,250	& 15,812 \\
	& 11,446	& 11,456	& 12,106	& 13,338	& 14,384	& 14,902	& 15,529	 \\
\hline
\label{tab:spindletemperatures}
\end{tabular}
\end{center}
\small{For each element the upper row is for model~A and the lower row is for model~B }
\end{table*}
\renewcommand{\baselinestretch}{1.5}

\begin{figure*}
\begin{center}
\begin{minipage}[t]{6.cm} 
\psfig{file=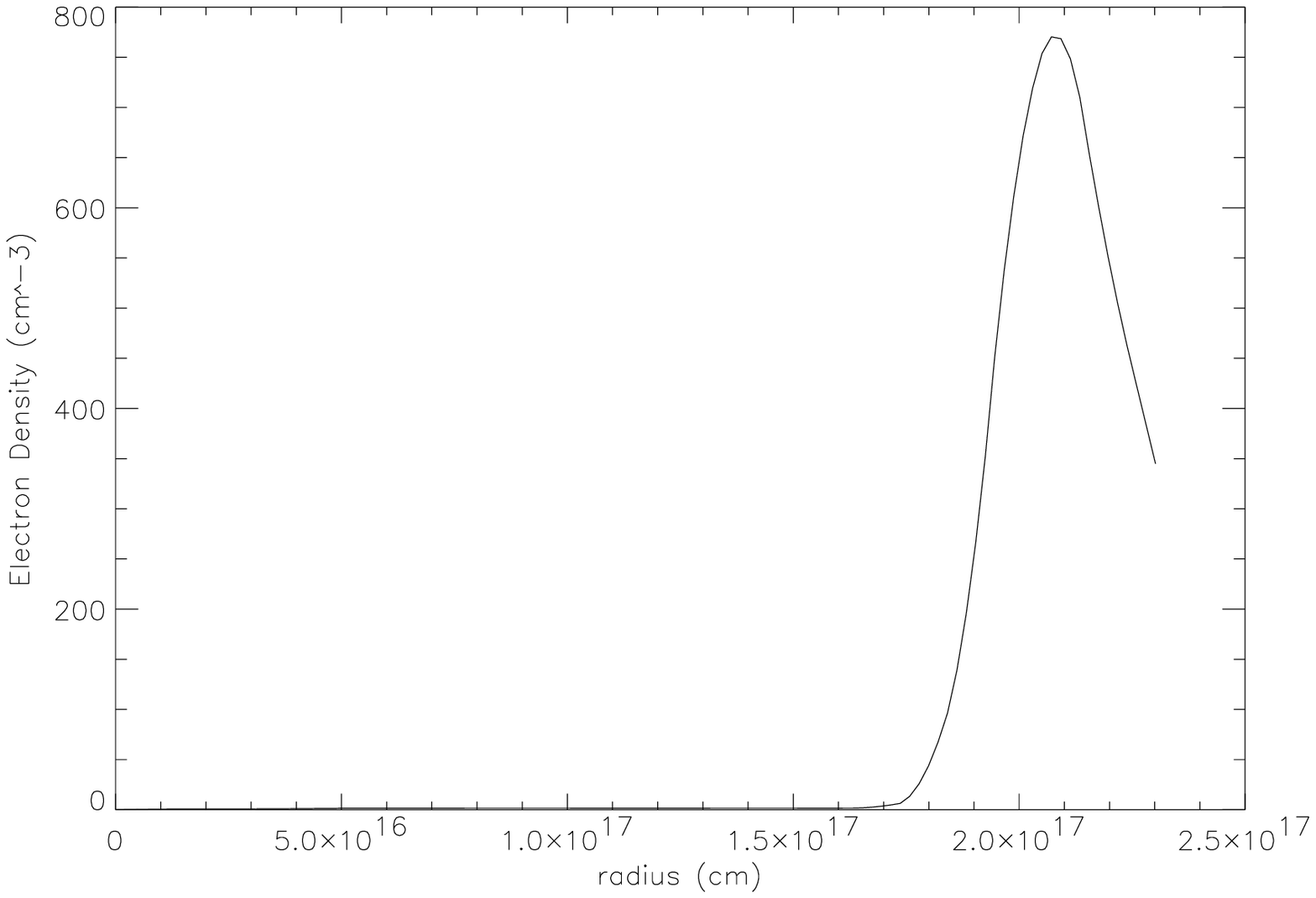, height=60mm, width=60mm}
\end{minipage}
\begin{minipage}[t]{6.cm}
\psfig{file=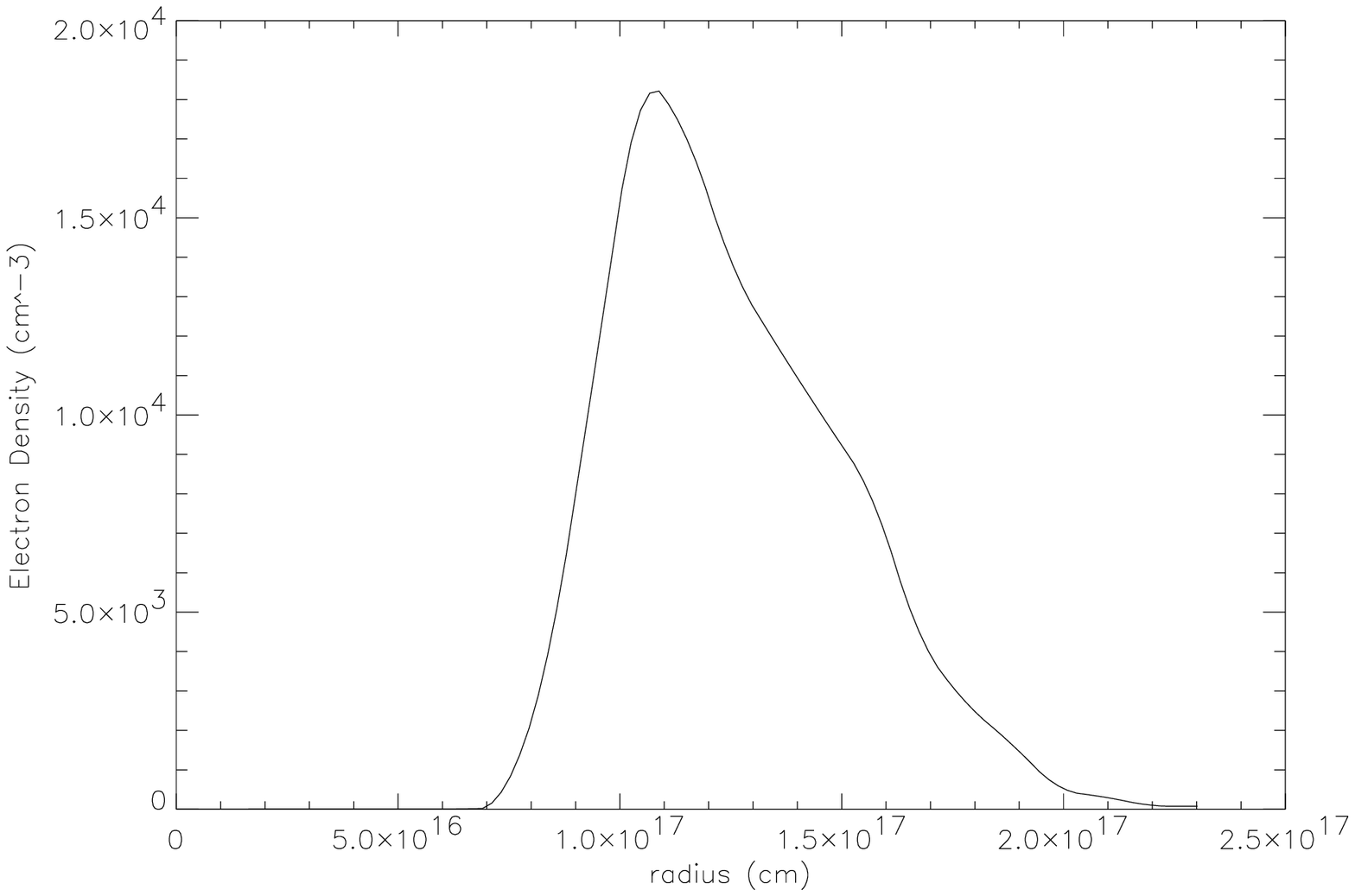, height=60mm, width=60mm}
\end{minipage}
\begin{minipage}[t]{6.cm} 
\psfig{file=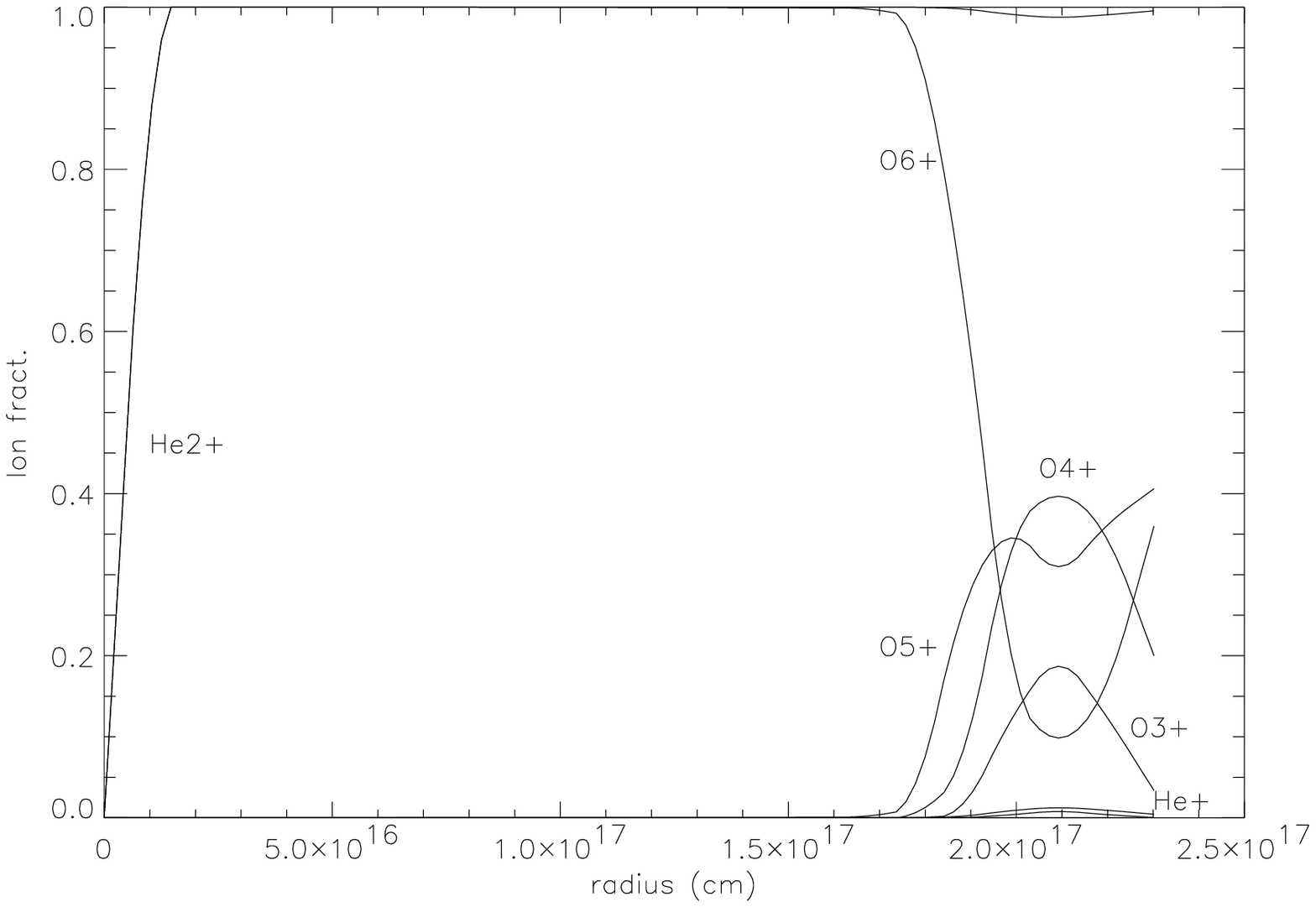, height=60mm, width=60mm}
\end{minipage}
\begin{minipage}[t]{6.cm}
\psfig{file=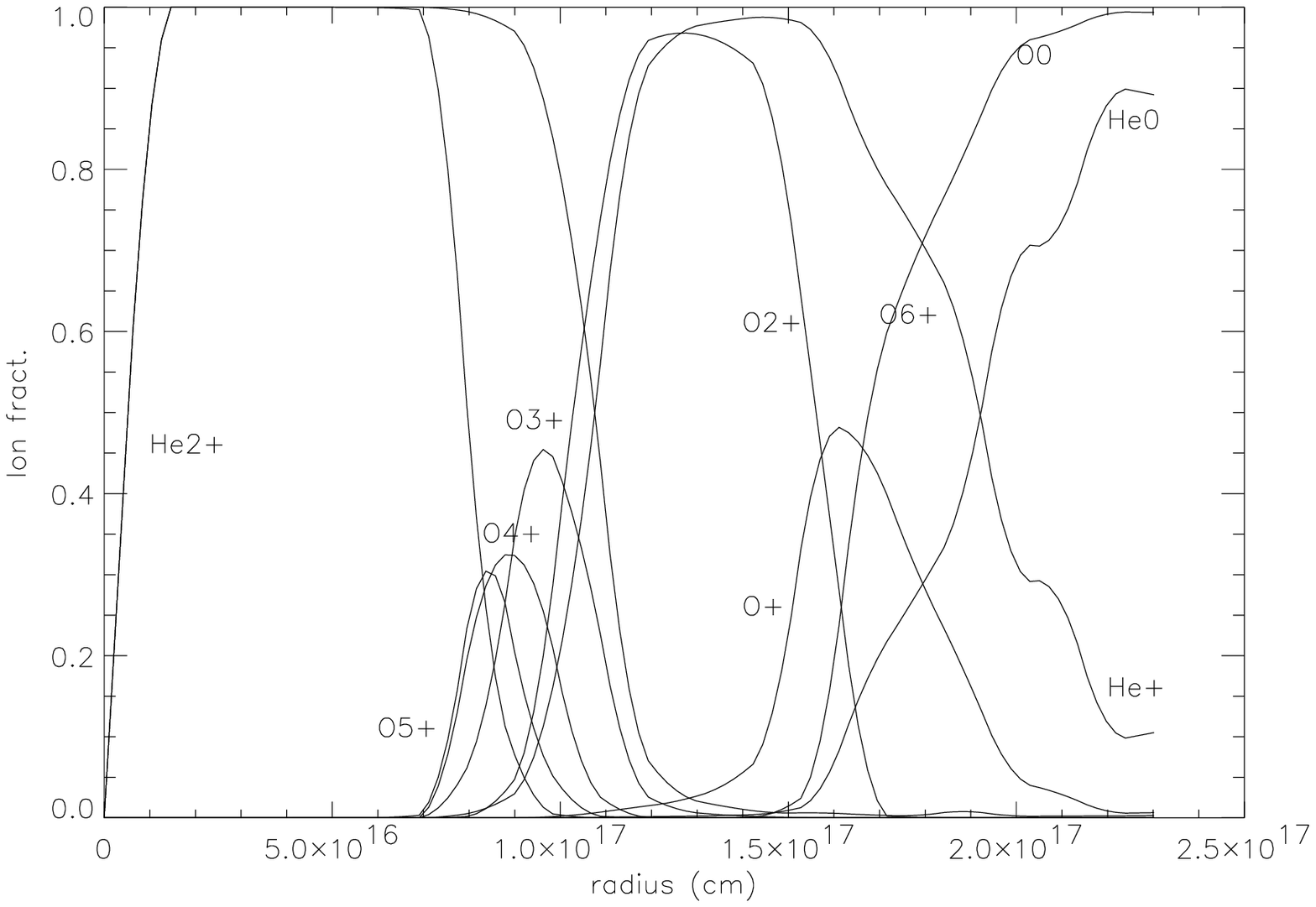, height=60mm, width=60mm}
\end{minipage}
\begin{minipage}[t]{6.cm} 
\psfig{file=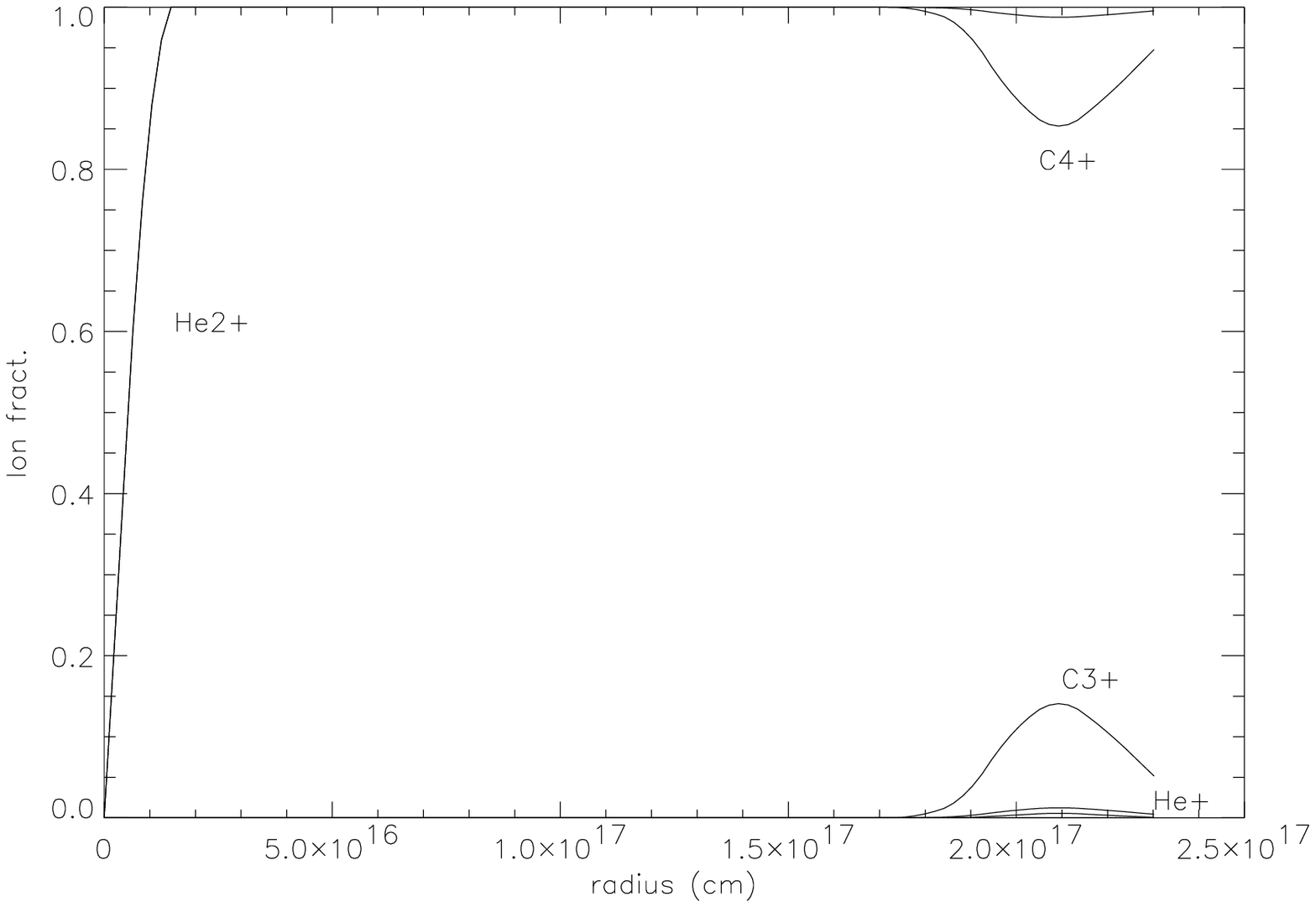, height=60mm, width=60mm}
\end{minipage}
\begin{minipage}[t]{6.cm}
\psfig{file=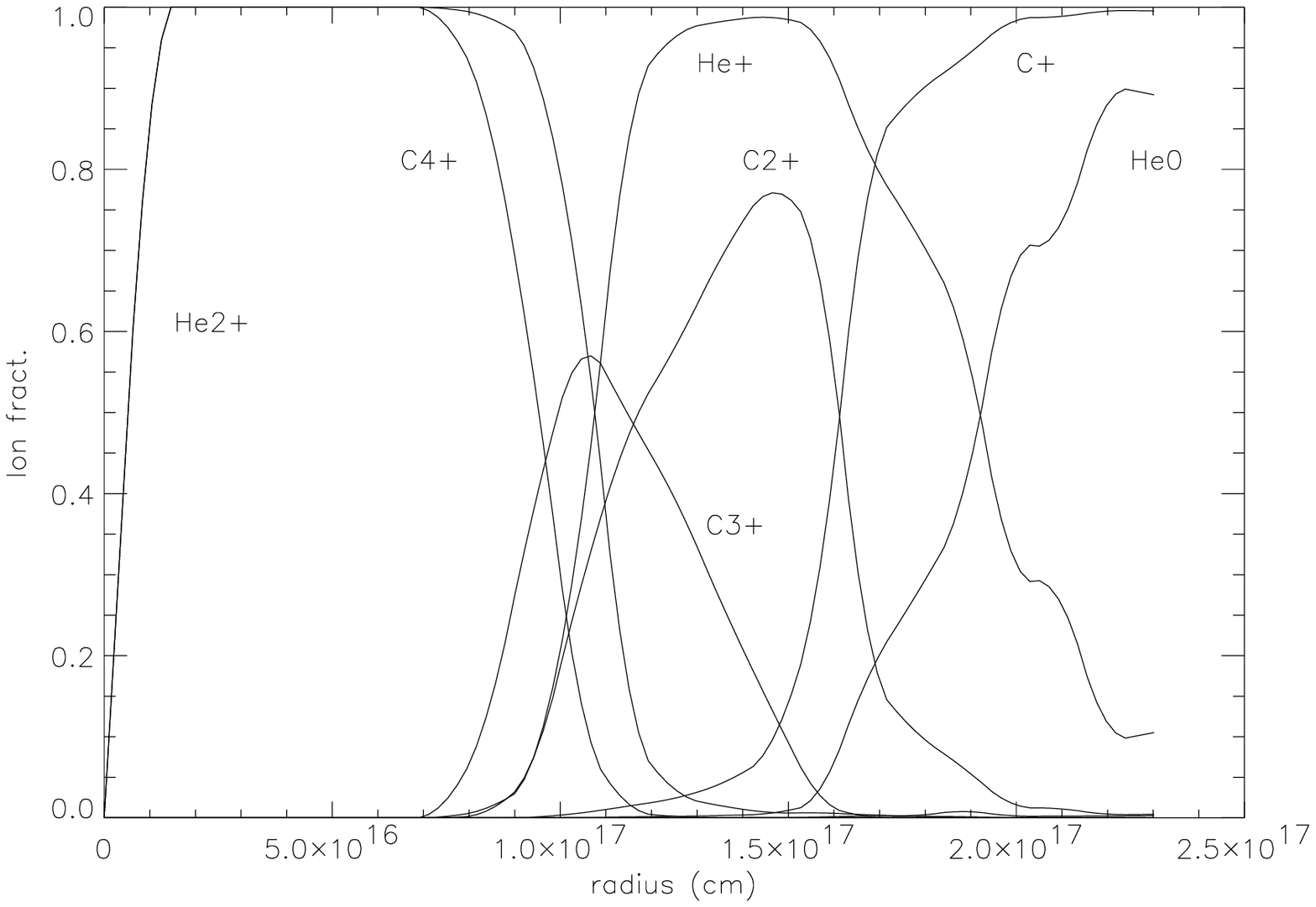, height=60mm, width=60mm}
\end{minipage}
\caption[The ionic fractions of oxygen and carbon along the polar and equatorial directions in NGC3918. Spindle-like model~A.]{Ionic fractions for oxygen (middle panels) and carbon (bottom panels) along the polar (left panels) and equatorial (right panels) directions. The ionization structure of helium is also plotted in each panel.The top panels show the electron density plotted as a function of radius along the polar (top left panel) and equatorial (top right panel) directions. Spindle-like model~A for NGC~3918.}
\label{fig:ionizationA}
\end{center}
\end{figure*}

\begin{figure*}
\begin{center}
\begin{minipage}[t]{6.cm} 
\psfig{file=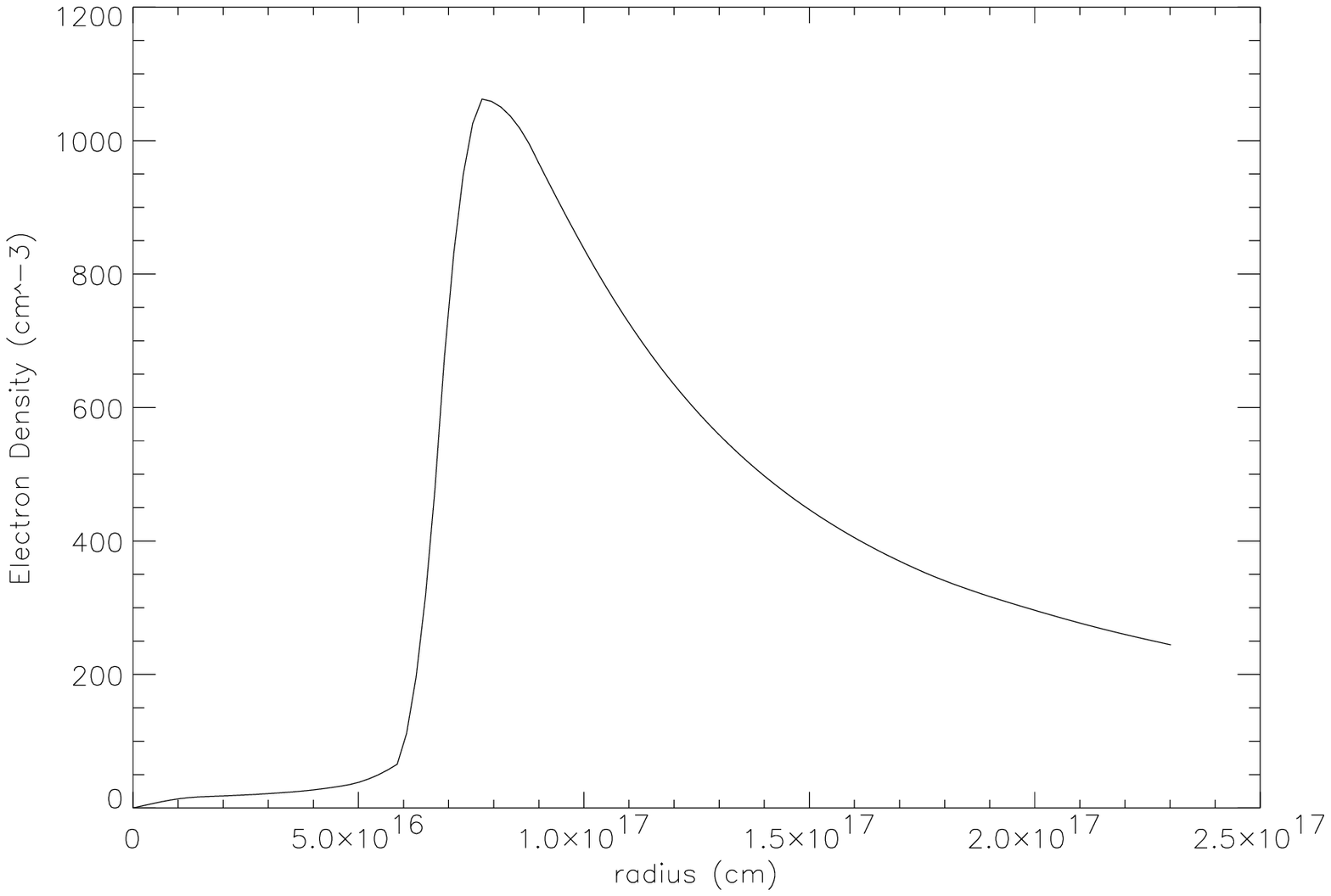, height=60mm, width=60mm}
\end{minipage}
\begin{minipage}[t]{6.cm}
\psfig{file=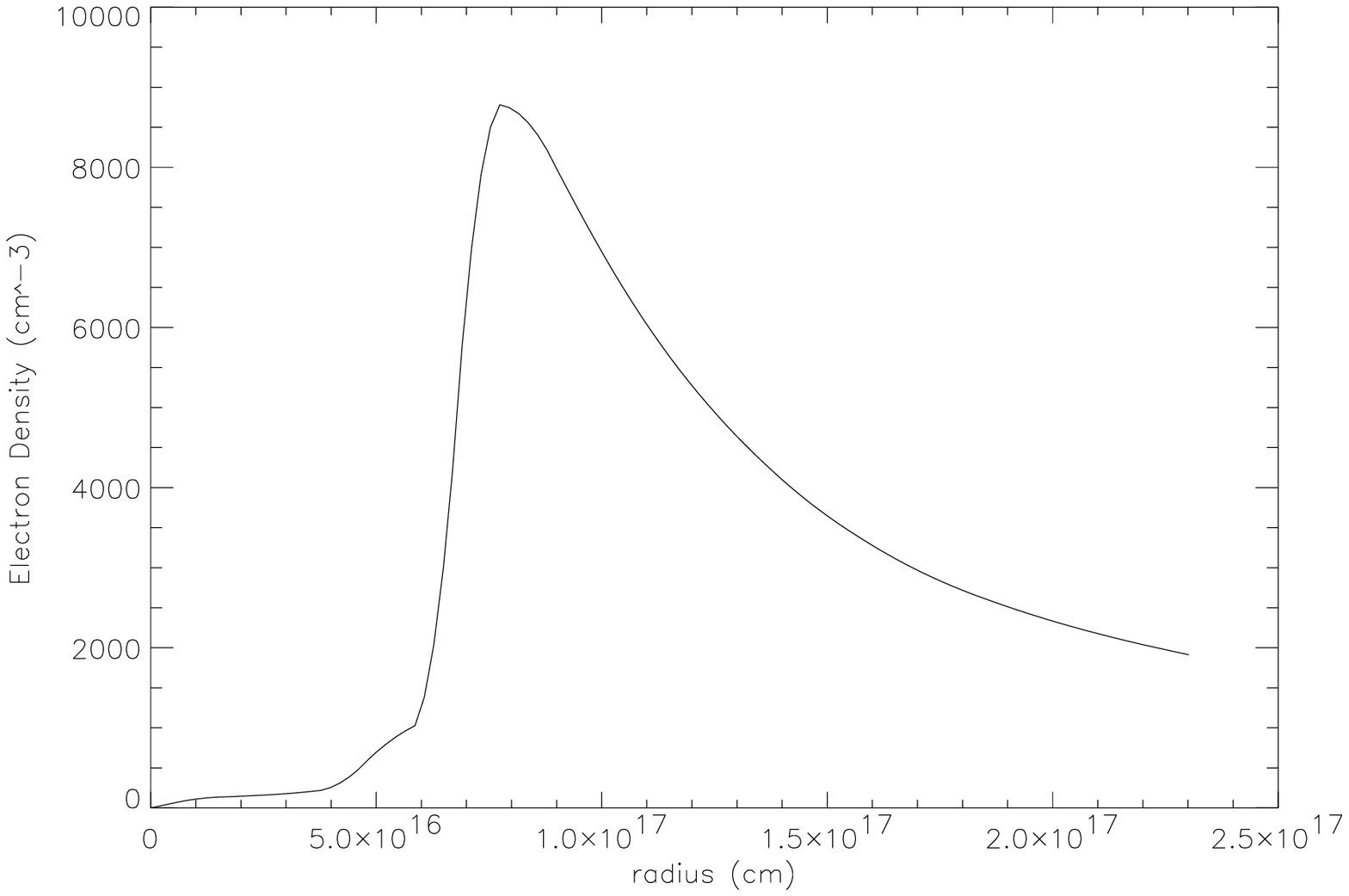, height=60mm, width=60mm}
\end{minipage}
\begin{minipage}[t]{6.cm} 
\psfig{file=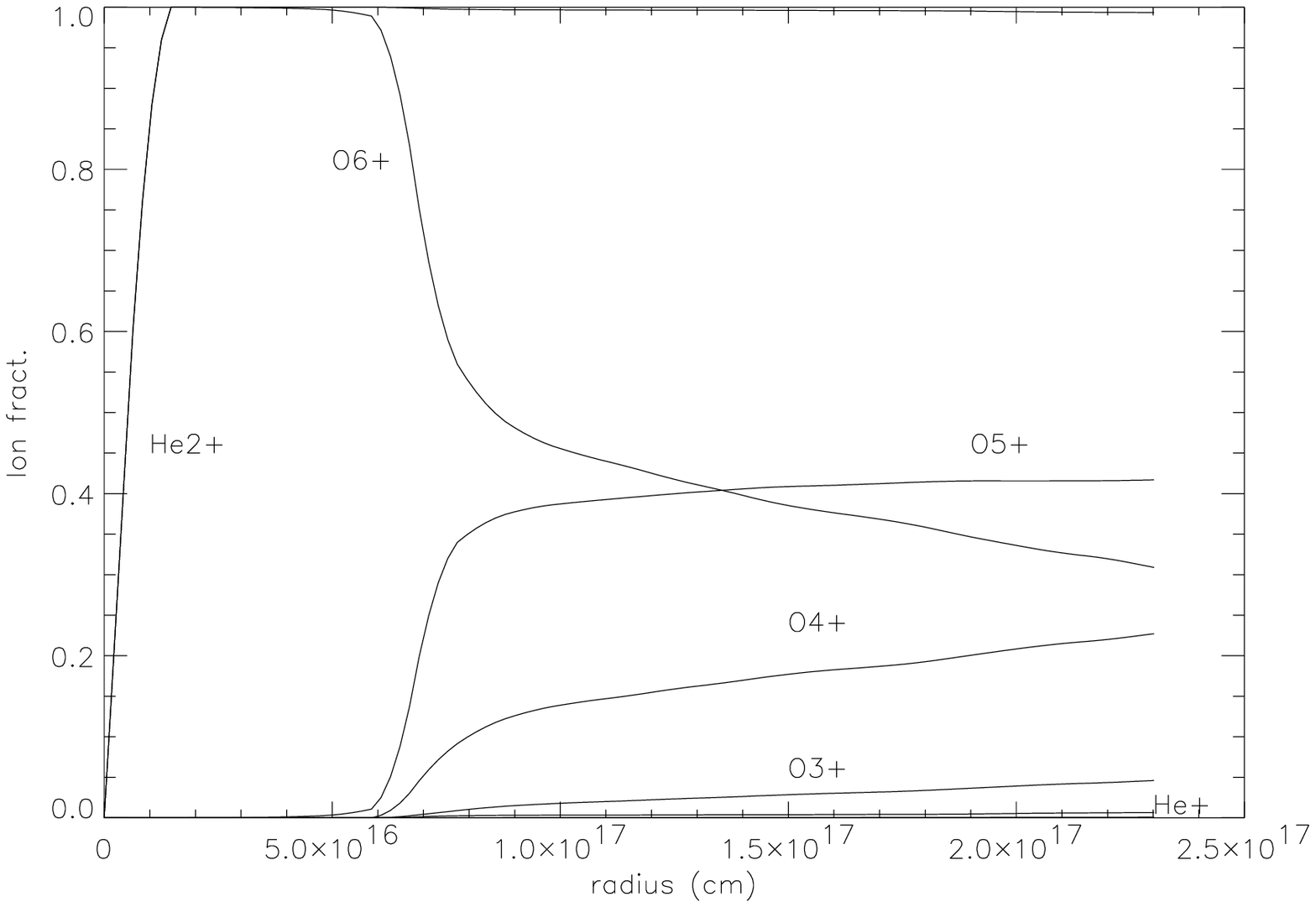, height=60mm, width=60mm}
\end{minipage}
\begin{minipage}[t]{6.cm}
\psfig{file=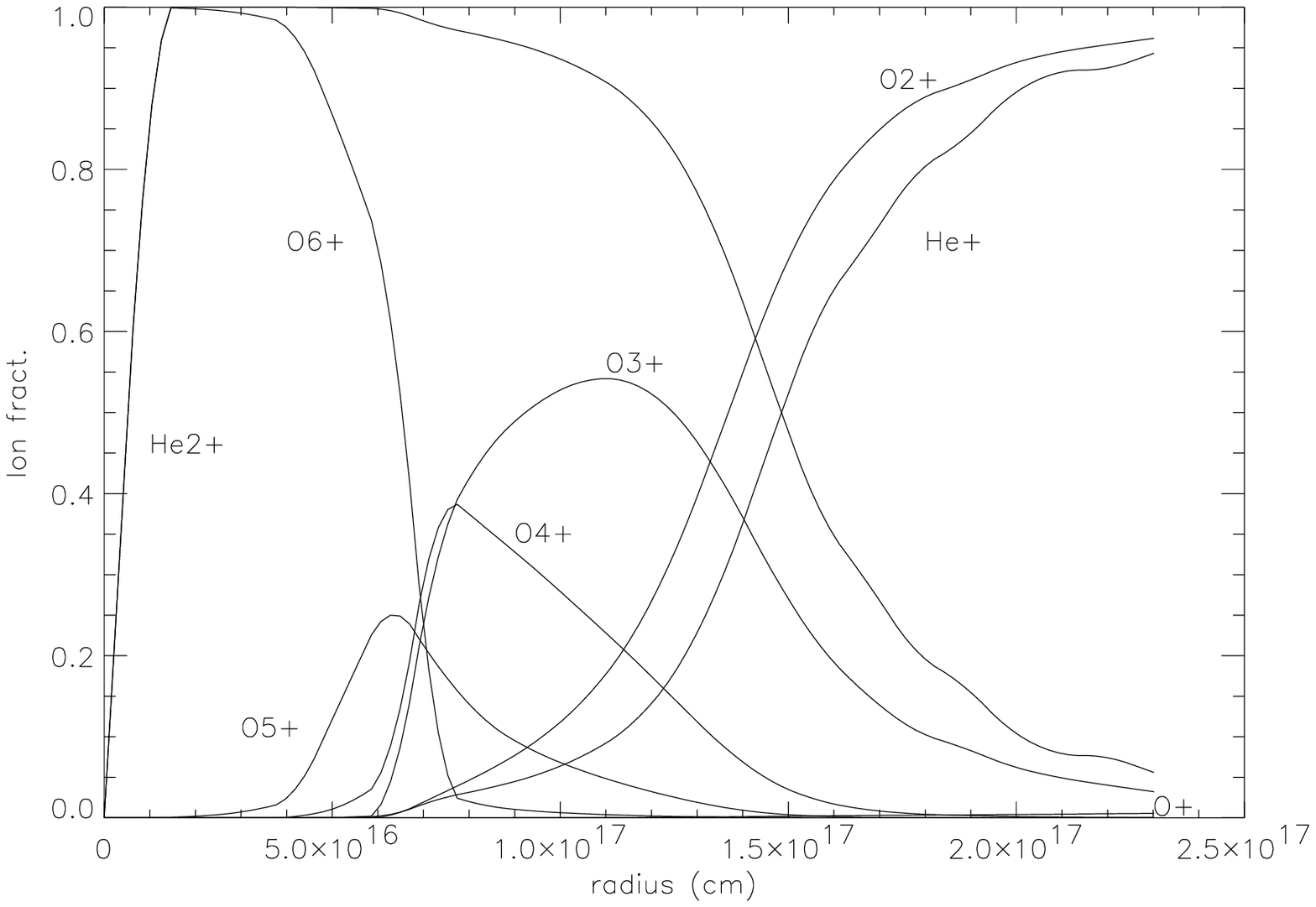, height=60mm, width=60mm}
\end{minipage}
\begin{minipage}[t]{6.cm} 
\psfig{file=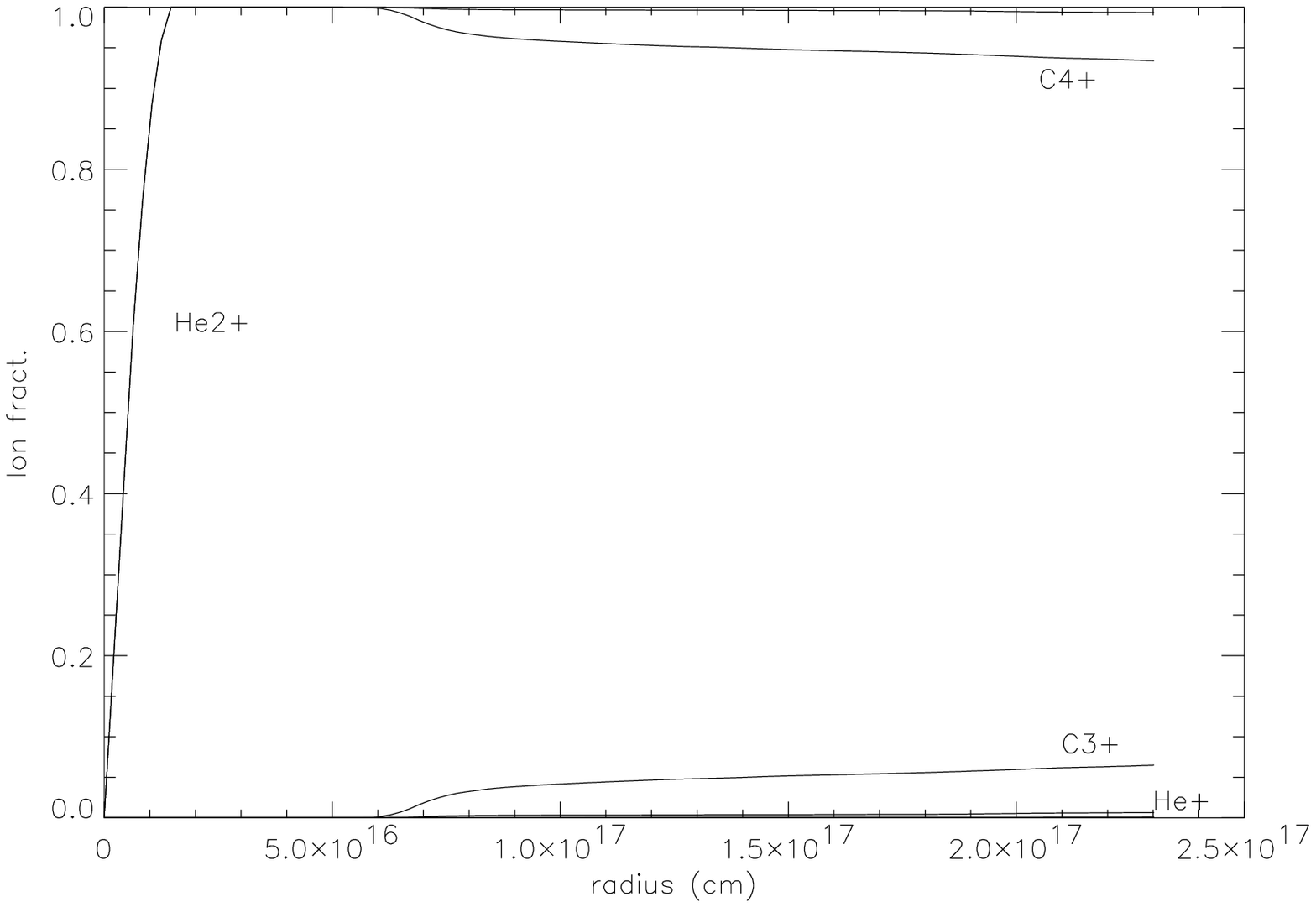, height=60mm, width=60mm}
\end{minipage}
\begin{minipage}[t]{6.cm}
\psfig{file=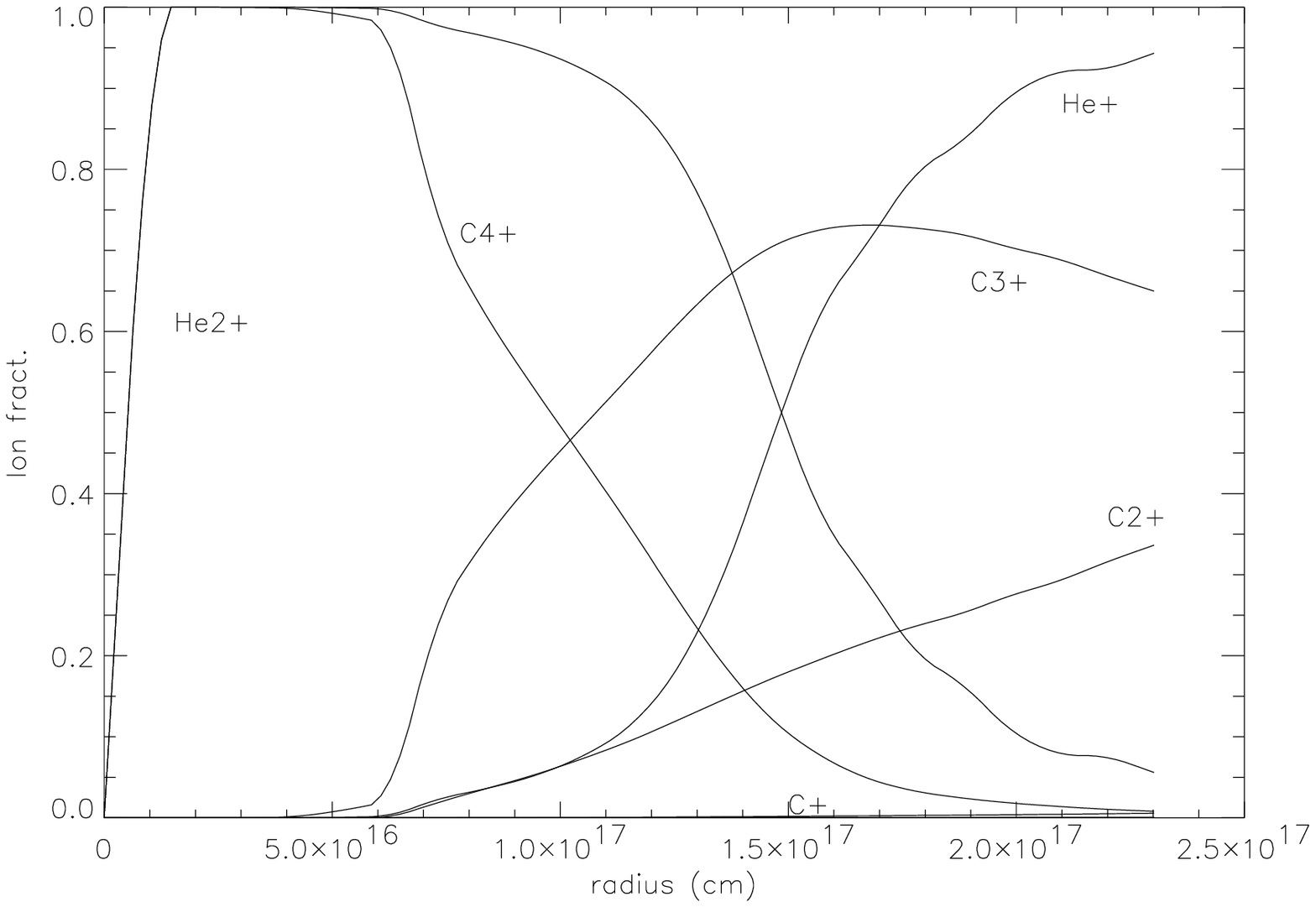, height=60mm, width=60mm}
\end{minipage}
\caption[The ionic fractions of oxygen and carbon along the polar and equatorial directions in NGC3918. Spindle-like model~B.]{Ionic fractions for oxygen (middle panels) and carbon (bottom panels) along the polar (left panels) and equatorial (right panels) directions. The ionization structure of helium is also plotted in each panel. The top panels show the electron density plotted as a function of radius along the polar (top left panel) and equatorial (top right panel) directions. Spindle-like model~B for NGC~3918.}
\label{fig:ionizationB}
\end{center}
\end{figure*}

Figure~\ref{fig:mellemadist}, from C99, shows  a qualitative representation of a density distribution obtained from a hydrodynamical model by \citet{mellema96} that was adopted by C99 in their observational study of the spatio-kinematical properties of NGC~3918. The model chosen was one from a set that Dr. G. Mellema computed for his original thesis work (parameters $B=6$ and $A=0.9$ in his calculations, page 126). 

The exact density distribution for the model used in the C99 paper was kindly provided by G.~Mellema, and a first attempt to construct a photoionization model for this distribution was carried out. The density distribution, however, had to be scaled up by a factor of 1.6 in order to match the observed H$\beta$ flux. However, the integrated emission line spectrum still did not match the observations, as the emission line strengths from the lower ionization species were heavily underestimated.  From the images, however, it is evident that NGC~3918 has an inner shell of denser material which {\it screens} the gas behind it from direct starlight, thus allowing the existence of a lower ionization region. In Mellema's simulations such {\it shells} can only be made by considering an evolving central star and having a slow H ionization front. This, however, was not the case for the model used in C99 analysis. In order to simulate such a dense inner shell, the density was artificially raised in the equatorial region to form a thick torus between the two lobes. The torus had to be thick enough to effectively screen the material behind it, and the latitudinal extension of the density enhancement was determined such that the model would return the observed H$\beta$ flux. This trade-off between the density enhancement factor and its spatial extension resulted in the creation of a thick equatorial waist. The three-dimensional iso-density plot (obtained for a high density threshold) of the final density distribution used for this model is shown in Figure~\ref{fig:garreltdensity}. This model will be referred to as {\it model~A}.

A second attempt to reconstruct the density distribution of this object was to seek an analytical expression which could reproduce the structure shown in Figure~\ref{fig:mellemadist} and in the {\it HST} and {\it NTT} images published by C99. The parameters of the central star and the adopted nebular abundances were the same as in the biconical model. The density distribution was obtained by combining a sphere with an ellipsoid, using a set of analytical epressions as described in Appendix~A. The best fit to the observed integrated spectrum and to the images of NGC~3918 was obtained using the parameters listed in Table~\ref{tab:spindleparametersan}. Figure~\ref{fig:chrisdensity} shows the three-dimensional iso-density plot corresponding to this combination. In this work the spindle-like model constructed by using the analytical description will be referred to as {\it model~B}.

 C99's study primarily focused on the kinematics of NGC~3918 and therefore they did not propose any new values for the central star parameters or for the nebular elemental abundances. Although derived for a different density distribution, the parameters derived by C87 were used by us for these models too. As given by C87, the dereddened stellar flux at 1300~{\AA} is equal to 2.4$\times$10$^{-12}$ergs\,cm$^{-2}$s$^{-1}${\AA}$^{-1}$. For an assumed distance of 1.5~kpc and a stellar effective temperature of T$_{eff}$=150,000\,K, this flux corresponds to a stellar luminosity, L$_*$, of 5780\,L$_{\odot}$. 

The stellar luminosity in the C87 model, however, was finally adjusted by them to a slightly higher value of F$_*$(1300{\AA})\,=\,2.9$\times$10$^{-12}$ergs\,cm$^{-2}$s$^{-1}${\AA}$^{-1}$, corresponding to L$_*$~=~6900\,L$_{\odot}$. In this work both values for F$_*$(1300{\AA}) were used, in turn, for both models A and B. Finally, F$_*$(1300{\AA})~=~2.9$\times$10$^{-12}$ergs\,cm$^{-2}$s$^{-1}${\AA}$^{-1}$ (L$_*$~=~6900\,L$_{\odot}$) was chosen for model~A and F$_*$(1300{\AA})~=~2.4$\times$10$^{-12}$ergs\,cm$^{-2}$s$^{-1}${\AA}$^{-1}$ (L$_*$~=~5780L$_{\odot}$) was chosen for model~B, as these produced the best results. Table~\ref{tab:garreltpar} summarises the nebular and stellar parameters used for the models described in this section. 

\subsection{Spindle-like Density Distribution Models: Results}
\label{sub:spindleresults}

Table~\ref{tab:garreltlines} lists the predicted emission line fluxes integrated over the nebular volume for the spindle-like models~A and~B for NGC~3918; for comparison, the observed values are also reported in the last column of this table. The observed spectrum is reasonably reproduced by both models. The  H$\beta$ fluxes predicted by models A and B are slightly different, but they are both within 10\% of the observed value. The emission line predictions for helium seem to indicate that in both models the abundance of He$^{2+}$ is slightly overestimated, hence leaving too little He$^+$. However both results are comparable to those obtained using the biconical model and indeed to the C87 results presented in Section~\ref{sub:biconicalresults}. The major discrepancies between the two models and the observations are the same as for the biconical model case. The C~{\sc iv}~1549~{\AA} and the Mg~{\sc ii}~2800~{\AA} resonance doublets are overestimated by large factors by the Mocassin models, which do not take dust absorption into account. The Si~{\sc iv}~1393~{\AA} and 1401~{\AA} resonance line doublet, which should also suffer from absorption by dust, does not seem to be affected. In fact if the 30\% attenuation predicted by \citet{harrington88} were to be applied to the flux of this doublet, the fit to the observations would be made worse. However, as has already been discussed in section~\ref{subsub:resonance}, Si~{\sc iv}~1393~{\AA} and 1401~{\AA} are quite weak and the flux measured might not be accurate. As before, large discrepancies with the observations, of factors of about ten, are also observed for the [Ne~{\sc v}] fine structure lines, namely [Ne~{\sc v}]~14.3~$\mu$m and [Ne~{\sc v}]~24.3$\mu$m; this is discussed in Appendices~B and~C. The large discrepancies found between the spindle-like model predictions for the infrared line fluxes on the one hand, and the {\it ISO~SWS} measurements of \citet{bower01} reported in Table~\ref{tab:garreltlines}, on the other hand, are most probably due to the {\it ISO~SWS} aperture having been offset from the centre of the nebula. This is discussed in more detail in Section~\ref{sec:infrared} and Appendix~C. In general, the integrated spectra obtained from the spindle-like models are in good agreement with those obtained from Mocassin's biconical model, and, therefore, comments made during the discussion of the latter (Section~\ref{sub:biconicalresults}) largely apply to the models A and B discussed in this section. The diagnostic line ratios calculated for models A and B are given in Table~\ref{tab:diagnosticsgarrelt}. They are in satisfactory agreement with the observed ratios, also reported in the same table, with model~B providing the slightly better fit. 

Tables~\ref{tab:spindleionratio} and~\ref{tab:spindletemperatures} list, respectively, the fractional ionic abundances, ${\it f}(X^i)$, and the mean electron temperatures weighted by ionic species, for the spindle-like distribution models of NGC~3918. For each element listed in the two tables, the upper entry is for model~A and the lower entry is for model~B. It is difficult to make a comparison with Mocassin's biconical model results or with those of C87, since both of the spindle-like models consist of a continuously varying gas density distribution where clear edges do not exist. The values of the ionic fractions reported in Table~\ref{tab:spindleionratio} were obtained from Equation~(\ref{eq:fnij}), and represent an average over the whole ionized volume of the local ionic fractions weighted by the local density. Figures~\ref{fig:ionizationA} and~\ref{fig:ionizationB} show the ionic fractions of oxygen (middle panels) and carbon (bottom panels) as a function of radius along the polar (left panels) and equatorial (right panels) directions for the spindle-like models A and B, respectively. The electron density distribution along the two directions is also shown in the top panels of Figures~\ref{fig:ionizationA} and~\ref{fig:ionizationB} for the polar (left panels) and equatorial (right panles) directions. It is clear from the figures that the ion distributions are very different along the two directions of each model, according to the gas density structure. Integrated ionic fractions, ${\it f}(X^i)$, such as those listed in Table~\ref{tab:spindleionratio},  are very useful when discussing the large-scale ionization structure of models with density variations, however plots of the type shown in Figures~\ref{fig:ionizationA} and~\ref{fig:ionizationB} can provide a better insight into the actual ionization topography of a model.

As shown in Table~\ref{tab:spindletemperatures}, the weighted mean temperatures calculated by Mocassin for model~B (lower entries) are consistently lower than those calculated for model~A (upper entries), apart from the third and fourth column values (for doubly and triply ionized species), which are, on the contrary, higher for model~B. It is also evident that the scatter of values for the same column of ionic stage is larger for model~B even for the lower ionization stages, while the scatter is very small for model~A. The effects of density distributions on the final values obtained for the mean weighted temperatures have already been discussed in Section~\ref{sub:test}. It is clear that in this case the behaviour of the two models depends heavily on their density distributions. 

\section{The infrared fine-structure line flux discrepancies}
\label{sec:infrared}

\begin{table}
\begin{center}
\caption{Predicted infrared line fluxes from NGC~3918 corrected for the {\it ISO~SWS} beam profile and pointing position. }
\label{tab:infraredCorrections}
\begin{tabular}{lccc}
\multicolumn{4}{c}{} \\
\hline
\hline
		& \multicolumn{2}{c}{Biconical Model$^a$} &			\\
\hline
Line		& \multicolumn{2}{c}{Model predictions} & Observed$^d$	\\
\cline{2-3}
		& Uncorr. & Corr. &  \\
\hline
$[$O~{\sc iv}$]$ 25.9~$\mu$m   & 352 & 96.1 & 89.0 \\
$[$Ne~{\sc iii}$]$ 15.6~$\mu$m & 131 & 33.6 & 46.3 \\
$[$Ne~{\sc v}$]$ 14.3~$\mu$m   & 286 & 69.7 & 19.6 \\
$[$Ne~{\sc v}$]$ 24.3~$\mu$m   & 220 & 60.1 & 23.1 \\
$[$S~{\sc iii}$]$ 18.7~$\mu$m  & 48  & 14.3 & 8.7 \\
$[$S~{\sc iv}$]$ 10.5~$\mu$m   & 220 & 86.1 & 35.0 \\
\hline
\hline
		& \multicolumn{2}{c}{Spindle-like A$^b$} &			\\
\hline
Line		& \multicolumn{2}{c}{Model predictions} & Observed$^d$	\\
\cline{2-3}
		& Uncorr. & Corr. &  \\
\hline
$[$O~{\sc iv}$]$ 25.9~$\mu$m   & 352 & 100  & 89.0 \\
$[$Ne~{\sc iii}$]$ 15.6~$\mu$m & 122 & 21.8 & 46.3 \\
$[$Ne~{\sc v}$]$ 14.3~$\mu$m   & 268 & 68.9 & 19.6 \\
$[$Ne~{\sc v}$]$ 24.3~$\mu$m   & 170 & 49.4 & 23.1 \\
$[$S~{\sc iii}$]$ 18.7~$\mu$m  &  27 & 5.9  & 8.7 \\
$[$S~{\sc iv}$]$ 10.5~$\mu$m   & 255 & 91.5 & 35.0 \\
\hline
\hline
		& \multicolumn{2}{c}{Spindle-like B$^c$} &			\\
\hline
Line		& \multicolumn{2}{c}{Model predictions} & Observed$^d$ \\
\cline{2-3}
		& Uncorr. & Corr. &  \\
\hline
$[$O~{\sc iv}$]$ 25.9~$\mu$m   & 364 & 104  & 89.0 \\
$[$Ne~{\sc iii}$]$ 15.6~$\mu$m & 126 & 40.8 & 46.3 \\
$[$Ne~{\sc v}$]$ 14.3~$\mu$m   & 256 & 58.0 & 19.6 \\
$[$Ne~{\sc v}$]$ 24.3~$\mu$m   & 146 & 40.9 & 23.1 \\
$[$S~{\sc iii}$]$ 18.7~$\mu$m  &  27 & 10.0 & 8.7 \\
$[$S~{\sc iv}$]$ 10.5~$\mu$m   & 302 & 114  & 35.0 \\
\hline
\hline
\end{tabular}
\\
\small{$^a$ The fluxes are given on a scale relative to H$\beta$\,=\,100, where $I({\rm H}\beta)$\,=\,2.33\,$\times$\,10$^{-10}$ ergs\,cm$^{-2}$\,s$^{-1}$.\\
       $^b$ The fluxes are given on a scale relative to H$\beta$\,=\,100, where $I({\rm H}\beta)$\,=\,2.20\,$\times$\,10$^{-10}$ ergs\,cm$^{-2}$\,s$^{-1}$.\\
       $^c$ The fluxes are given on a scale relative to H$\beta$\,=\,100, where $I({\rm H}\beta)$\,=\,2.56\,$\times$\,10$^{-10}$ ergs\,cm$^{-2}$\,s$^{-1}$.\\
       $^d$ The fluxes are given on a scale relative to H$\beta$\,=\,100, where $I({\rm H}\beta)$\,=\,2.34\,$\times$\,10$^{-10}$ ergs\,cm$^{-2}$\,s$^{-1}$.}
\end{center}
\end{table}

\begin{figure*}
\begin{center}
\begin{minipage}[t]{4.7cm} 
\epsfig{file=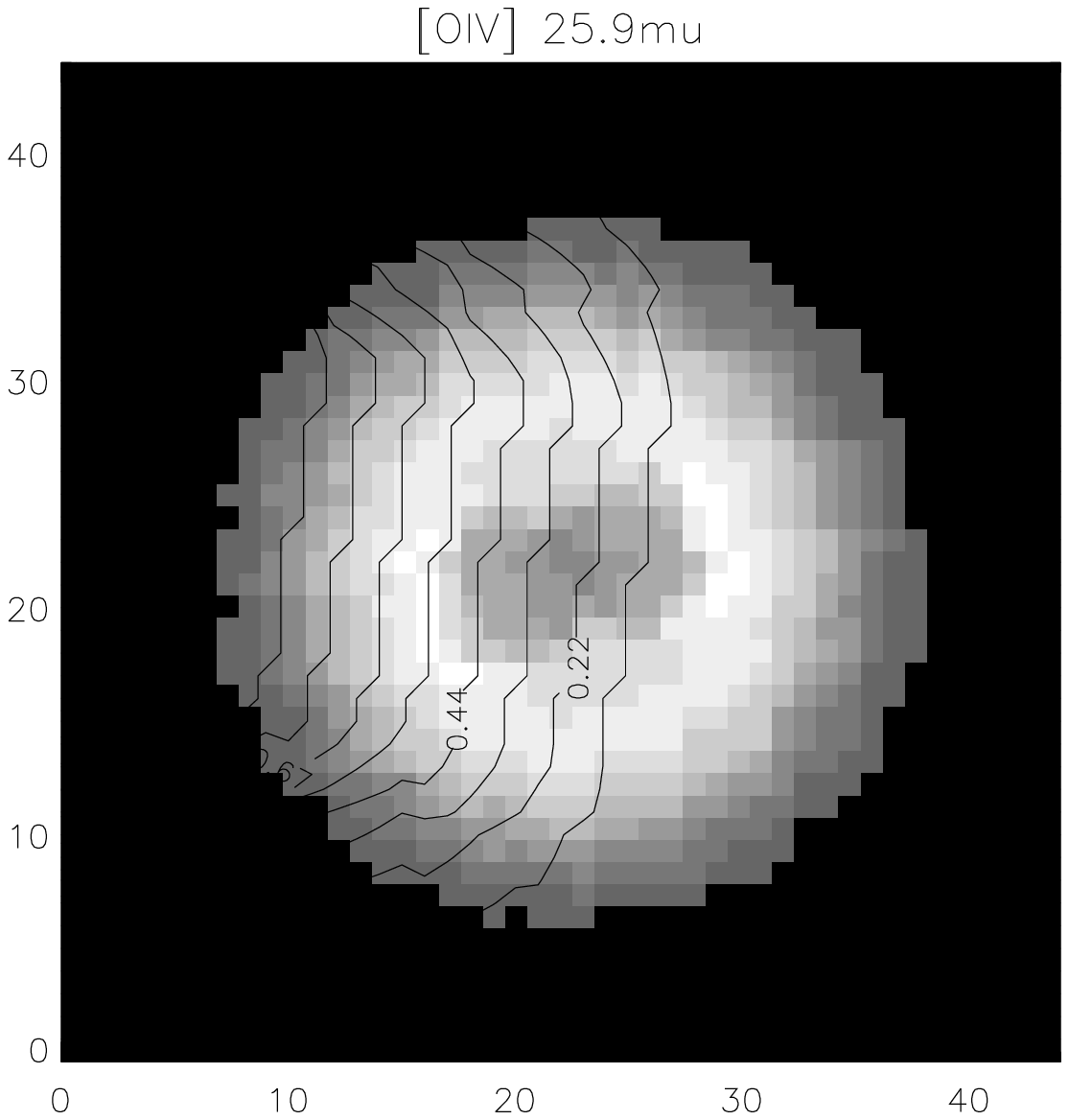,width=4.7cm,clip=,bbllx=144, bburx=487,bblly=381,bbury=720}
\end{minipage}
\begin{minipage}[t]{4.7cm} 
\epsfig{file=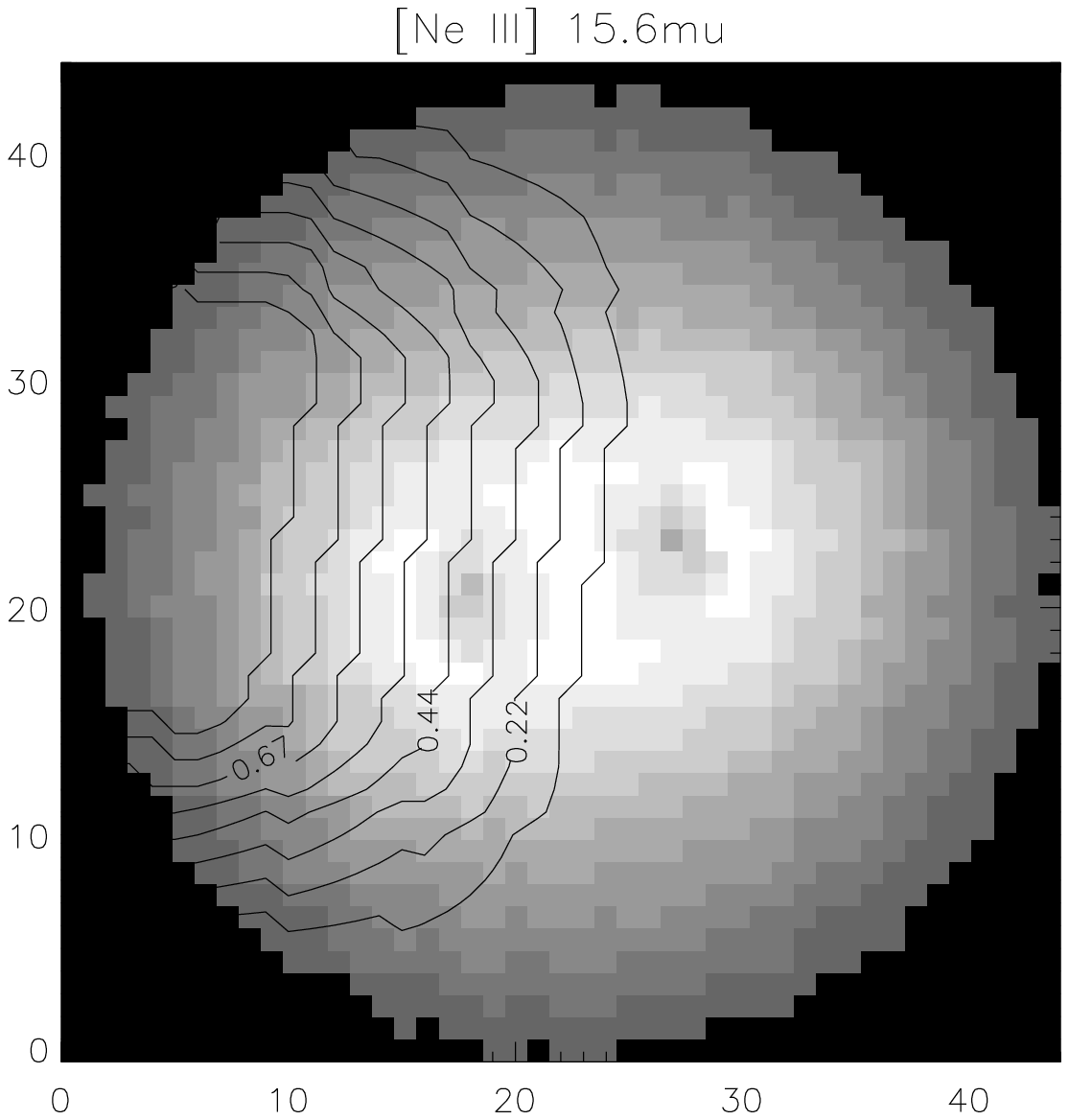,width=4.7cm,clip=,bbllx=144, bburx=487,bblly=381,bbury=720}
\end{minipage}
\begin{minipage}[t]{4.7cm} 
\epsfig{file=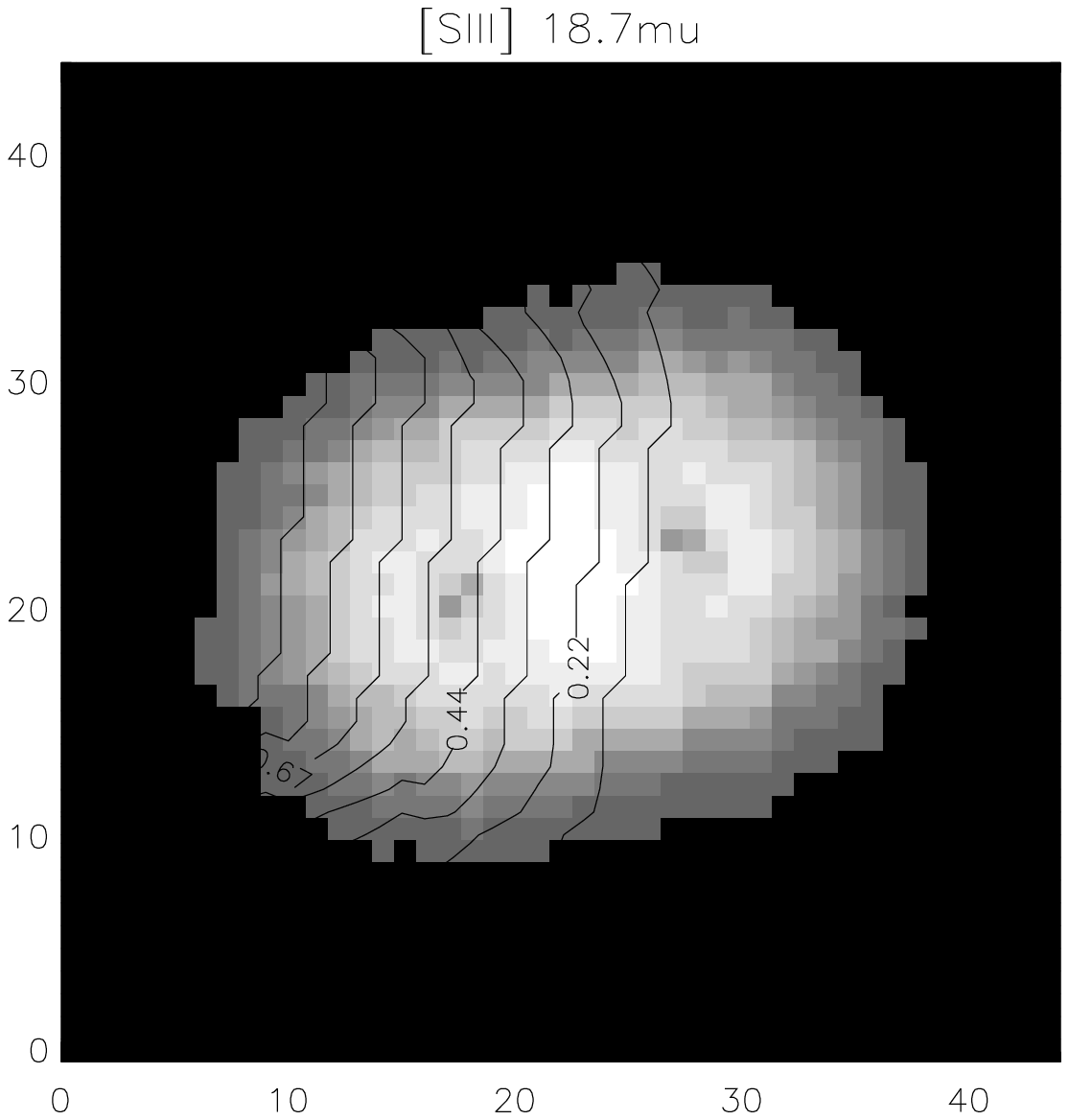,width=4.7cm,clip=,bbllx=144, bburx=487,bblly=381,bbury=720}
\end{minipage}
\begin{minipage}[t]{4.7cm} 
\epsfig{file=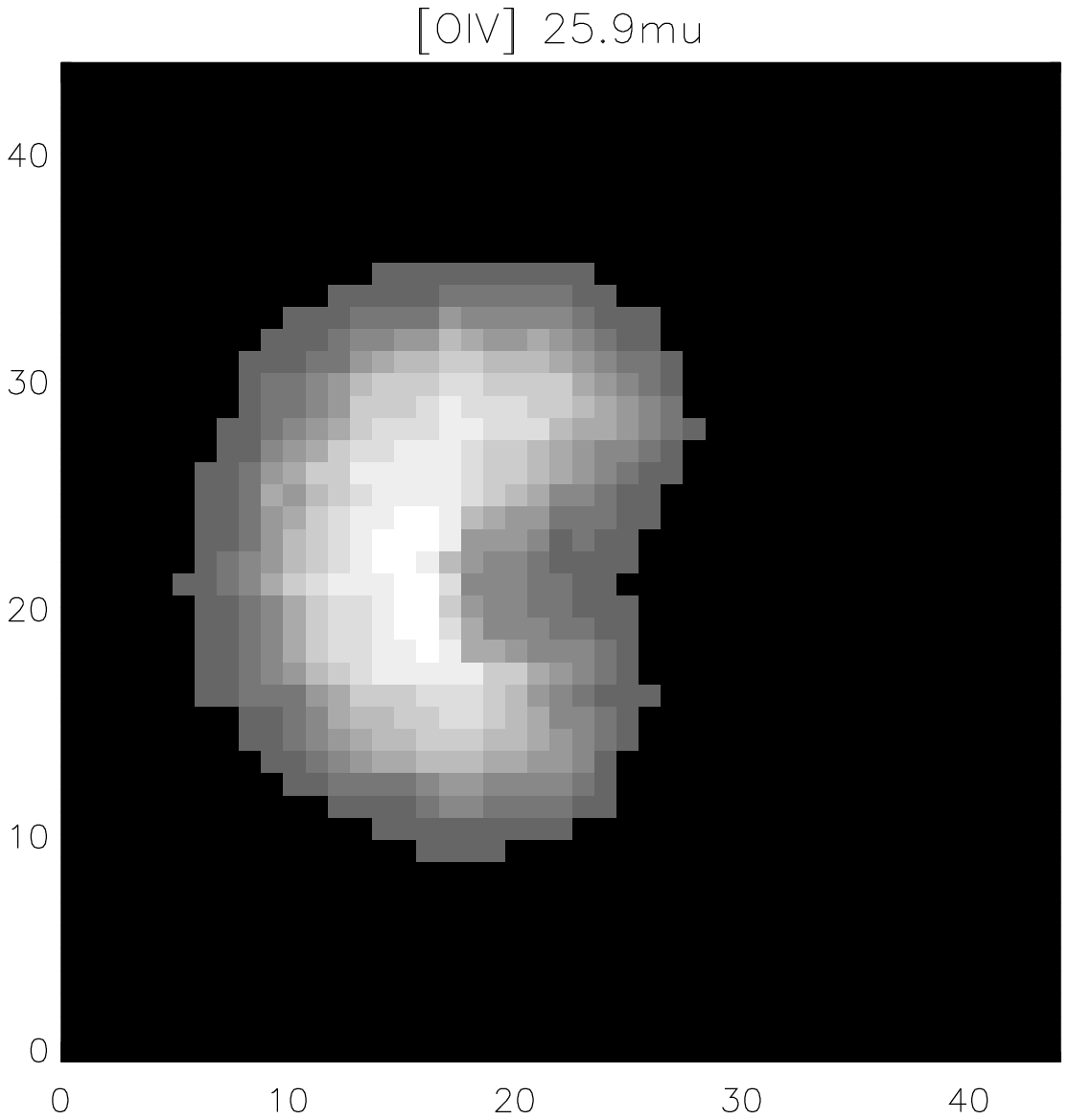,width=4.7cm,clip=,bbllx=144, bburx=487,bblly=381,bbury=720}
\end{minipage}
\begin{minipage}[t]{4.7cm} 
\epsfig{file=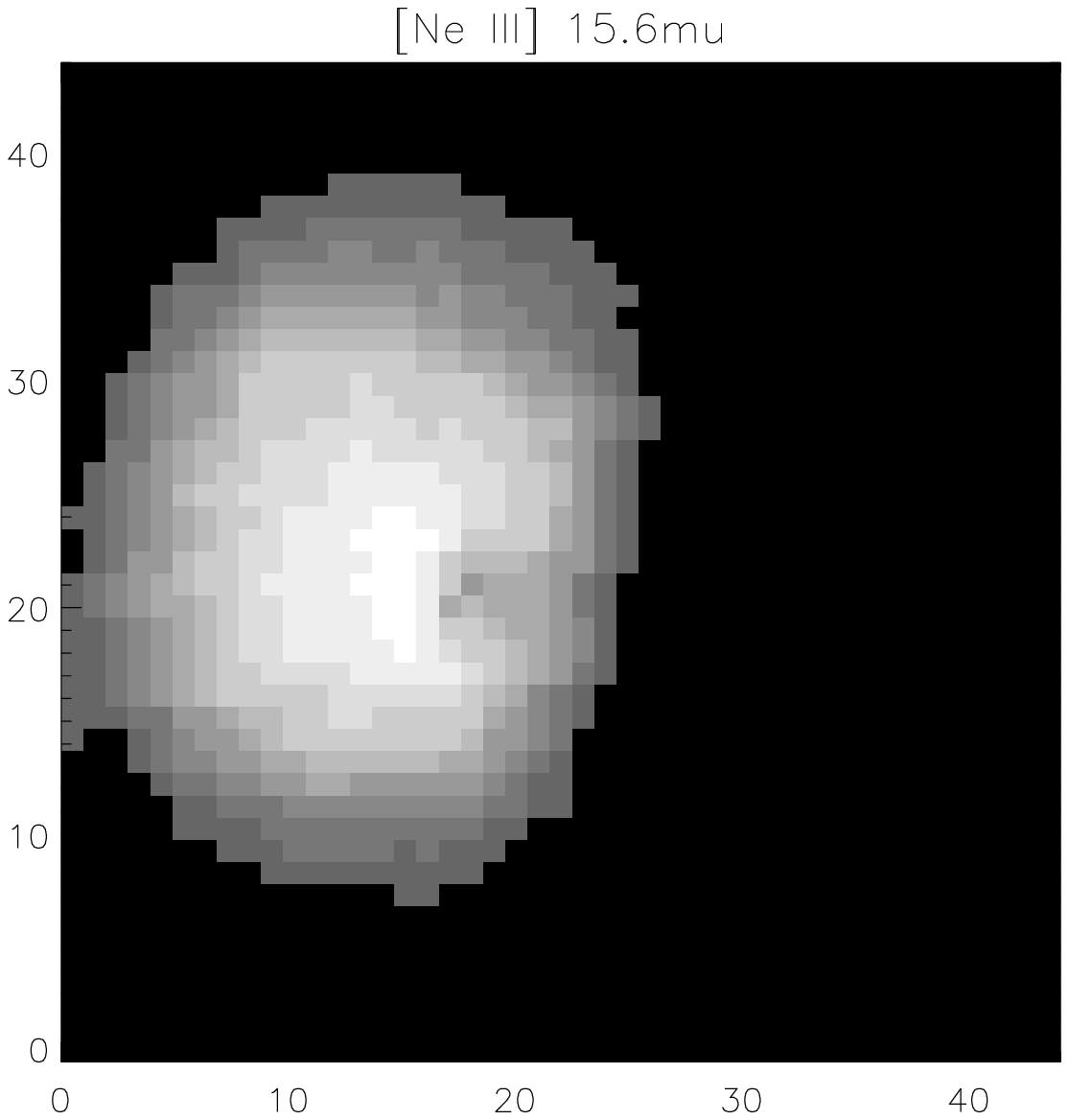,width=4.7cm,clip=,bbllx=144, bburx=487,bblly=381,bbury=720}
\end{minipage}
\begin{minipage}[t]{4.7cm} 
\epsfig{file=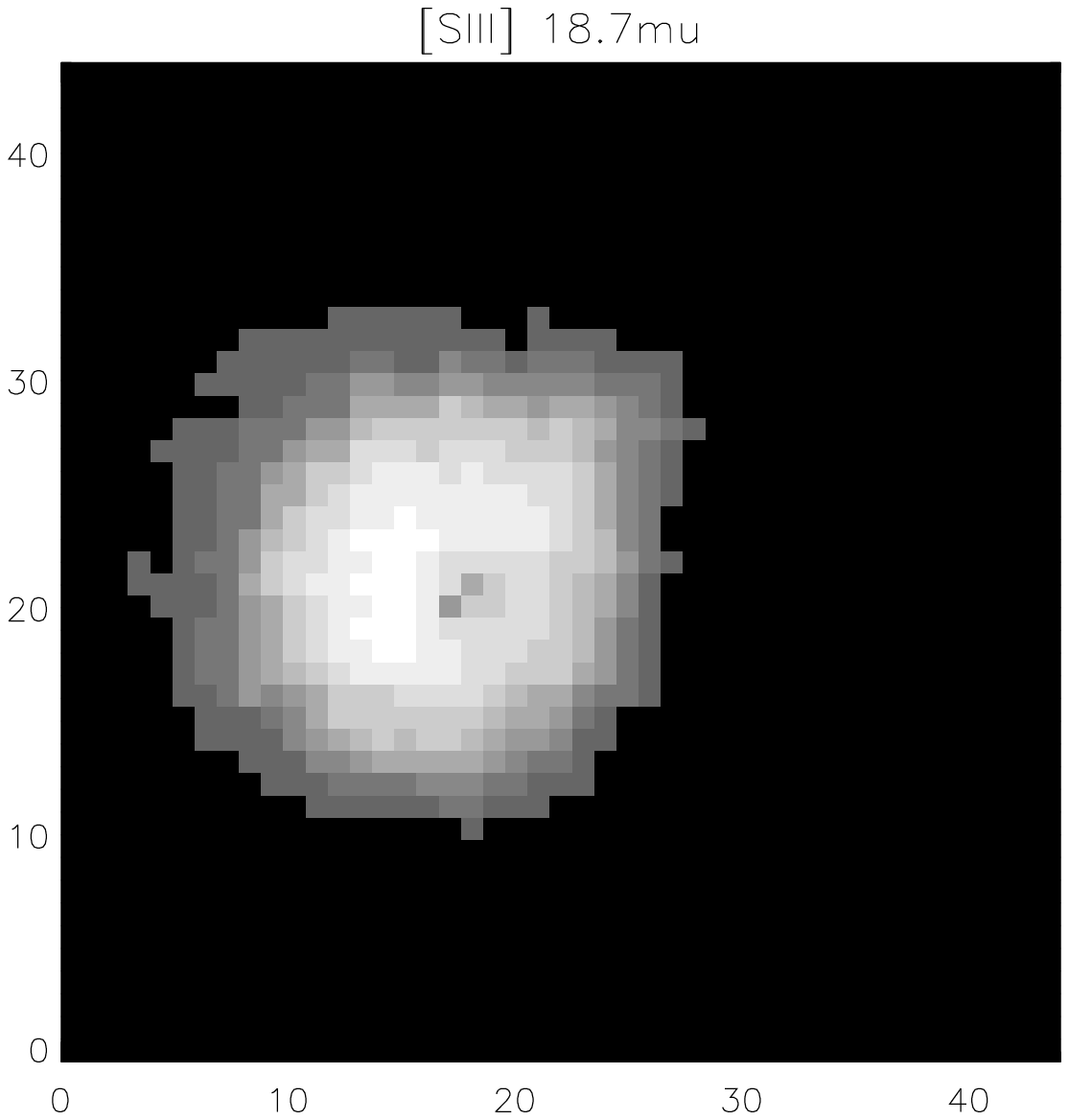,width=4.7cm,clip=,bbllx=144, bburx=487,bblly=381,bbury=720}
\end{minipage}
\caption[Infrared line maps of NGC~3918 through {\it ISO~SWS} aperture masking (Spindle-like Model~B)]{Predicted projected images of NGC~3918 in three infrared fine-structure lines observed by the {\it ISO~SWS}, for the spindle-like model~B. The upper panels show contour plots of the {\it ISO-SWS} apertures' transmission efficiency superimposed onto the map in the given emission line. The lower panels contain the projected maps of the nebula in each line after convolution with the aperture response maps. All the images are oriented such that north is up and east to the left.}
\label{fig:chrismini}
\end{center}
\end{figure*}

The main reason for the discrepancy observed between the models' predictions of the infrared line fluxes and the Bower (2001) measurements from the {\it ISO~SWS} data for NGC~3918 is that the nebular coordinates used for the {\it ISO} observations were incorrect by approximately 14~arcsec (see Appendix~C). A {\it masking} IDL routine \citep{morisset02}, which makes use of the measured wavelength-dependent {\it SWS} beam profiles, was applied to the projected maps obtained from the final converged model grids, in the six infrared fine-structure lines. This routine returns a {\it mask} which takes into account the wavelength-dependent aperture size and beam profile, and which can then be applied to a map of the object in the required emission line to predict the emergent flux that would be measured. The position of the aperture centre was offset from the centre of the nebula by the amounts specified above, and the aperture was rotated by an angle of 80$^{\circ}$ East of North\footnote{From the {\it ISO} archive, CINSTROLL~=~350$^{\circ}$, where CINSTROLL is the instrument's roll angle, defined as the angle anticlockwise between north and the spacecraft z-axis. Since the long axis of the {\it SWS} apertures is in the y-direction and the short axis is in the z-direction, then P.A.~=~CINSTROLL~+~90$^{\circ}$.} to correspond to the spacecraft roll-angle. 
Resultant predicted fluxes were obtained by convolution of the offset, rotated mask with the correctly oriented map of NGC~3918 in the required emission line. 

Table~\ref{tab:infraredCorrections} provides a summary of the line fluxes predicted for the six infrared fine structure lines, first for the whole nebula, i.e. without the aperture corrections (column 2) and then after the corrections 
for the aperture profile, offset and orientation (column 3). Figure~\ref{fig:chrismini} illustrates the effects of the aperture corrections on the final transmitted flux for three of the the six infrared fine-structure lines for the spindle-like model~B of NGC~3918, namely [O~{\sc iv}]~25.9$\mu$m, [Ne~{\sc iii}]~15.6$\mu$m and [S~{\sc iii}]~18.7$\mu$m. The upper panel of each pair shows a map of NGC~3918 in a particular infrared fine-structure emission line, oriented such that north is up and east is to the left; a contour map of the appropriate wavelength-dependent aperture transmission profile is superimposed on each map. The lower panel of each pair shows the map from the corresponding upper panel {\em after} convolution with the appropriate {\it ISO~SWS} aperture profile.

It is clear from the images, and from the corrected line fluxes reported in Table~\ref{tab:infraredCorrections}, that the SWS aperture offset is indeed the main reason for the SWS infrared line flux discrepancy. However, even after these corrections the fluxes for the predicted [Ne~{\sc v}] fine structure lines are still a factor of two to three too strong compared to the observed values; this may be due to the collision strengths by \citet{lennon91} used in the Mocassin code, which are much larger than those used in the Harrington code by C87. This has already been discussed in Section~\ref{sub:biconicalresults}. A discrepancy also remains between the SWS flux in the [S~{\sc iv}]~10.5~$\mu$m line measured by \citet{bower01} and the corrected fluxes predicted by Mocassin, which are a factor of 2 to 3 stronger. C87 also predicted a flux for this line which was a factor of 3.6 stronger than the IRAS LRS measurements of \citet{pottasch86}.  The origin of this discrepancy still remains unclear. 

It is worth noting that although the aperture offset of 14~arcsec also applied to the LWS observations made with {\it ISO} \citep{liu01}, this would not affect the integrated flux measurements, due to the much larger aperture size of the LWS ($\approx$70--80~arcsec~FWHM).

\section{Visualization of the Model Results}
\label{sec:visualization}

A large amount of observational data has been obtained for NGC~3918 over the years. Some of the most recent data published include the optical images and the high resolution, long slit spectra presented by C99. In the same paper they also presented an image of NGC~3918, retrieved from the {\it HST} archives, and obtained in 1995 with the Wide Field Planetary Camera~2 (WFPC2) in the F555W filter and shown here in Figure~\ref{fig:hstimage}. C99 claimed that the emission in this image (central wavelength\,=\,5252\,{\AA}, FWHM\,=\,1223\,{\AA}) is dominated by [O~{\sc iii}] 4959+5007{\AA}, however many other nebular lines, including H$\beta$, He~{\sc i}~5876~{\AA}, He~{\sc ii}~4686{\AA}, [Ar~{\sc iv}]~4711,4740~{\AA} and [N~{\sc ii}]~5755~{\AA}, fall within the transmission range of this filter and , although weaker than the [O~{\sc iii}] lines, can be important at particular positions in the nebula. 

\subsection{Projected Images}

\begin{figure}
\begin{center}
\psfig{file=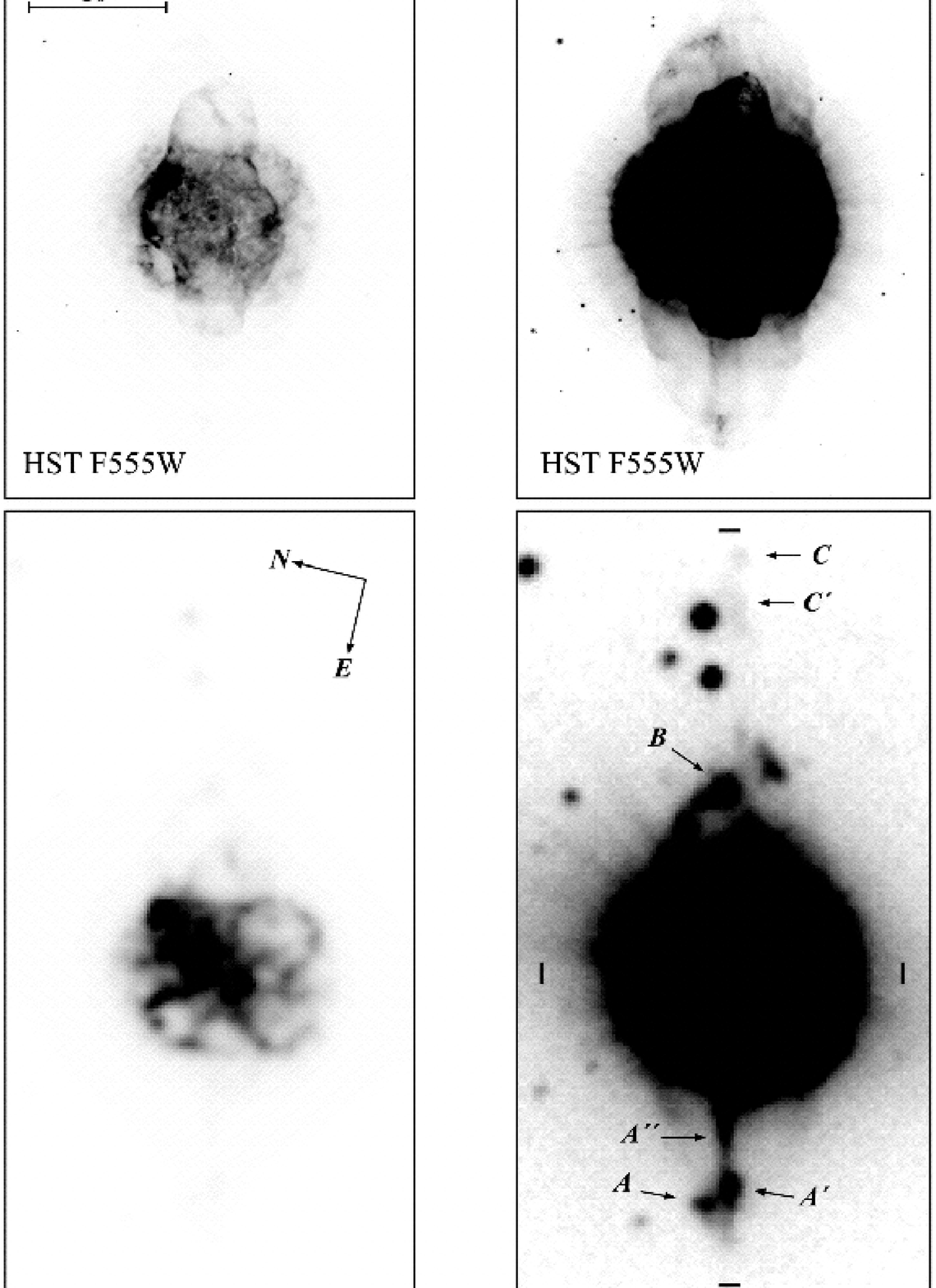, height=119mm, width=80mm}
\caption[{\it HST} and {\ it NTT} images of NGC~3918]{{\it HST} and {\it NTT} images of NGC~3918. The images are rotated so as to have the main jet-like feature oriented along the vertical direction. {\it Left:} Displayed to show the highest intensity levels by using a linear scale. {\it Right:} Displayed to show the faintest structures on a logarithmic scale. \citep{corradi99}.}
\label{fig:hstimage}
\end{center}
\end{figure}

\begin{figure*}
\begin{center}
\begin{minipage}[t]{4.7cm} 
\epsfig{file=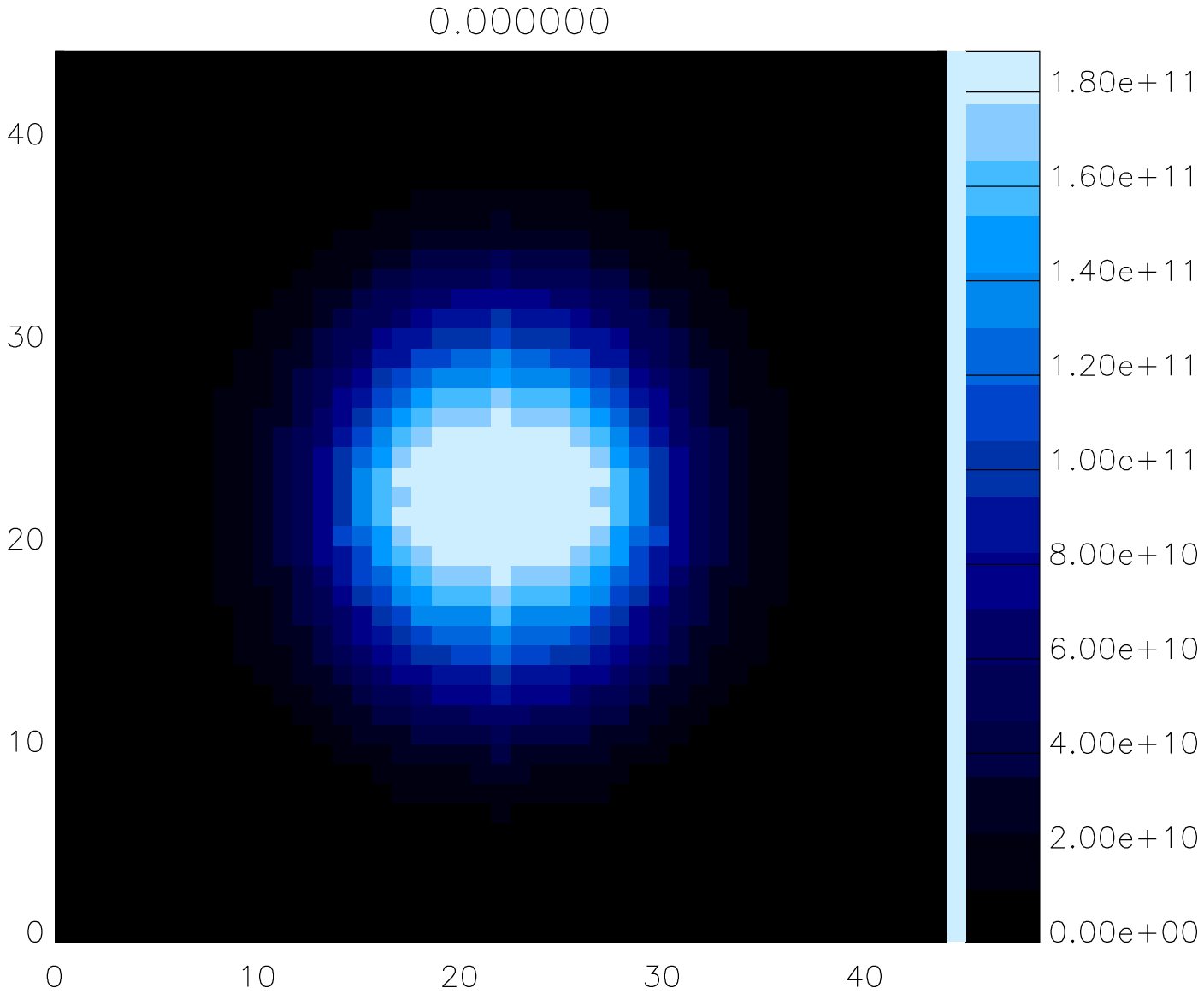,width=4.7cm,clip=,bbllx=132, bburx=509,bblly=385,bbury=701}
\end{minipage}
\begin{minipage}[t]{4.7cm} 
\epsfig{file=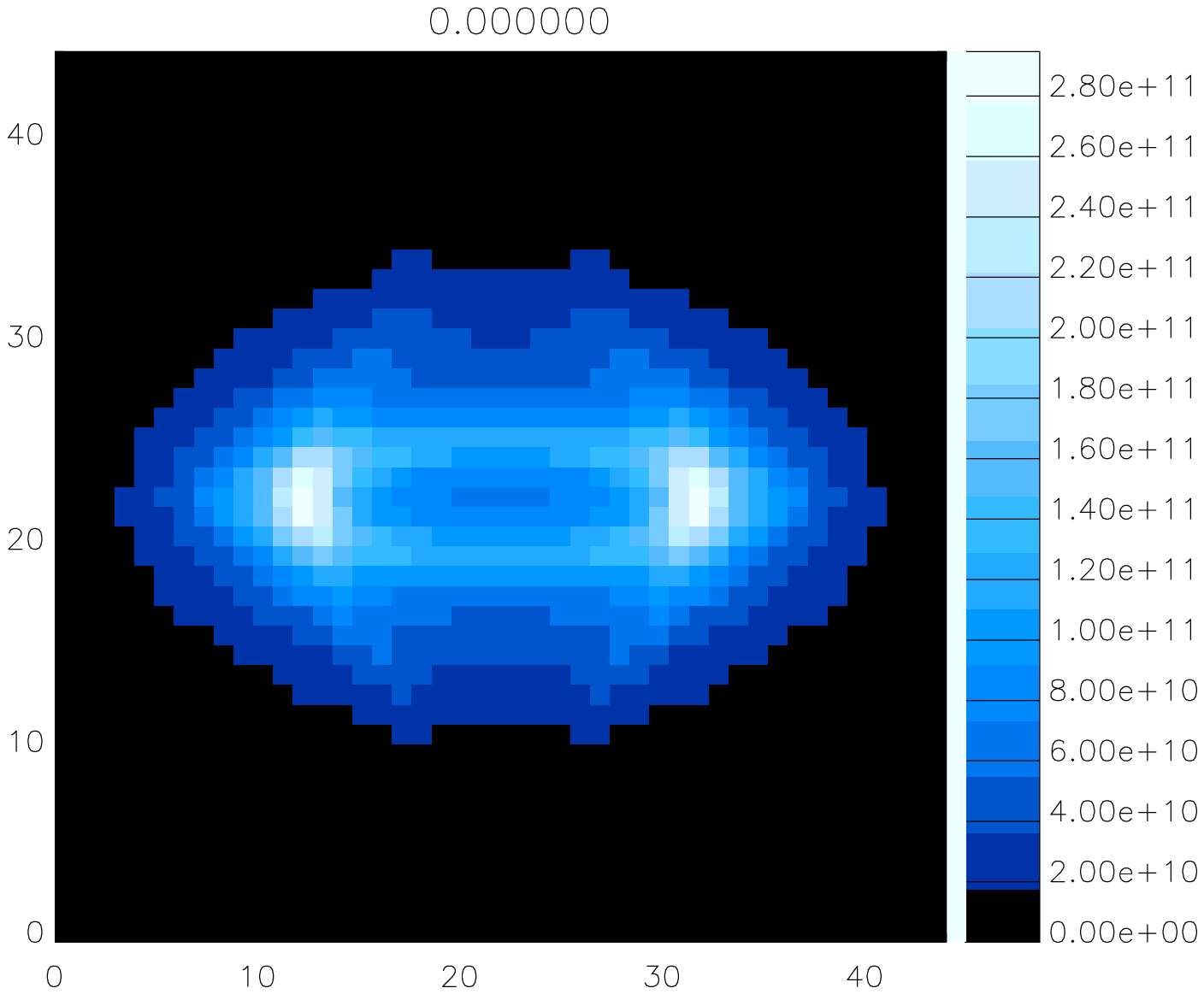,width=4.7cm,clip=,bbllx=132, bburx=509,bblly=385,bbury=701}
\end{minipage}
\begin{minipage}[t]{4.7cm} 
\epsfig{file=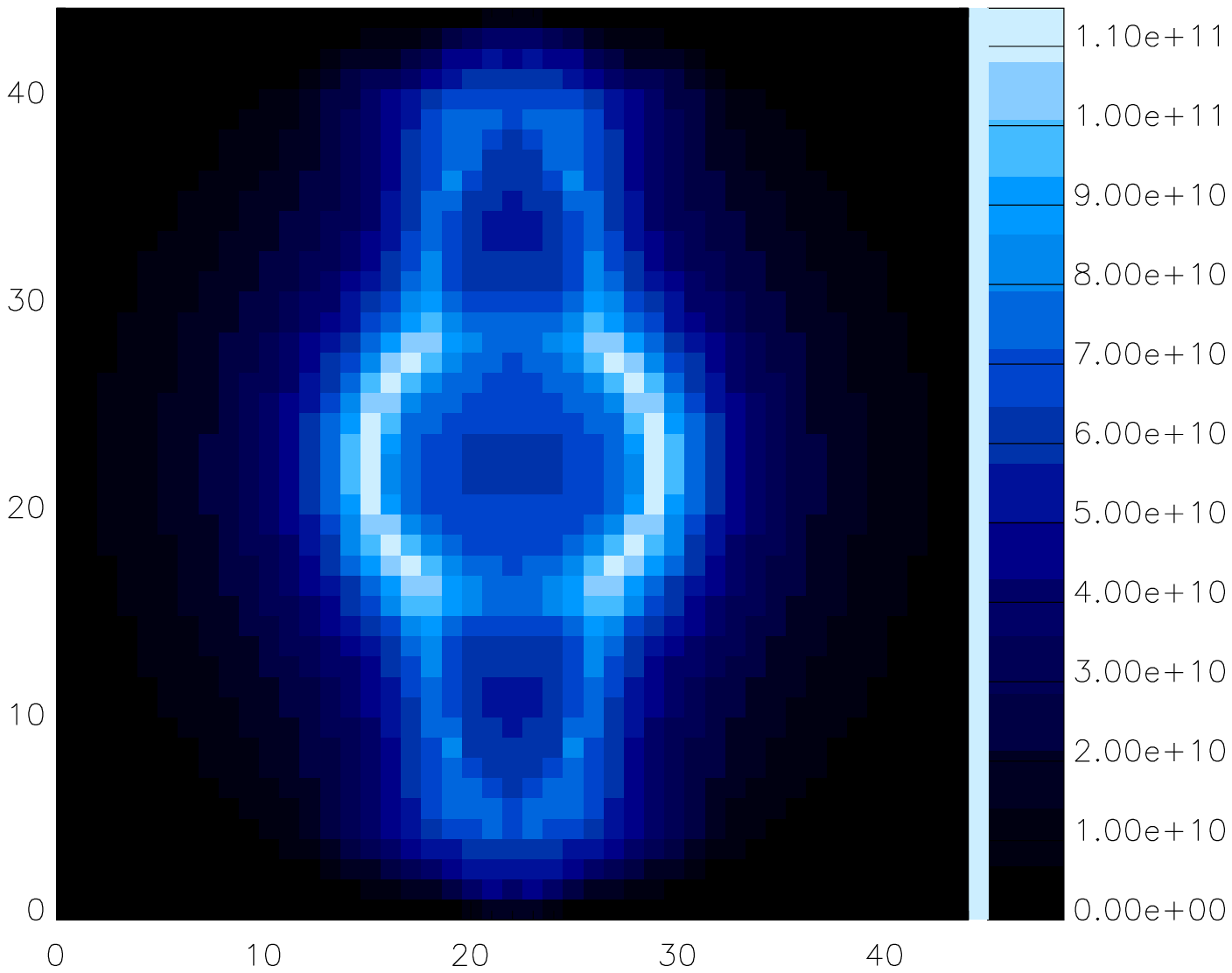,width=4.7cm,clip=,bbllx=132, bburx=509,bblly=385,bbury=701}
\end{minipage}
\begin{minipage}[t]{4.7cm} 
\epsfig{file=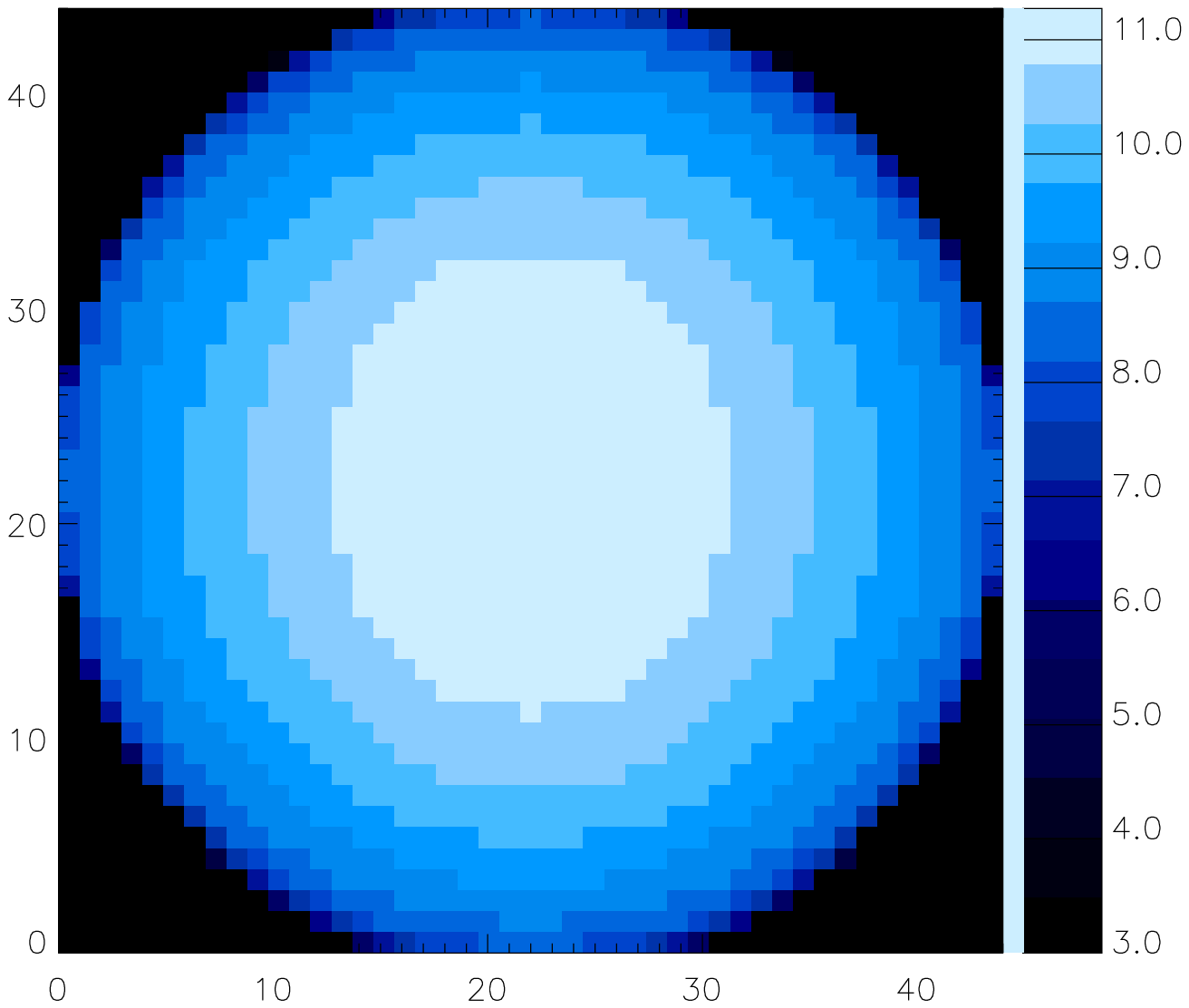,width=4.7cm,clip=,bbllx=132, bburx=509,bblly=385,bbury=701}
\end{minipage}
\begin{minipage}[t]{4.7cm} 
\epsfig{file=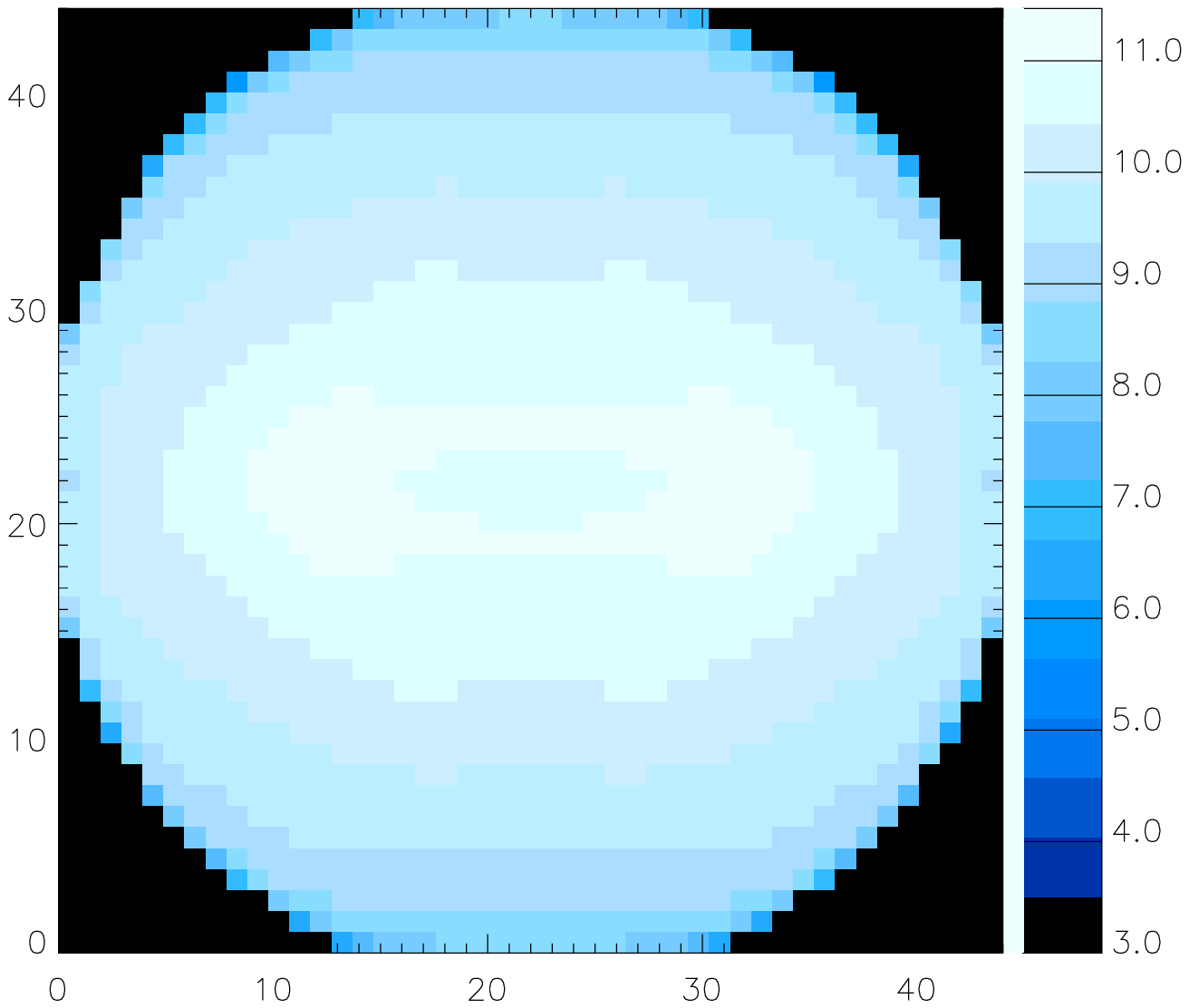,width=4.7cm,clip=,bbllx=132, bburx=509,bblly=385,bbury=701}
\end{minipage}
\begin{minipage}[t]{4.7cm} 
\epsfig{file=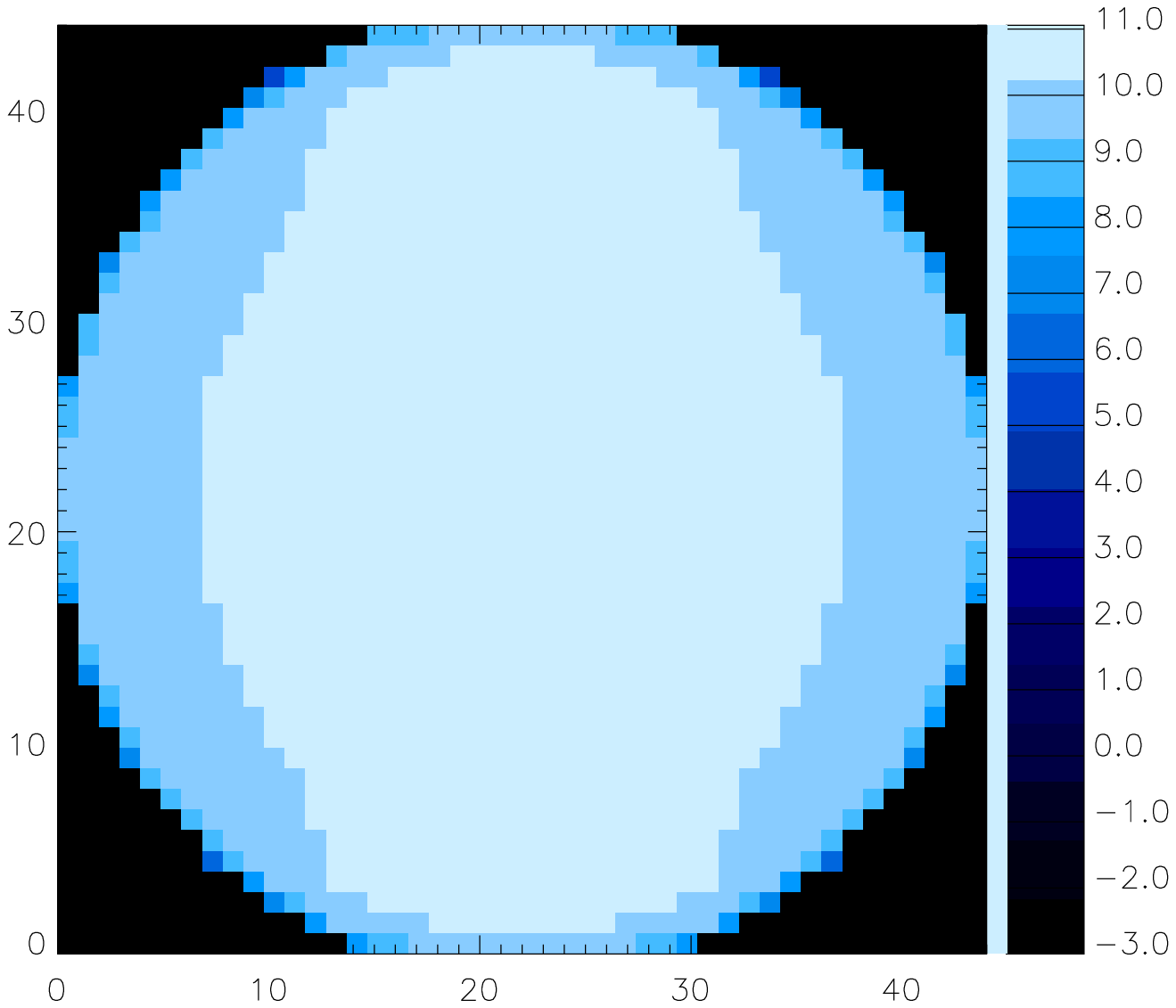,width=4.7cm,clip=,bbllx=132, bburx=509,bblly=385,bbury=701}
\end{minipage}
\caption[Projected maps in H$\beta$ of NGC~3918]{Projected maps of NGC~3918 in H$\beta$ for the biconical model (left-hand panels), the spindle-like model~A (central panels) and for the spindle-like model~B (right-hand panels). The top panels are displayed with linear intensity scales and the bottom panels are displayed on logarithmic scales. The spatial units are pixels and the intensity range is in arbitrary units. Note that a different greyscale has been used for the the spindle-like model~A maps, in order to highlight some of the features.}
\label{fig:projectionsChris}
\end{center}
\end{figure*}

In order to compare the final model results for the biconical distribution and for the spindle-like distribution, IDL visualization routines, described by \citet{morisset00} and \citet{monteiro00}, were employed. Figure~\ref{fig:projectionsChris} shows predicted maps of NGC~3918 in the H$\beta$ emission line, obtained from the three models developed in this work; namely, the biconical model (left-hand panel) and the spindle-like models~A (central panel) and B (right-hand panel). The bottom panels of Figure~\ref{fig:projectionsChris} display the maps on a logarithmic scale. 

The inclination and spatio-kinematical parameters of NGC~3918 were derived by C99 using the spatio-kinematical model of \citet{solf85}. They obtained the best fit to the observed shape and kinematics of the inner shell by assuming an inclination angle $\theta$ of 75$^{\circ}$, where $\theta$ is defined as the angle between the polar axis and the line of sight. The data cubes calculated in this work for the spindle-like models A and B were, therefore, rotated through the same angle to produce projected maps in various emission lines, including those shown in Figure~\ref{fig:projectionsChris}. C87 derived some kinematical information from a Coud\'e spectrum of the [O~{\sc ii}]~$\lambda\lambda$3726, 3729 doublet. From this they derived an inclination angle, $\theta$, for the biconical density distribution model, of 15$^{\circ}$. Therefore the projected images for the biconical model (including the H$\beta$ maps shown in Figure~\ref{fig:projectionsChris}) were obtained by rotating the grid by the same angle. The images of NGC~3918 published by C99 also show some low intensity small scale features associated with this object. In our work, however, no attempt was made to reproduce these features. They are discussed by more detail by C99, including several low-ionization structures located roughly along the major axis of the nebula and clearly visible in the logarithmic [N~{\sc ii}] image (Figure~1 of C99, bottom right), where they are labeled with the letters A, A$'$, A$''$, B, C and C$''$. The shape of the inner shell, described in Section~\ref{sec:ngc3918intro}, is more clearly visible in the {\it HST} images (e.g. Figure~\ref{fig:hstimage}), while the lower angular resolution {\it NTT} image in [N~{\sc ii}] (Figure~1 of C99, bottom left) shows an almost filamentary distribution. 

From the projected maps obtained in several emission lines for the biconical density distribution model (the H$\beta$ maps are shown in the left-hand panels of Figure~\ref{fig:projectionsChris}), it became obvious that the model proposed by C87, although successful in reproducing the emission spectrum of NGC~3918, cannot be used to actually describe the geometric structure of this object. C87, however, were right to assume that this object must be optically thick in some directions and optically thin in others, in order to obtain such a complex spectrum, exhibiting strong emission lines from a wide range of ionization stages, e.g. [Ne~{\sc v}] and He~{\sc ii} to [O~{\sc i}] and [N~{\sc i}]. A number of combinations of thick and thin phases  could possibly have been used to construct models to fit the emission lines and, without additional spatial information, it  would have been virtually impossible to distinguish between them. 

Projected maps were also obtained in several emission lines for the density distribution given by the \citet{mellema96} model (spindle-like model~A), discussed in Section~\ref{sec:spindle}. Once again, for the sake of space only the H$\beta$ maps are shown in Figure~\ref{fig:projectionsChris} (central panel). As discussed in the previous section, the original distribution had to be enhanced in the equatorial region in order to correctly reproduce the observed spectrum. Although the shape of the inner shell of NGC~3918 is better reproduced than by the biconical model, the polar protrusions were still missing and the enhancement seemed to be too localised in a small equatorial torus. 

The best matches to the observed images of NGC~3918 were obtained for the spindle-like model B, where the density distribution of the inner shell was reproduced using the analytical expression given in Appendix~A. Figure~\ref{fig:projectionsChris} (right-hand panels) shows the H$\beta$ maps obtained for this model on a linear scale (top-right) and on a logarithmic scale (bottom-right). The detailed shape and density distribution given by the analytical description depends on ten parameters and it is possible that a number of combinations of these could be used to reproduce the observations. However the combination used in this model (see Table~\ref{tab:spindleparametersan}) produces images (Figure~\ref{fig:projectionsChris}) and spectroscopic results (see below) which are consistent with the observations. It might be possible to obtain even better fits by further adjusting the density distribution parameters and, possibly, also the central star parameters, considering that those obtained by C87 were based on a completely different geometrical model. The central star parameters, however, were based on global nebular properties which shouldn't change as much. 

\subsection{Position-Velocity diagrams and line profiles}
\label{sub:PV}

\begin{figure*}
\begin{center}
\psfig{file=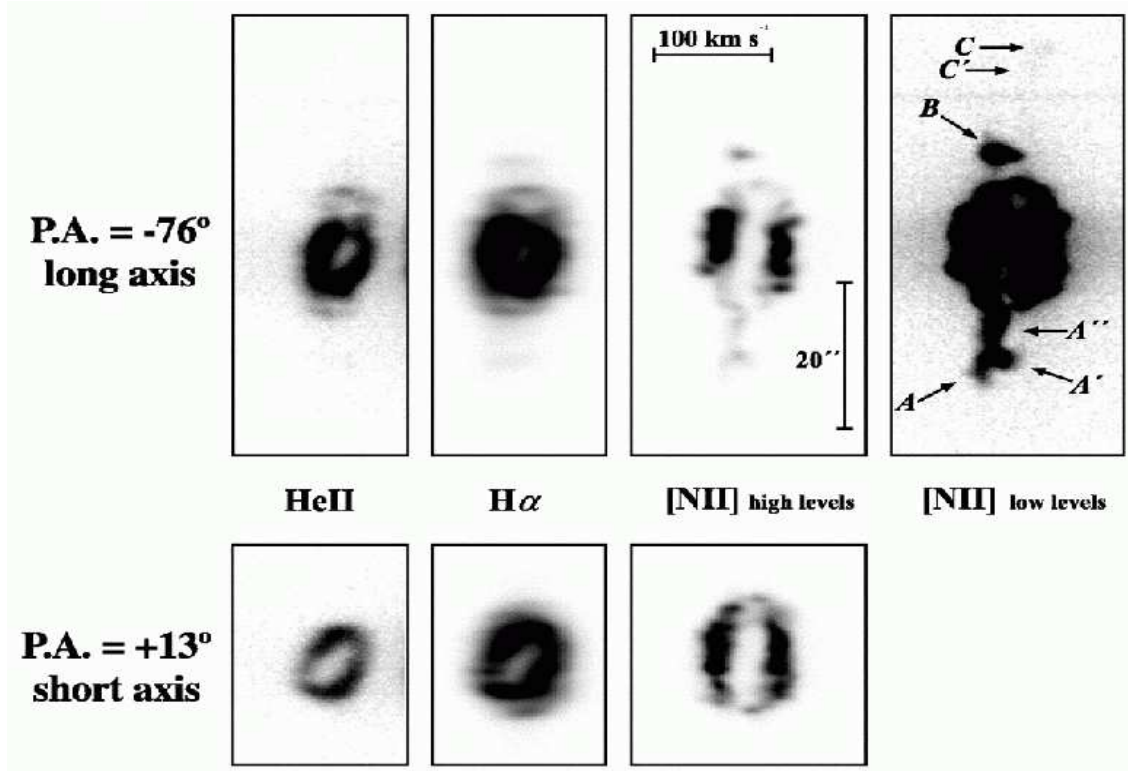, height=103mm, width=150mm}
\caption[{\it NTT} long-slit spectra of NGC~3918]
{{\it NTT} long-slit spectra of NGC~3918. Images are in a logarithmic intensity scale with different cuts for the He~{\sc ii} 6560{\AA}, H$\alpha$, and $[$N~{\sc ii}$]~$6584{\AA} (C99)}
\label{fig:longslit}
\end{center}
\end{figure*}

\begin{figure*}
\begin{center}
\begin{minipage}[t]{4.7cm} 
\epsfig{file=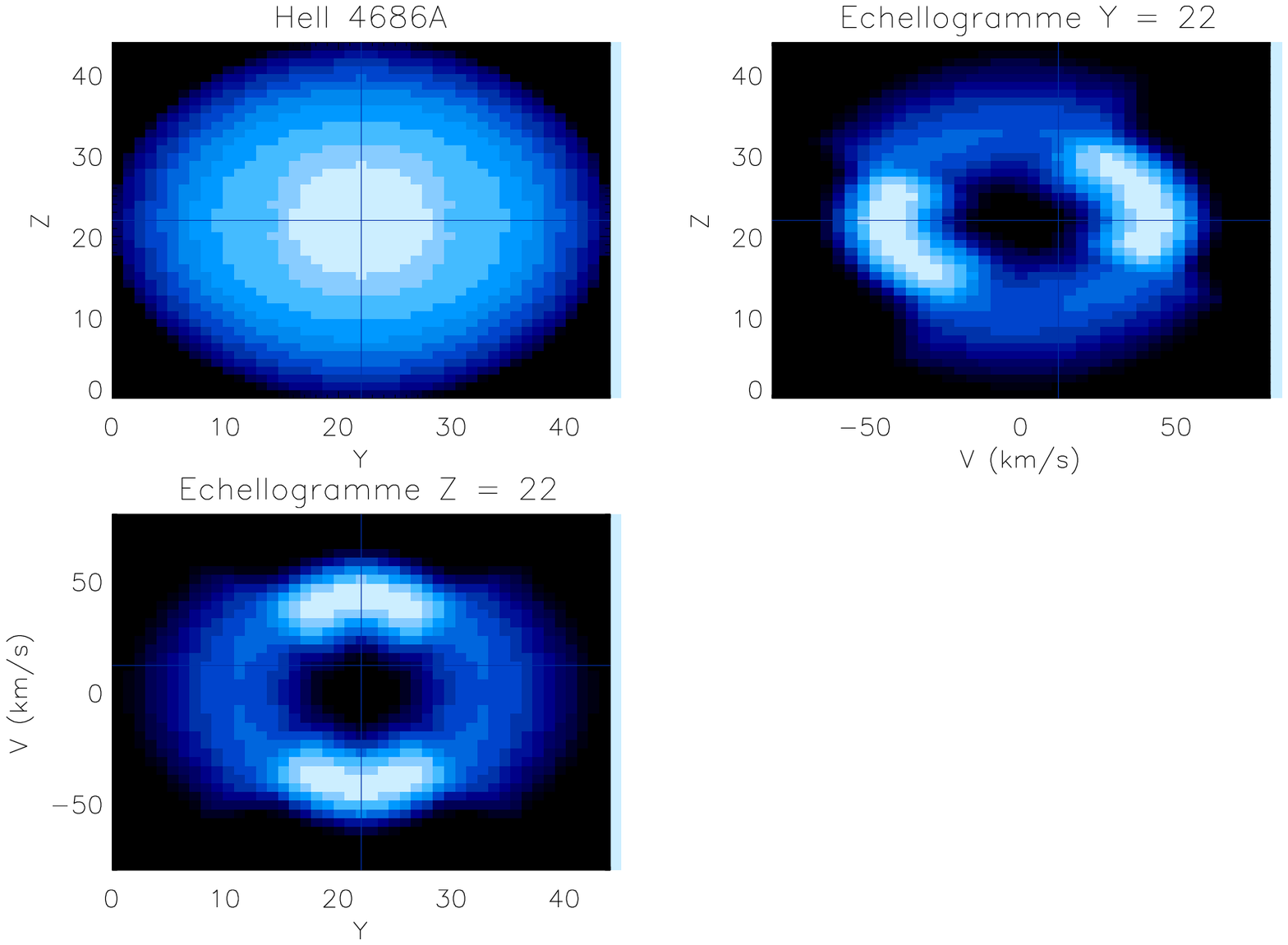,width=4.7cm,clip=,bbllx=295, bburx=533,bblly=528,bbury=704}
\end{minipage}
\begin{minipage}[t]{4.7cm} 
\epsfig{file=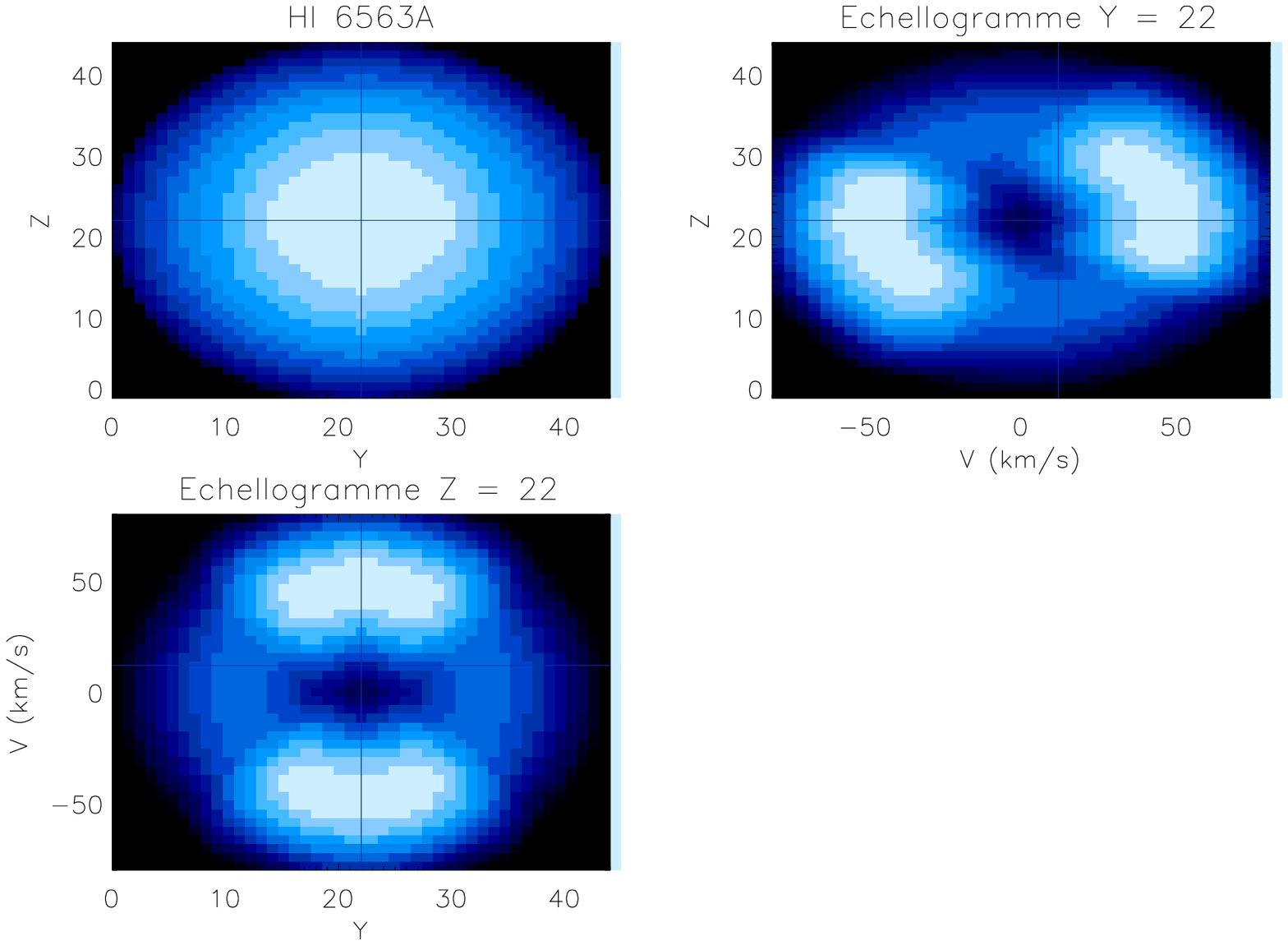,width=4.7cm,clip=,bbllx=295, bburx=533,bblly=528,bbury=704}
\end{minipage}
\begin{minipage}[t]{4.7cm} 
\epsfig{file=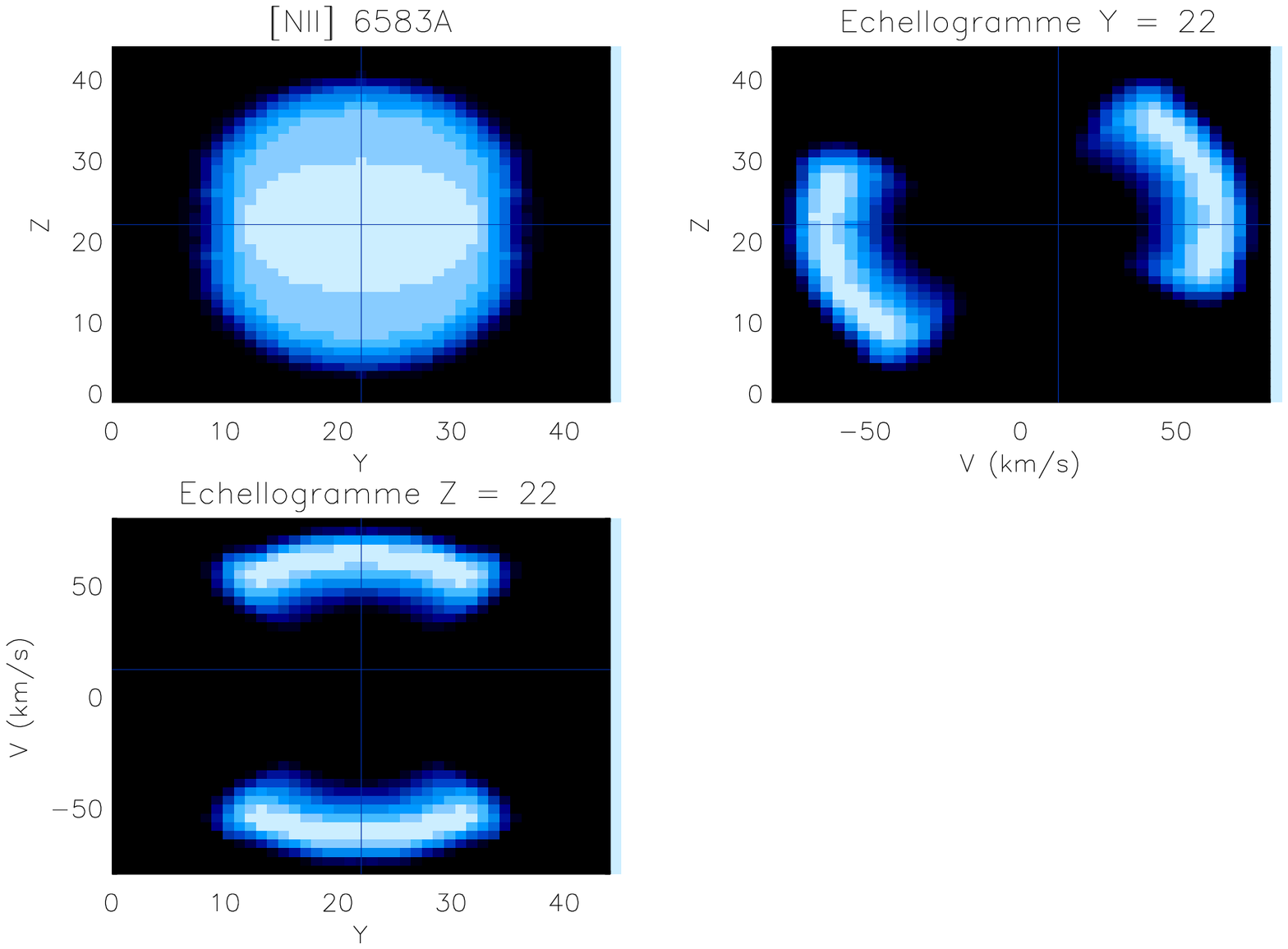,width=4.7cm,clip=,bbllx=295, bburx=533,bblly=528,bbury=704}
\end{minipage}
\begin{minipage}[t]{4.7cm} 
\epsfig{file=heiiecheloclegg.ps,width=4.7cm,clip=,bbllx=20, bburx=265,bblly=338,bbury=514}
\end{minipage}
\begin{minipage}[t]{4.7cm} 
\epsfig{file=halphaecheloclegg.ps,width=4.7cm,clip=,bbllx=20, bburx=265,bblly=338,bbury=514}
\end{minipage}
\begin{minipage}[t]{4.7cm} 
\epsfig{file=niiecheloclegg.ps,width=4.7cm,clip=,bbllx=20, bburx=265,bblly=338,bbury=514}
\end{minipage}
\begin{picture}(400, 10)(1,1)
\put(10,10){\line(400,0){390}}
\end{picture}
\begin{minipage}[t]{4.7cm} 
\epsfig{file=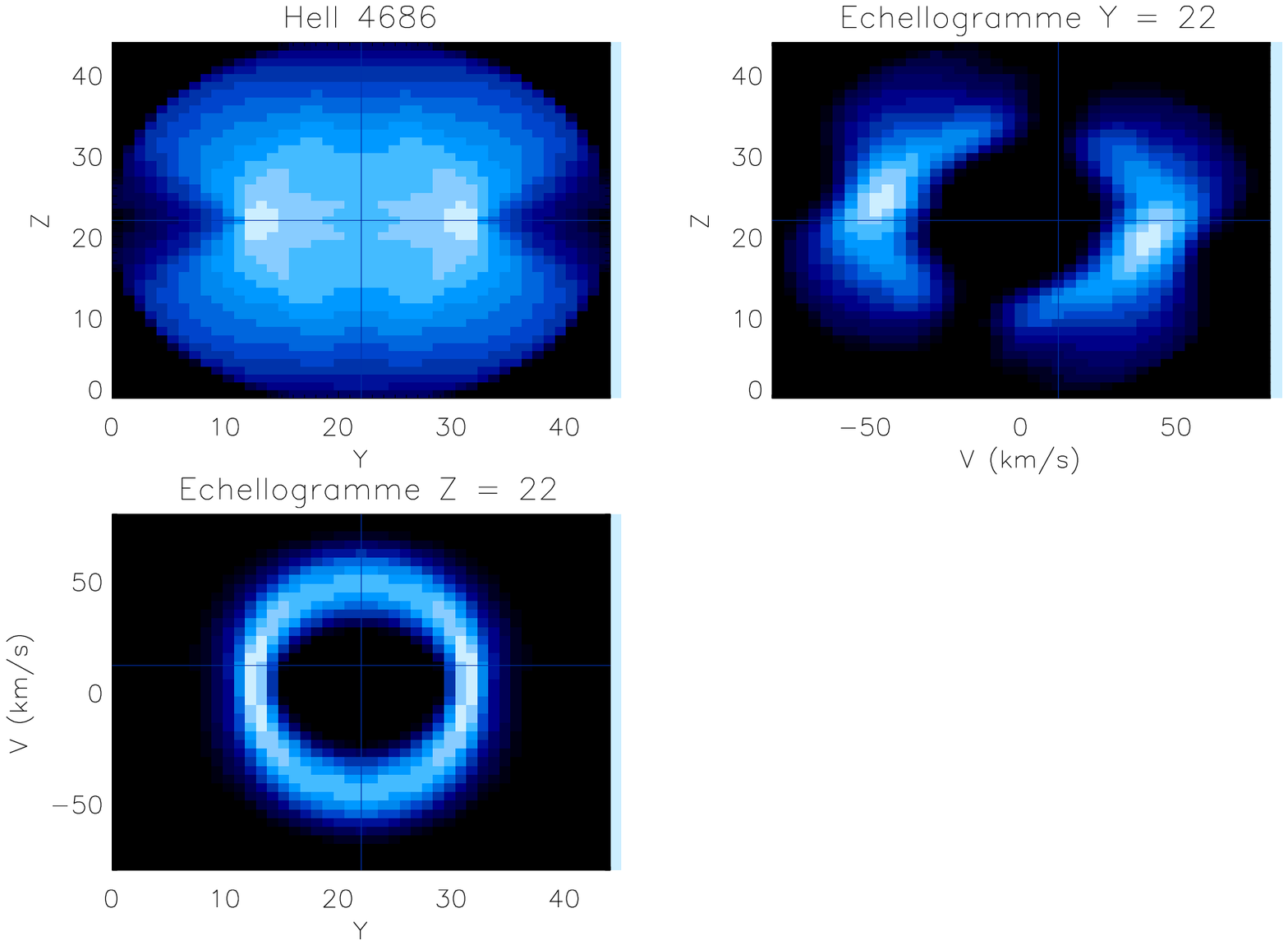,width=4.7cm,clip=,bbllx=295, bburx=533,bblly=528,bbury=704}
\end{minipage}
\begin{minipage}[t]{4.7cm} 
\epsfig{file=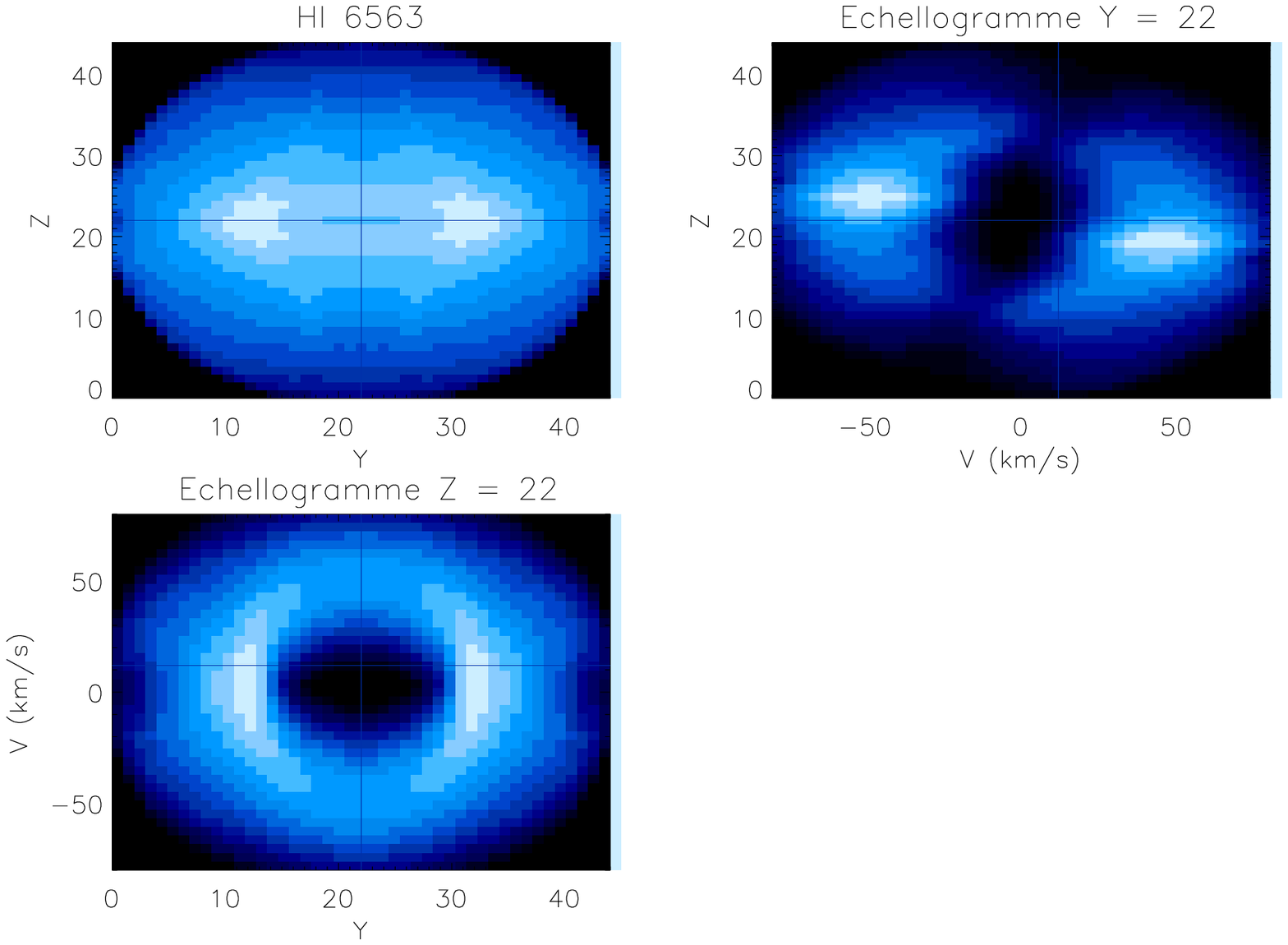,width=4.7cm,clip=,bbllx=295, bburx=533,bblly=528,bbury=704}
\end{minipage}
\begin{minipage}[t]{4.7cm} 
\epsfig{file=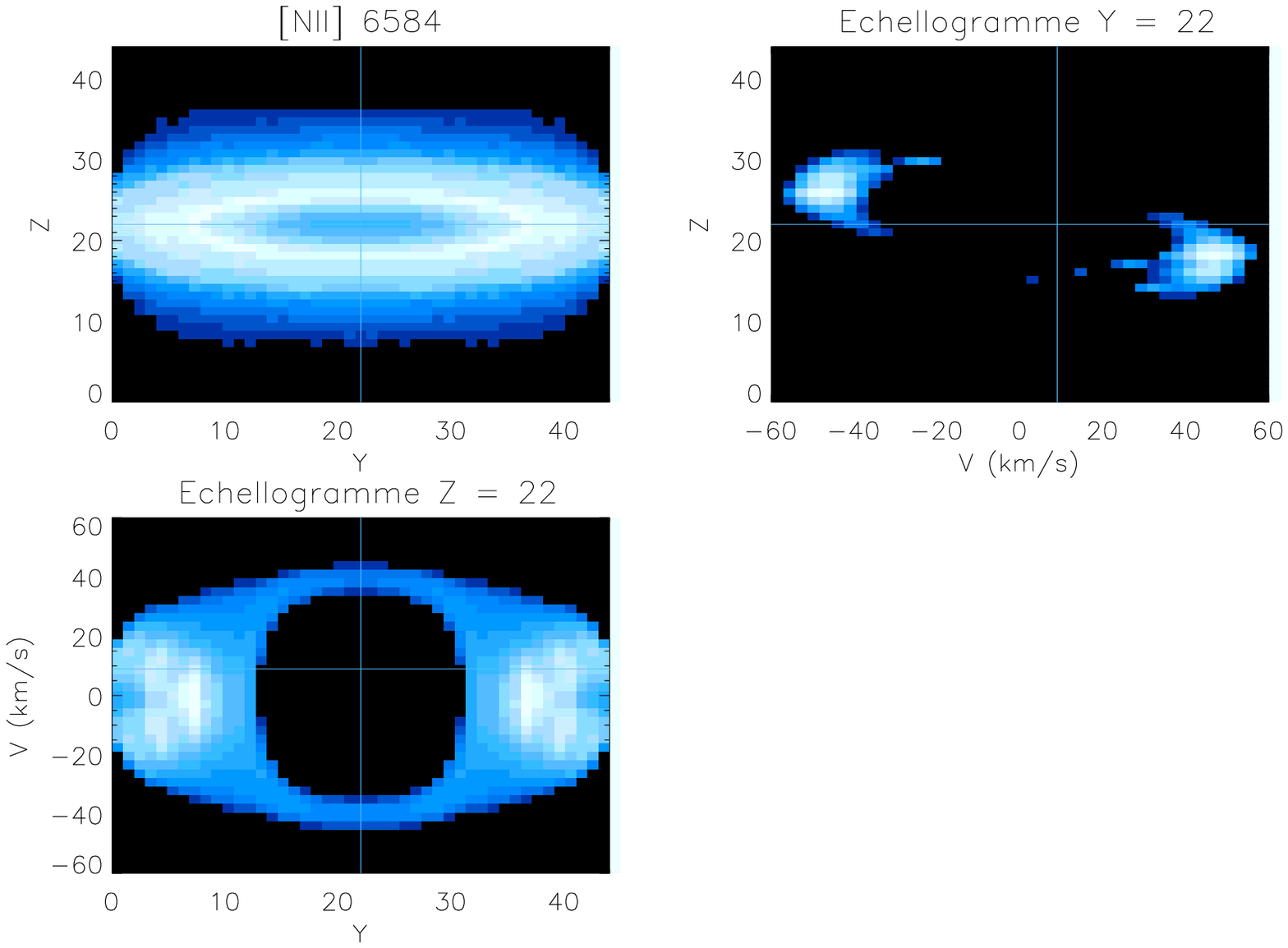,width=4.7cm,clip=,bbllx=295, bburx=533,bblly=528,bbury=704}
\end{minipage}
\begin{minipage}[t]{4.7cm} 
\epsfig{file=heiiechelogarrelt.ps,width=4.7cm,clip=,bbllx=20, bburx=265,bblly=338,bbury=514}
\end{minipage}
\begin{minipage}[t]{4.7cm} 
\epsfig{file=halphaechelogarrelt.ps,width=4.7cm,clip=,bbllx=20, bburx=265,bblly=338,bbury=514}
\end{minipage}
\begin{minipage}[t]{4.7cm} 
\epsfig{file=niiechelogarrelt.ps,width=4.7cm,clip=,bbllx=20, bburx=265,bblly=338,bbury=514}
\end{minipage}
\begin{picture}(400, 10)(1,1)
\put(10,10){\line(400,0){390}}
\end{picture}
\begin{minipage}[t]{4.7cm} 
\epsfig{file=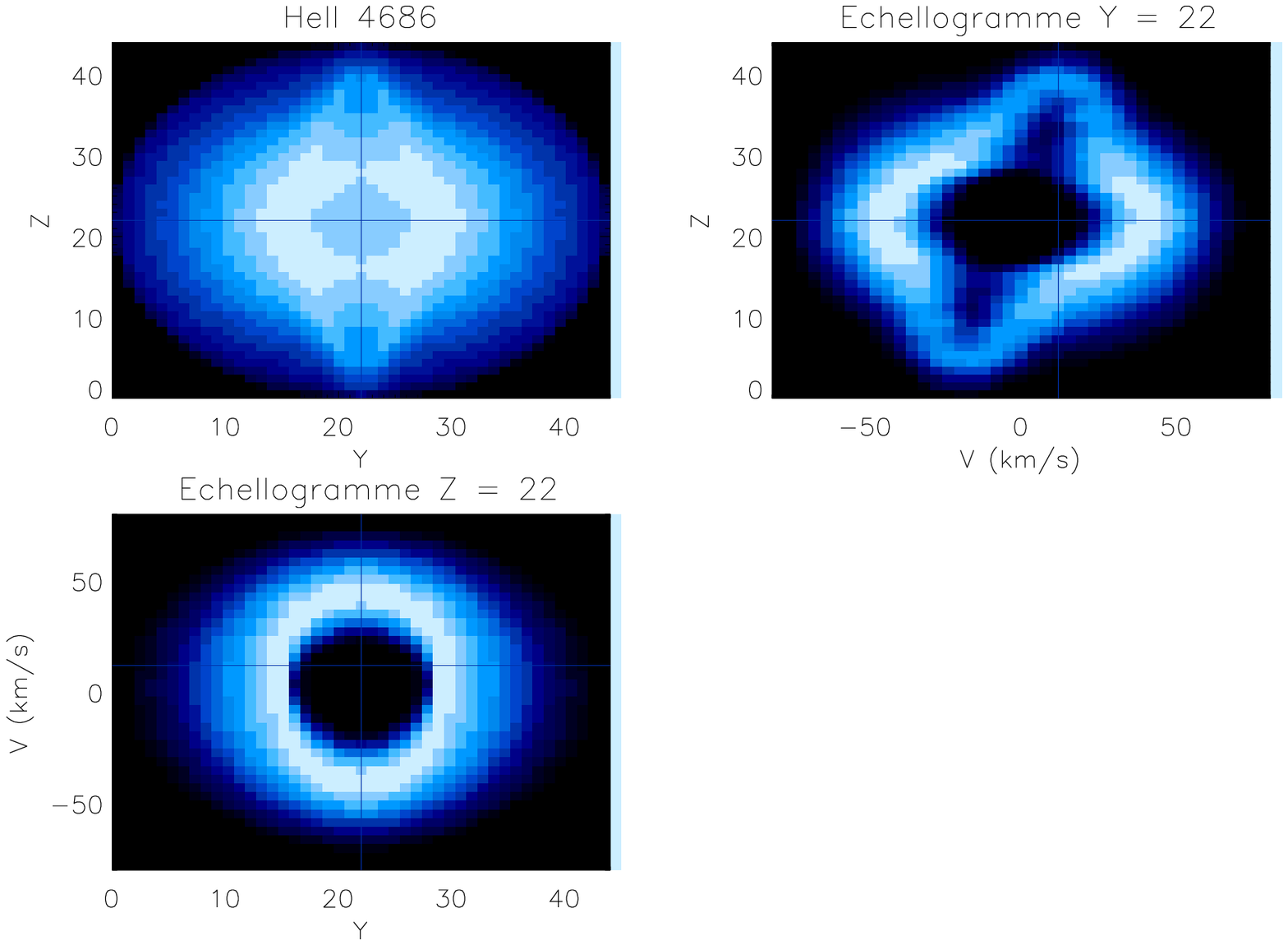,width=4.7cm,clip=,bbllx=295, bburx=533,bblly=528,bbury=704}
\end{minipage}
\begin{minipage}[t]{4.7cm} 
\epsfig{file=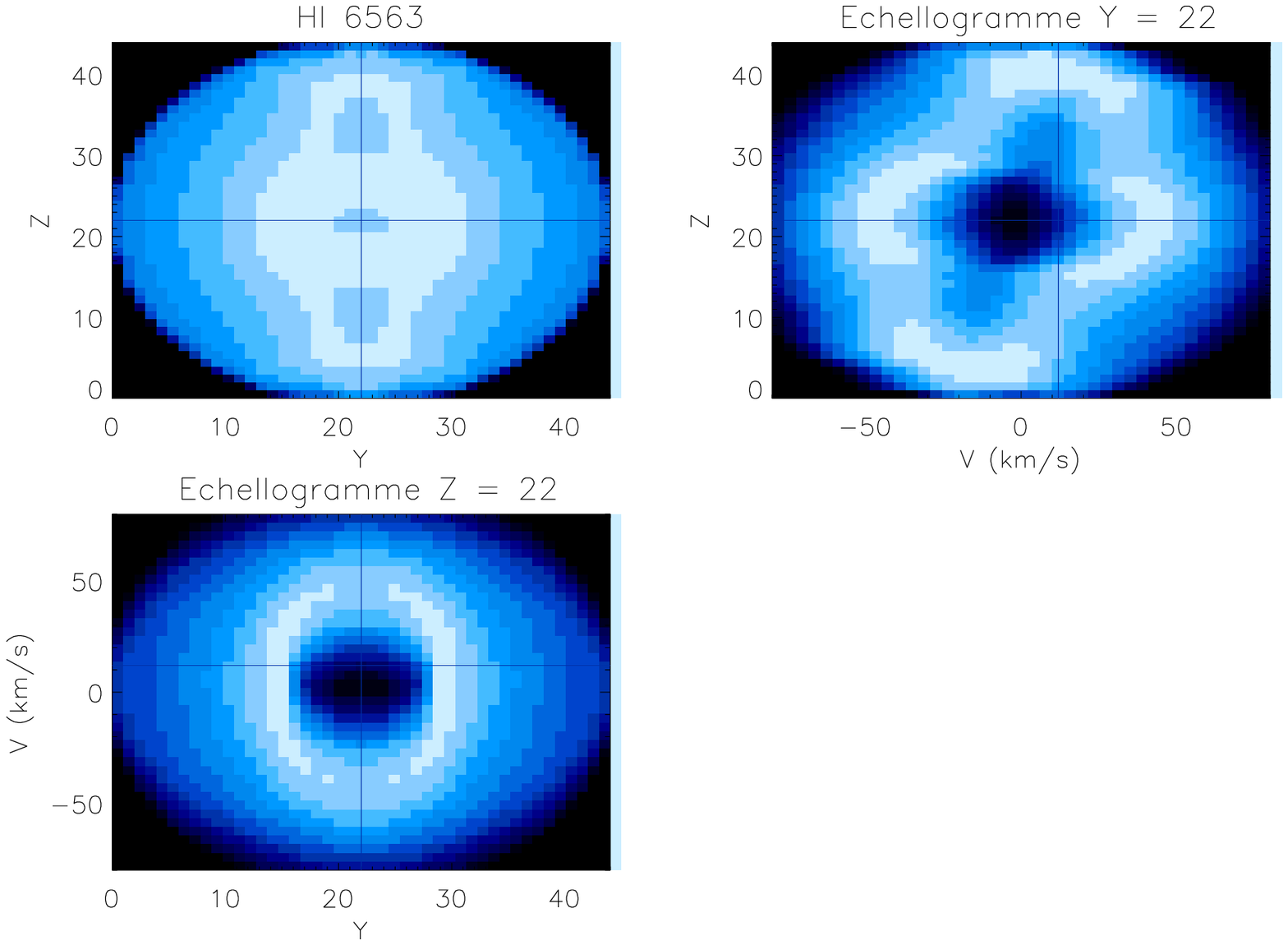,width=4.7cm,clip=,bbllx=295, bburx=533,bblly=528,bbury=704}
\end{minipage}
\begin{minipage}[t]{4.7cm} 
\epsfig{file=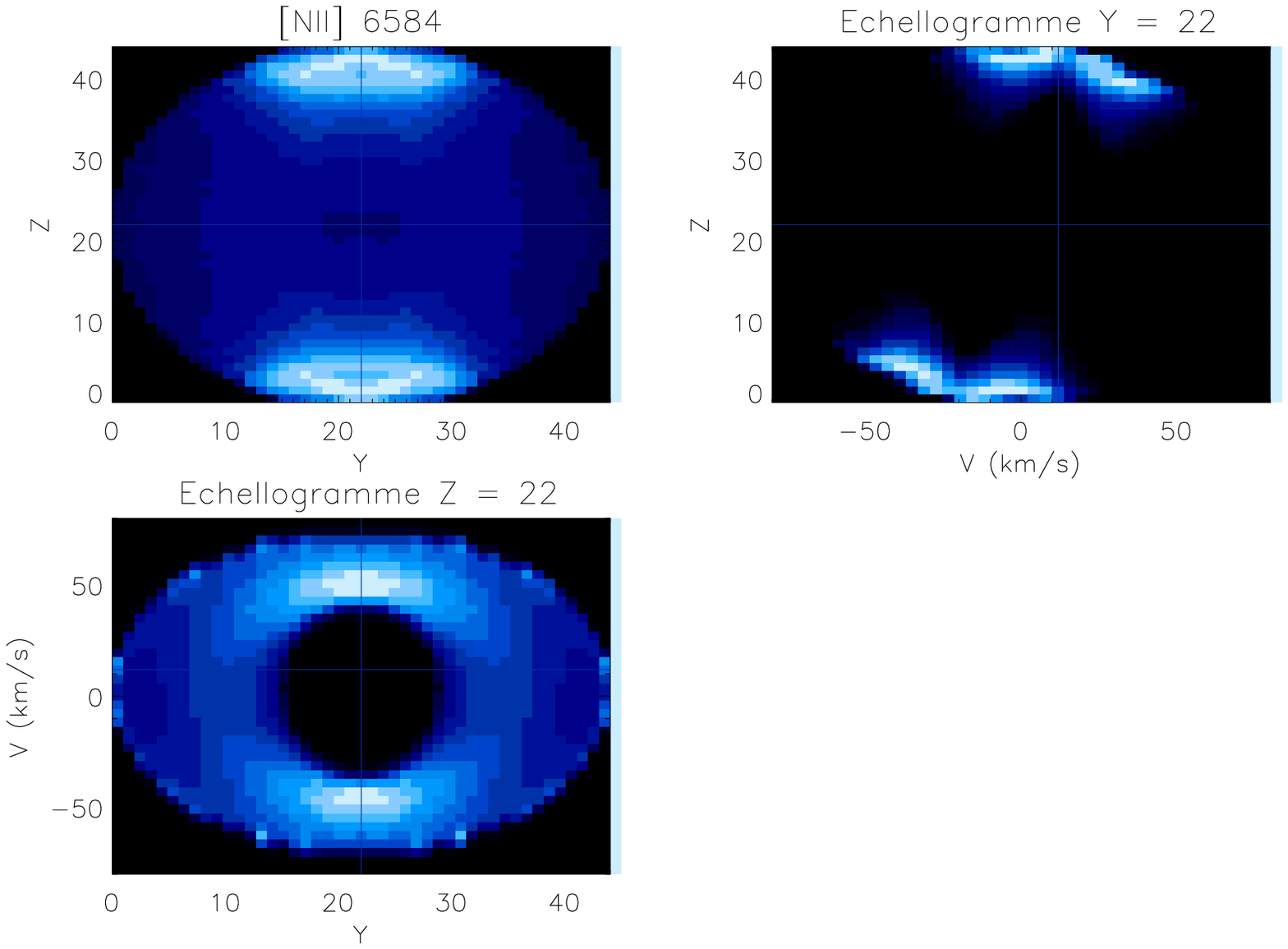,width=4.7cm,clip=,bbllx=295, bburx=533,bblly=528,bbury=704}
\end{minipage}
\begin{minipage}[t]{4.7cm} 
\epsfig{file=heiiechelochris.ps,width=4.7cm,clip=,bbllx=20, bburx=265,bblly=338,bbury=514}
\end{minipage}
\begin{minipage}[t]{4.7cm} 
\epsfig{file=halphaechelochris.ps,width=4.7cm,clip=,bbllx=20, bburx=265,bblly=338,bbury=514}
\end{minipage}
\begin{minipage}[t]{4.7cm} 
\epsfig{file=niiechelochris.ps,width=4.7cm,clip=,bbllx=20, bburx=265,bblly=338,bbury=514}
\end{minipage}
\caption[Position-velocity diagrams for the spindle like model~B of NGC~3918.]{Position-velocity diagrams in He~{\sc ii}~4686~{\AA} ({\it left-hand panels}), H$\alpha$ ({\it central panels}) and [N~{\sc ii}]~6584~{\AA} ({\it right-hand panels}) for the the biconical model (top two rows) and spindle-like models~A (middle two rows) and~B (bottom two rows) of NGC~3918. The top panels in each pair of rows show the echellograms obtained with slits oriented at PA~=~--76$^{\circ}$ while the bottom diagrams show echellograms obtained in the same lines, but with slits positioned at PA~=~+13$^{\circ}$.}
\label{fig:pv}
\end{center}
\end{figure*}

As mentioned above, C99 used the spatio-kinematical empirical model of \citet{solf85} to describe the nebular expansion velocity, V$_{\rm exp}$, in order to fit their observations of position velocity plots of H$\alpha$ and the shape of the inner shell of NGC~3918. In this model V$_{\rm exp}$ increases from the equatorial plane towards the polar axis according to the following equation \citep{solf85}
\begin{equation}
V_{\rm exp}(\phi)=V_{\rm e}+(V_{\rm p}-V_{\rm e}){\rm sin}^{\gamma}\mid\phi\mid
\label{eq:solf}
\end{equation}
where $\phi$ is the latitude angle, defined such that $\phi$=0$^{\circ}$ in the equatorial plane and $\phi$=90$^{\circ}$ in the polar direction; $V_{\rm p}$ and $V_{\rm e}$ are the equatorial and polar velocity components, respectively ($V_{\rm p}$~=~50\, km~s$^{-1}$ and $V_{\rm e}$~=~23\, km~s$^{-1}$ in their work); $\gamma$ is a shape parameter ($\gamma$=12 in their work). Here the velocity field described above was used for the spindle-like density distribution model A in order to produce  position-velocity diagrams along slits positioned at P.A.=--76$^{\circ}$ and P.A.=+13$^{\circ}$, in order to compare with Corradi et al.'s observations, shown in Figure~\ref{fig:longslit}. Position-velocity diagrams were obtained for each of the lines with long slit spectra presented by C99, and include He~{\sc ii}~$\lambda$4686, H$\alpha$ and $[$N~{\sc ii}$]~\lambda$6584. These are shown in Figure~\ref{fig:pv} (middle two rows).

Position-velocity maps were also produced for the biconical density distribution model and for spindle-like model~B (Figure~\ref{fig:pv}, top two rows and bottom two rows, respectively) using the same slit positions as above. A simple radial velocity field was adopted, namely 
\begin{equation}
V(r) = V_0 + V_1 \cdot r
\label{eq:radfield}
\end{equation}
 where r is the radial distance; V$_0$ and V$_1$ and constants (V$_0$~=~23~km~s$^{-1}$ and V$_1$~=~50~km~s$^{-1}$ in this work). Originally, the \citet{solf85} description of the velocity field was also used for spindle-like model~B; however it was found that better matching results were obtained by using the simple radial velocity field described above, which was applied to obtain the echellograms shown in Figures~\ref{fig:pv}. The \citet{solf85} kinematical model does not seem to work well with the analytical density distribution used for model~B. The disagreement is particularly evident in the major axis position-velocity diagrams, where the polar enhancement of the velocity field creates two bright spots which have not been observed. The same does not happen when the \citet{solf85} velocity field is applied to the spindle-like model~A, since the two polar protrusions (which are created in model~B by embedding an elliptical shell within a spheroid) are {\it not closed} at the poles (see Figure~\ref{fig:mellemadist}). However, from the {\it HST} image shown of the left hand panel in Figure~\ref{fig:hstimage}, it seems that the polar protrusions are indeed {\it closed}, although not very dense, and therefore the density distribution used in model~B seems more appropriate. 
        
After inspection of all the position-velocity diagrams obtained for the three models (Figure~\ref{fig:pv}) and comparison with C99's Figure~4, it is clear that the observations are most closely reproduced by the spindle-like model~B (bottom two rows in Figure~\ref{fig:pv}). All the synthetic position-velocity diagrams are plotted on a logarithmic intensity scale. It is worth noting at this point that all minor axis position-velocity diagrams (bottom panels in each pair of rows) are plotted such that the ordinate is the velocity (in [km~s$^{-1}$]) and the abscissa is the spatial variable; this has to be taken into account when comparing the synthetic spectra to the observations in Figure~\ref{fig:longslit}, where the abscissa and the ordinate are reversed. The [N~{\sc ii}] position velocity diagrams show the largest discrepancies and will be discussed below. Before that, however, we will focus on the H$\alpha$ and He~{\sc ii} position-velocity diagrams. The projected maps (Figure~\ref{fig:projectionsChris}) had already anticipated, and the synthetic echellograms in Figure~\ref{fig:pv} (top two rows) confirmed, that the biconical distribution is not a realistic representation of NGC~3918. However, as already mentioned, the imaging information available for NGC~3918 was very limited at the time when this model was devised by C87. Model~A (middle two rows of the position-velocity diagrams in Figure~\ref{fig:pv})  fails to reproduce the polar protrusions in the observed long slit spectra (Figure~\ref{fig:longslit}), as was also clear from the projected maps obtained from this model (Figure~\ref{fig:projectionsChris}, central panels). Moreover, the thick waist in model~A, which had to be artificially created in order to obtain the observed line strengths for the lower ionization species, is not observed in NGC~3918. 

None of the models presented in this work succeeded in reproducing the [N~{\sc ii}]~$\lambda$6584 echellograms shown in Figure~\ref{fig:longslit}. There are two main reasons for the observed discrepancies: firstly, N$^+$ is abundant in the FLIERs of NGC~3918 (C99), which are not accounted for in the models presented here; secondly, N$^+$ also seems to exist in the inner regions of the shell; the {\it NTT} image in Figure~1 of C99 shows [N~{\sc ii}] emission to be present in almost a {\it filamentary} form. All the models presented in this work have {\it smooth} density distributions and do not include any filaments or knots. In spindle-like model~A, N$^+$ is mainly concentrated in the toroidal region behind the thick equatorial waist, which is screened from direct starlight, whereas in model~B, N$^+$ is mostly concentrated in small polar regions beyond the elliptical shell which creates the protrusion, and they are also screened from direct starlight by this shell and are further away from the star. In fact, it is quite hard to have N$^+$ {\it survive} in the very hard stellar radiation field ($T_{eff}$~=~140,000~K) in the inner shell, unless some sort of density enhancement is postulated. This shows that, although spindle-like model~B appears to be quite successful in reproducing the main spectroscopic and kinematic characteristics of NGC~3918, there is still scope for improvement.

 A three-dimensional reconstruction technique for studying the morphology, physical condition, ionization and  spatial structure of planetary nebulae has been introduced by \citet{sabbadin00a}, \citet{sabbadin00b} and \citet{ragazzoni01}. In this appproach, the relative density distribution of an emitting region along the cross-section of nebula covered by the slit can be obtained from the radial velocity,  the FWHM and the intensity profile. F. Sabaddin and other members of the Padova group have recently obtained observations for several other planetary nebulae, including NGC~3918, consisting of long slit echellograms at various position angles on the nebulae, and are planning to carry out detailed studies of them by applying their three-dimensional reconstruction technique to them. This means that in the near future a {\it data cube} containing a three-dimensional density distribution obtained using the method above could be available for NGC~3918. It would be very interesting then to map such a data cube onto a Mocassin grid and construct a new photoionization model for this object, since as discussed earlier, although spindle-like model~B produces satisfactory results, there are still some discrepancies remaining with the observational data. 

\section{Conclusions}
In this paper three photoionization models were constructed for the planetary nebula NGC~3918, a biconical model and two spindle-like models. The first spindle-like model (A) used the density distribution of the \citet{mellema96} model already applied to this object by C99. The second spindle-like model (B) instead used an analytical expression for the density distribution, which aimed to mimic, by means of an ellipsoid embedded in a sphere, the shape of the inner shell of NGC~3918. The aim was to find a model which could not only reproduce the main spectroscopic features observed for this object, but also the spatio-kinematical properties of the nebula which recent observations (C99) have uncovered. 

The integrated emission line spectra obtained from the three different models were all in fair agreement with the observations. Discrepancies exist between the observed line fluxes of the C~{\sc iv}~$\lambda\lambda$1548,1550 and Mg~{\sc ii}$\lambda\lambda$2796,2803 resonance doublets, which can largely be explained by dust absorption in the nebula \citep{harrington88}. Large discrepancies were also found between the models' predictions for the fluxes of the infrared fine-structure lines and the measurements of the {\it ISO~SWS} spectra. The main cause for this discrepancy is a pointing error which affected {\it ISO} observations of NGC~3918, causing an offset of approximately 14~arcsec from the centre of the nebula. The corrected predicted line fluxes, obtained by convolving the {\it ISO~SWS} aperture profiles with projected nebular maps in the relevant emission lines, confirmed that most of the discrepancy between the observed and model fluxes could be attributed to the pointing error, although there is still a factor of three discrepancy for the [S~{\sc iv}]~10.5~$\mu$m line.

A diffuse radiation field consistency test was also carried out in this work, which showed that the interaction of the diffuse radiation fields from two adjacent regions of different densities is not negligible, even in the relatively uncomplicated case of the biconical density distibution used by C87. We found that the low ionization species in the optically thick cones were particularly effected by the diffuse radiation field coming from the optically thin sector. 

Although the volume-integrated emission line spectra obtained by the three models of NGC~3918 were in agreement with each other, the projected maps and the synthetic long-slit spectra obtained in several emission lines were, however, very different from one model to the other. Spindle-like model~B produced the best fits to the observations, although some discrepancies still exist, particularly in the [N~{\sc ii}] maps and long slit spectra, as discussed in Section~\ref{sub:PV}. 

Confirming the conclusions of \citet{monteiro00}, from their three-dimensional modelling of NGC~3132, this work has indicated that a detailed model of a nebula cannot be verified just by comparison of the observed integrated spectrum with model predictions. In fact, in the case of NGC~3918 approximately the same spectrum can be obtained with a number of different geometries and density distributions. For this reason, three-dimensional models are necessary in order to allow a spatio-kinematical analysis to be carried out by comparing predicted images and position-velocity diagrams in several lines to available observational data. Mocassin provides all the tools needed for such simulations and for the visualization of the final results. 

\vspace{7mm}
\noindent
{\bf Acknowledgments}
We are most grateful to Dr. G. Mellema for providing the density distribution file used by C99 and in our spindle-like model~A of NGC3918. We thank the anonymous referee for useful comments. BE aknowledges support from PPARC Grant PPA/G/S/1997/00728 and the award of a University of London Jubber Studentship.

\bibliographystyle{mn2e}

\bibliography{references}

\vspace{5mm}

\noindent
{\bf Appendix A: Analytical description of spindle-like model~B} \\

We define the following {\it cut} functions:

\begin{eqnarray}
C_{01}(x,x_0,\Delta{x}) &=& 0.5 + \frac{1}{\pi}\,tan^{-1}\left(\frac{x-x_0}{\Delta{x}}\right) \nonumber \\
C_{10}(x,x_0,\Delta{x}) &=& 0.5 - \frac{1}{\pi}\,tan^{-1}\left(\frac{x-x_0}{\Delta{x}}\right) 
\end{eqnarray}

The functions above can be used to describe a smooth transition from 0 to 1 and from 1 to zero, respectively, about the value $x_0$, with a typical size for the transition of $\Delta\,x$.

The equatorial spherical shell can then be described as follows:

\begin{equation}
equat(r, \theta) = C_{01}(\theta,\theta_1,\Delta\theta){\cdot}C_{01}(r,r_{eq},\Delta{r_{eq}^{in}}){\cdot}C_{10}(r,r_{eq},\Delta{r_{eq}^{out}})
\label{eq:equator}
\end{equation}

where $r$ is the radial distance and $\theta$ the polar angle ($\theta=90^{\circ}$ at the equator and $\theta=0^{\circ}$ at the two poles). Values for $\theta_1$, $\Delta\,\theta$, $r_{eq}$, $\Delta{r_{eq}^{in}}$ and $\Delta{r_{eq}^{out}}$ are given in Table~\ref{tab:spindleparametersan}. 

We then define $r_e$, the radius of the ellipsoid, as $r_e~=~\sqrt{x^2+y^2+(e\cdot{z})^2}$, where $e$ is the eccentricity (given in Table~\ref{tab:spindleparametersan}), then the polar protrusions can be created using the following expression

\begin{equation}
pole(r, \theta) = C_{10}(\theta, \theta_1, \Delta\theta){\cdot}C_{01}(r_e,r_{po},\Delta{r_{po}^{in}}){\cdot}C_{10}(r_e,r_{po},\Delta{r_{po}^{out}})
\label{eq:pole}
\end{equation}

where $r_{po}$, $\Delta{r_{po}^{in}}$ and $\Delta{r_{po}^{out}}$ are also given in Table~\ref{tab:spindleparametersan}. Finally, the density distribution at any point in the grid can be obtained by combining Equations~\ref{eq:equator} and~\ref{eq:pole}, as follows

\begin{equation}
N_H(r, \theta) = n_1 {\times} (equator(r, \theta) + pole(r, \theta)).
\end{equation}

\vspace{5mm}

\noindent
{\bf Appendix B: The [Ne~{\sc v}] fine-structure lines } \\

\label{sub:nev}

The largest discrepancies between this work and that of C87 were obtained for the predictions of the infrared fine structure line fluxes of Ne$^{4+}$, namely  $[$Ne~{\sc v}$]$~14.3\,$\mu$m and $[$Ne~{\sc v}$]$~24.3\,$\mu$m, which disagree by factors of almost 10 and 5, respectively with C87's predictions and indeed with the observations. The discrepancy between the models is due to differences in the collision strengths used for these lines. The collision strengths for these transitions used by Mocassin are the calculations by \citet{lennon91}, based on the R-matrix method described by \citet{berrington87}. These collision strengths are much larger than the rates given by \citet{mendoza82} based on the results of \citet{saraph70}, which were used by C87 in their model. In 1987, however,  the collision strengths of \citet{aggarwal83} were already available for $[$Ne~{\sc v}$]$, and these were also much higher than those given by \citet{mendoza82}, since they took into account the large resonances existing at energies just above the threshold for excitation of the fine structure lines. C87 decided against using those results as they were not in agreement with the observed ratios of $[$Ne~{\sc v}$]~3426\,{\AA}/~14.3\,\mu$m. However, \citet{vanhoof00} later argued that the collision strengths calculated by \citet{lennon91}, which were similar to the \citet{aggarwal83} values, are reliable, and that the discrepancies found by C87 can be explained  by the inaccuracy of the $[$Ne~{\sc v}$]$~3426\,{\AA} flux that they adopted. Moreover, \citet{berrington01} in his contribution to the November 2000 Lexington workshop, discussed near-threshold resonances in the calculation for collision strengths for fine-structure transitions, taking [Ne~{\sc v}] as a case study. He compared the effective collision strengths obtained using different methods, showing that they all largely agree with the \citet{lennon91} results. During the same Lexington meeting, a new R-matrix calculation of [Ne~{\sc v}] was independently published by \citet{griffin00} who also obtained similar results. In the current work the decision to use the collision strengths of \citet{lennon91}, although producing results which are in disagreement with C87's observations, was finally taken, since the benchmarking carried out for the Lexington optically thick planetary nebula, which has a central star with effective temperature, $T_{eff}=150,000$\,K (similar to the central star of NGC~3918),  clearly showed that most photoionization codes (all but Sutherland's {\it Mappings}) use the new collision strengths \citep{ercolanoI}.

The likely cause of the large discrepancy between the predicted and observed [Ne~{\sc v}] fine-structure line fluxes is addressed in Section~\ref{sec:infrared} and in Appendix~C below.

\vspace{5mm}

\noindent
{\bf Appendix C: The ISO pointing error} \\

Most of the infrared line fluxes measurements reported in Table~\ref{tab:clegglines} are taken from \citet{bower01} (reference $d$ in Table~\ref{tab:clegglines}), measured from spectra taken with the {\it ISO} Short-Wavelength Spectrometer ({\it SWS}) between 1996 and 1998. The fluxes measured by \citet{bower01} for the infrared lines of NGC~3918 are systematically smaller, by approximately a factor of two, than those measured by \citet{pottasch86}, who used IRAS Low-Resolution Spectrometer (LRS) observations and are smaller by factors of 3 to 6 for most lines, apart from the [Ne~{\sc v}] fine structure lines, which are between 10-15 times smaller, compared to both Mocassin and C87 predictions. The reason for the [Ne~{\sc v}] fine structure line discrepancy between the Mocassin and C87 predictions has already been discussed above. This {\it ISO~SWS} discrepancy is also evident in the spindle-like models of NGC~3918 (see Section~\ref{sub:spindleresults}) and is believed to be due to the combined effects of incorrect object coordinates used for the {\it ISO} observations and the small size of the {\it ISO~SWS} aperture. The {\it ISO~LWS} fluxes for the [N~{\sc iii}]~57.3\, $\mu$m and [O~{\sc iii}]~51.8\, $\mu$m and 88.4\, $\mu$m lines are in good agreement with the model predictions. Although the same pointing error of 14~arcsec occured, the much larger LWS aperture of 70--80~arcsec minimised the effects on the observed fluxes. Corrections to the predicted {\it ISO~SWS} line fluxes are calculated and  discussed in detail in Section~\ref{sec:infrared}. 

The position given by the Strasbourg-CDS Simbad database for NGC 3918, in the J2000 FK5 reference frame, is $\alpha$(2000)~=~11~50~17.2 and $\alpha$(2000)~=~--57~10~53, while the position used for the {\it ISO~SWS} and {\it LWS} observations of NGC 3918 was that measured by \citet{milne76}, namely $\alpha$(1950)~=~11~47~50.1 and $\delta$(1950)~=~--56~54~10, equivalent to $\alpha$(2000)~=~11~50~18.9 and $\delta$(2000)~=~--57~10~51. This resulted in the SWS apertures being offset from the nebular centre by 1.8 arcsec north in declination and by 13.6 arcsec (=~1.68~sec) east in right ascension, giving a total offset of 13.8~arcsec from the centre of the nebula. The effects of this SWS aperture offset, together with the particular spacecraft roll angle at the time of the the observations (retrieved from the {\it ISO} archives) and the SWS subspectra beam profiles, are sufficient to largely explain the discrepancies initially found between the model predictions and the observations. The wavelength dependent aperture sizes of the {\it ISO~SWS} range from 14$''$$\times$20$''$ for 2.38$\mu$m\,$\leq\,\lambda\leq$\,12.0$\mu$m, 14$''$$\times$27$''$ for 12.0$\mu$m\,$\leq\,\lambda\leq$\,27.5$\mu$m, 20$''$$\times$27$''$ for 27.5$\mu$$m\,\leq\,\lambda\leq$\,29.0$\mu$m, to 20$''$$\times$33$''$ for 29.0$\mu$m\,$\leq\,\lambda\leq$\,45.2$\mu$m \citep{degraaw96}. Five of the six infrared fine-structure lines listed in Table~\ref{tab:infraredCorrections} were observed by the {\it ISO~SWS} through a 14$''$$\times$27$''$ aperture, while the [S~{\sc iv}]~$\lambda$10.5$\mu$m line was observed through a 14$''$$\times$20$''$ aperture. As mentioned above, the aperture dimensions are not the only factor to be taken into account, there is also the transmission efficiency profile across the spatial aperture, which, in some cases, is far from flat \citep{leech02}.

\end{document}